\documentstyle[12pt,epsf,colordvi]{article}
\newcommand{\et}{{\em et al.}}

\newcommand{\de}{deuteron}

\newcommand{\fcz}{$F_{C0}(q)$}
\newcommand{\fmo}{$F_{M1}(q)$}
\newcommand{\fct}{$F_{C2}(q)$}
\newcommand{\aq}{$A(q)$}
\newcommand{\bq}{$B(q)$}

\newcommand{\hef}{$^4 \hspace*{-0.5mm} He$}
\newcommand{\het}{$^3 \hspace*{-0.5mm} He$}
\newcommand{\hyt}{$^3 \hspace*{-0.5mm} H$}
\newcommand{\gen}{$G_{en}$}
\newcommand{\gep}{$G_{ep}$}
\newcommand{\gmn}{$G_{mn}$}
\newcommand{\gmp}{$G_{mp}$}
\newcommand{\ba}{\begin{eqnarray*}}
\newcommand{\ea}{\end{eqnarray*}}
\begin{document}           % End of preamble and beginning of text.
\LARGE
\begin{center}
Elastic electron scattering from light nuclei \\[1cm] 
\Large
 Ingo Sick \\[3mm]
Departement f{\"u}r Physik und Astronomie, \\ 
 Universit{\"a}t Basel, Basel, Switzerland \\[2cm]
\end{center} 
\normalsize
\begin{abstract}
The charge and magnetic form factors of light nuclei, mainly for mass 
number A$\leq$4, provide a sensitive test of our understanding of 
nuclei. A number of ''exact'' calculations of the wave functions starting
from the nucleon-nucleon interaction are available. The treatment of
two-body effects needed in the calculation of the electromagnetic form factors
has made significant progress. Many electron scattering experiments have 
provided an extensive data base from which the various (mainly elastic) form
factors can be extracted. This review discusses the data and the determination
 of the form factors, and compares them to the results of theory.
\end{abstract}

%
%\tableofcontents
%\newpage
%\setlength{\parskip}{5mm}
%\setlength{\parindent}{0cm}
%
\section{Introduction}
Few-body nuclei have long attracted particular attention in nuclear physics. 
This essentially is due to the fact that for these nuclei non-relativistic
and sometimes relativistic calculations
of the wave function with high accuracy starting from the nucleon-nucleon (N-N) interaction can 
be performed.  For heavier nuclei models, or 
approaches with various approximations, have to be employed as a consequence
of the complexity of many-body systems. In this case one
largely looses the connection to the underlying, microscopic, physics.

For these light nuclei one then can best test our understanding 
of nuclei. Can they be described as essentially non-relativistic systems 
of nucleons interacting via the known N-N force, or do we have to go 
beyond this picture which has become known as the ''standard model of nuclear 
physics''? To which degree are nuclei relativistic systems? Do we need to 
explicitly allow for  other constituents like excited nucleons, pions,..? 
Does the composite (quark) structure of the nucleon play an important role? 

The few-body nuclei provide the possibility to study this rather wide spectrum
of questions. While the \de\ is a very dilute system, the \hef\ nucleus is 
the nucleus with largest density. In contrast to heavy nuclei, these nuclei
have large gradients of the density which also can play an interesting role. 
 The A=3 nuclei offer the possibility to 
study near-identical systems that differ only by the exchange of the role of 
protons and neutrons.

Significant 
progress has been made during the last decade in quantitatively computing the
wave function of light nuclei. For many observables one today can be sure
that differences in the calculated results reflect genuine differences in
the physics included, and not simply approximations that needed to be made in
the course of the calculation. A good part of these advances obviously is
due to the vastly increased computational means that have become available.  

Electron scattering provides an excellent tool for a detailed check of theoretical
calculations of the wave functions of few-body nuclei, a test  which often 
is more
informative than what can be learned from integral observables such
as energies of states or moments. Provided the experimental $(e,e)$ data go to 
large
(four) momentum transfer $q$, the spatial structure of the one-body densities can 
(in Impulse Approximation IA) be studied with good spatial resolution, of the order of $1.5/q$. 

 As a consequence of
the great interest in few-body form factors, many experiments on few-body
nuclei have been performed over the last decades despite the experimental 
challenge they represent. We today have a rather
complete set of data to which one can compare. An important aspect of the 
recent progress in $(e,e)$ experiments is the ability to measure polarization
observables in the region of $q$ where they can help to discriminate between
different theoretical predictions. Much of this progress has been made 
possible by the modern  high-energy electron accelerators with high duty
cycle such as MAMI or JLAB and the development of polarized sources, targets and polarimeters.

The form factors of light nuclei are of particular interest for another
reason: Selected form factors, {\em e.g.} the isovector magnetic ones, 
exhibit a special
sensitivity to exchange current contributions. For heavier nuclei the 
(magnetic) form factors show diffraction features at much lower momentum
transfer already  due to the structure of the wave function, and the 
contribution of two-body  currents is much more difficult to isolate.

 There is quite a vast literature on the few-body form factors --- too
vast for an exhaustive list of references without omitting significant work.
These reviews have different areas of emphasis. The present review ---
written by an experimentalist --- gives more than most others attention
to data and ignores much of the detailed formalism of
prime interest to more theory-oriented reviews. For general reviews on electron 
scattering see \cite{DeForest66,Donnelly84,Frois91}. 

In this review, we first discuss the ingredients necessary for the calculation
 of the form factors. We then discuss for A=2,3,4,6,16 the data available, the
determination of the optimal set of 
experimental form factors and their comparison to theoretical calculations. 
For the A=2 case, we include the transition form factor to the $^1$S state. 

\section{Ingredients}
\subsection{Nucleon form factors \label{nucleon}}
The charge- and magnetic form factors of nuclei are calculated from the
wave function of the nucleons, obtained in most cases from a solution of
the Schr\"odinger or Bethe-Salpeter equation for pointlike nucleons. 
In order to obtain the  distribution of electric
charge and magnetization as measured by $(e,e)$, the nucleon electromagnetic 
structure must be folded in.
This can be achieved using the proton- and neutron electric and magnetic Sachs 
form  factors $G_e$ and $G_m$ \cite{Sachs62} known from elastic 
electron-nucleon scattering.  In the one-body charge- and magnetic form
factors, the 
(appropriate combination of) 
nucleon form factors come in as a multiplicative factor when going from the 
body form factor (calculated from the wave function) to the nuclear electromagnetic 
form factor. The tacit
assumption is that the binding of the nucleon in the nucleus does not 
change its electromagnetic structure. One also assumes that the form 
factors for off-shell nucleons are the same as the ones for the free
nucleon. For a reliable prediction of
nuclear form factors we then need to know the nucleon form factors with adequate
precision over the range of $q$ of interest for nuclear form factors.

The different nucleon form factors are known from $(e,e)$ experiments to a 
very different level of accuracy:

The {\em proton magnetic form factor $G_{mp}(q)$}  is well known from many 
experiments on 
electron-proton scattering. The data reach momentum transfers that go much beyond
what is needed for a calculation of {\em nuclear} form factors (typically
8 $fm^{-1}$). Several parameterizations
providing a very good fit to the data are available from Hoehler \et\ and 
Mergell \et\ \cite{Hoehler76,Mergel96};
other parameterizations can be found in \cite{Iachello73,Gari86}, modern 
references are given in \cite{Mergel96}.

The {\em proton electric form factor} $G_{ep}(q)$  is well known at low and medium 
momentum transfers. At the higher momentum transfers, the data up to recently were
somewhat contradictory. At high $q$, the magnetic contribution to the {e-p}
cross section increasingly dominates, and a determination of $G_{ep}(q)$
via the usual Rosenbluth separation became more and more difficult with increasing
$q$. Small systematic errors in the cross sections, which depend on a weighted sum 
of $G_{ep}^2(q)$ and $G_{mp}^2(q)$,  started to play a big role.

The situation for $G_{ep}$ has recently improved with  experiments that 
exploit {\em polarization}  observables; when measuring the polarization of the 
recoil proton in scattering of polarized electrons from unpolarized hydrogen, 
the {\em ratio} 
$G_{ep}(q)/G_{mp}(q)$ becomes accessible \cite{Jones00,Popischil00,Milbrath98}.
This ratio is now well 
measured up to $q \sim 9 fm^{-1}$. This covers the $q$-range of interest
for {\em nuclear} form factors. The ratio $\mu_p G_{ep}/G_{mp}$ does deviate 
considerably from the naive expectation of 1 as given by the dipole formula. 
As it turns out, however, the new data on $G_{ep}(q)$
 largely agree with the  parameterization of Hoehler \cite{Hoehler76} 
which represents an often used standard for e-p scattering. This is shown 
\begin{figure}[hbt]
\vspace*{1cm}
%Figur mit topp/sideways hergestellt, modif boundingBox: 66 206 516 556
\centerline{\mbox{\epsfysize=7.5cm \epsffile{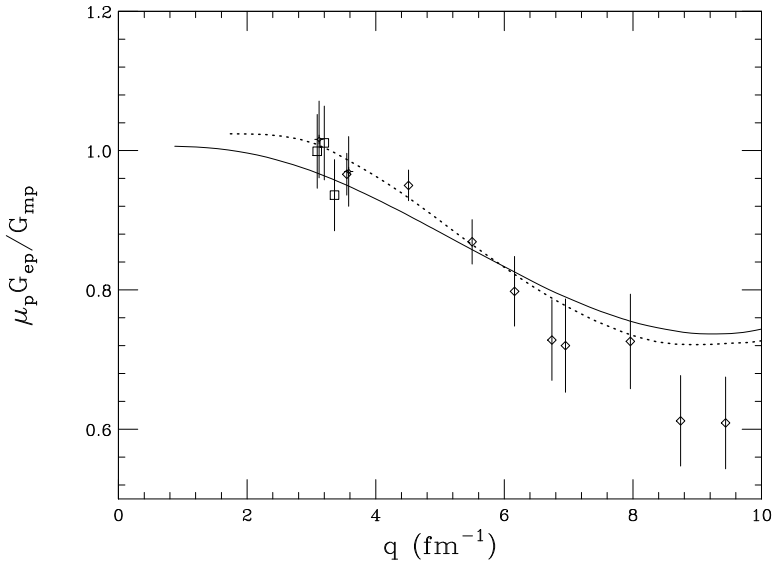}}}
\begin{center} \parbox{14cm}{\caption{
\label{gepgmp}
Ratio of $\mu_p G_{ep}/G_{mp}$ \protect{\cite{Jones00,Popischil00,Milbrath98}}
together with the standard Hoehler parameterization (solid line, 
\protect{\cite{Hoehler76}}) and the more
recent fit of Mergell (dotted \protect{\cite{Mergel96}}).
} }  \end{center}
\end{figure}
in fig.~\ref{gepgmp} which displays the recent data together with the Hoehler
parameterization. Certainly up to $q$ = 8 $fm^{-1}$, where reasonably accurate 
nuclear charge form factors are available, the proton charge form factor 
$G_{ep}$, even when used in terms of the venerable Hoehler parameterization,
 does not introduce a significant uncertainty. 

The {\em neutron magnetic form factor $G_{mn}(q)$}, which is important 
when  calculating nuclear magnetic form factors, also is a quantity
that is rather difficult to determine experimentally. Due to the absence
of a free neutron target, $G_{mn}$ has to be extracted from experiments
on quasi-elastic electron scattering on the deuteron (or more complicated 
nuclei).  In the case of inclusive $(e,e')$ quasi-elastic scattering, the subtraction of the 
dominant proton contribution, and the corrections for deviations of the 
reaction mechanism from PWIA, lead to large systematic uncertainties.
As a consequence, the data showed a large scatter \cite{Jourdan96}.

Clean measurements, which do not depend unduly on the theory needed 
to interpret quasi-elastic scattering and to eliminate the dominating 
contribution from the proton, involve {\em coincidence} experiments 
of the type $d(e,e'n)$ where scattering
from the neutron in the deuteron can unambiguously be identified (save
for corrections due to non-IA processes). Such 
experiments are today feasible with the modern 
high-duty cycle electron beam facilities, and we in the following will, as far as 
possible, only consider this type of data. 

Several such  experiments have been carried out recently
\cite{Anklin94,Anklin98,Jourdan00} at NIKHEF and MAMI. The
determination of the absolute efficiency of the detector used to identify the 
recoil neutron was measured at the proton-beam facility PSI which allowed 
one to produce intense high-energy {\em tagged} neutron beam of variable energy.
 At very low $q$, 
polarization observables can also be exploited as the theoretical understanding
of the \het\ nucleus and the reaction mechanism has progressed to the point 
where polarized \het\ 
can serve as a polarized neutron target. In this case $G_{mn}$ can
be obtained from the asymmetry of inclusive, single-arm quasielastic scattering 
\cite{Xu00,Gao94}. 
%In fig. \ref{gmn} we show the data that are presently available from coincidence
%or polarization experiments.

 %
\begin{figure}[hbt]
\vspace*{1cm}
%Figur mit topp/sideways hergestellt, modif boundingBox: 66 206 516 556
\centerline{\mbox{\epsfysize=7cm \epsffile{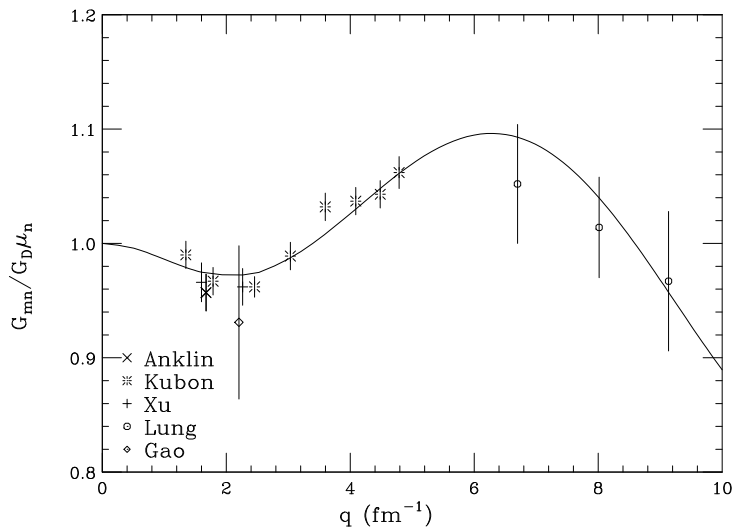}}}
\begin{center} \parbox{14cm}{\caption{
\label{gmn}
Data on the neutron magnetic form factor $G_{mn}$ divided by the dipole expression 
\mbox{ $G_D = (1+q^2/0.710(GeV/c)^2)^{-2}$.}
The curve represents a fit with a continued-fraction parameterization  
\protect{\cite{Jourdan00}}.
} }  \end{center}
\end{figure}

The status of our experimental knowledge  of $G_{mn}$ from coincidence
measurements  was clouded by 
two recent $(e,e'n)$ experiments that claimed to provide accurate data 
\cite{Bruins95,Markowitz93} which, however,
greatly differed from the ones shown in fig. \ref{gmn}.   Both
experiments used $d(e,e'n)$ to measure $G_{mn}$. Both relied, however, on 
a reaction with a {\em three}-body final state ($d(e,p)e'n$ or $p(e,\pi)e'n$)
to produce the ''tagged neutron'' which was needed to calibrate the neutron
detector efficiency. Use of a {\em three}-body final state for tagging 
--- without observation of the third particle, the scattered electron ---  
is not
possible as a matter of principle, and has been quantitatively shown to produce
erroneous results \cite{Jourdan97b,Kubon99}. The  respective results  should be ignored.

Fig.~\ref{gmn} shows the status of today's data on $G_{mn}$ 
\cite{Jourdan00}. Up to momentum 
transfers of $q \sim 5 fm^{-1}$ the neutron magnetic form factor is known
with excellent precision from coincidence experiments 
\cite{Anklin94,Kubon99,Jourdan00}, and these results have been confirmed by 
experiments exploiting polarization observables \cite{Gao94,Xu00}. At the 
larger transfers
one still has to rely on single-arm measurements \cite{Lung93}. At large $q$
the exploitation of $(e,e')$ actually gets a bit easier than at low $q$ as the 
\gep \ contribution becomes small, so that the $(e,e')$ cross section then 
depends mainly on $(G_{mp}^2+G_{mn}^2)$, with comparable contributions from 
\gmp \ and \gmn.   On the whole, 
$G_{mn}$ is now known with adequate precision over the entire $q$-range 
needed for the calculation of nuclear magnetic form factors.

Jourdan \et\ \cite{Jourdan00} have parameterized the modern data for $G_{mn}$  using a continued
fraction expansion
\begin{eqnarray*}
G_{mn}= 1/(1 + q^2 \cdot c_1 / (1 + q^2 \cdot c_2 / (1 + ...
\end{eqnarray*}
with the constants $c_1,... = 3.260, -0.272, 0.012, -2.5, 2.5$ and
$q$ in $GeV/c$. This fit also for the first time allows one to determine to better
than 10\% the magnetic radius of the neutron. It amounts  to $0.873 \pm 0.011 fm$
(systematic errors included).

The {\em neutron electric form factor $G_{en}$} is the quantity that is the poorest
known among the nucleon structure functions. Only the slope 
$dG_{en}(q^2)/d(q^2)$ of $G_{en}$ at 
$q=0$ is well known from neutron-electron scattering \cite{Kopecki97}.
 The lack of knowledge on $G_{en}$
can occasionally really hamper the understanding of nuclear form factors. It 
does not always do it, as the smallness of $G_{en}$,  which makes its experimental
determination so difficult, at the same time  leads often to a small contribution of 
the neutron to nuclear form factors. However, in interference response
functions $G_{en}$ can play a much more important role. 

\begin{figure}[hbt]
\vspace*{1cm}
%Figur mit topp/sideways hergestellt, modif boundingBox: 66 206 516 556
\centerline{\mbox{\epsfysize=8cm \epsffile{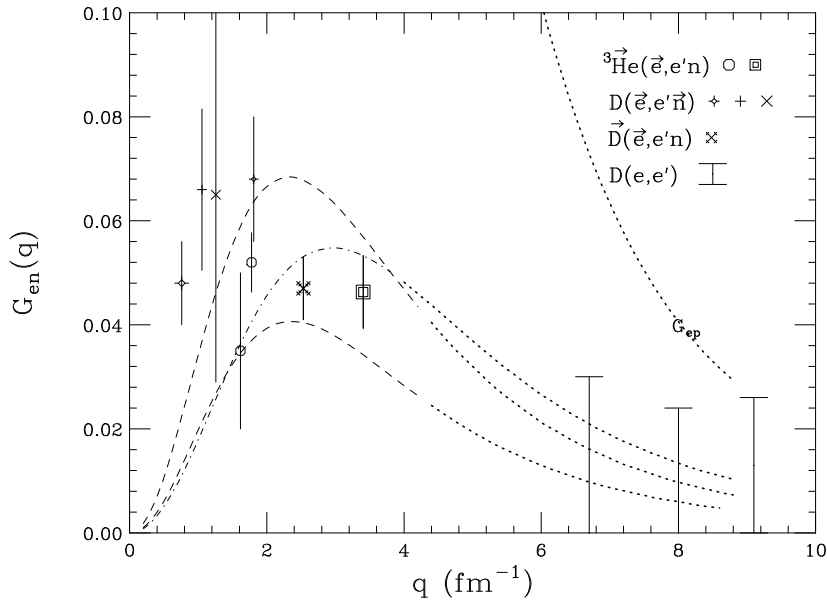}}}
\begin{center} \parbox{14cm}{\caption{
\label{gen}
 $G_{en}$ from  double-polarization experiments of type $\vec{d}(\vec{e},e'n)$,
$\vec{^3 \hspace*{-0.5mm} He}(\vec{e},e'n)$
or $d(\vec{e},e'\vec{n})$ (for references see text). 
The parameterizations by Galster \et\ and Platchkov 
\et\ (for the deuteron $A(q)$ calculated with Paris and AV14 potentials,
respectively) are given by  dash-dotted  and dashed lines. The extrapolation of
the parameterizations to large $q$ are shown as dotted lines. The 
Hoehler \gep\ is also indicated as a  dotted line. Also shown are some upper limits
of \gen\ from $d(e,e')$ \protect{\cite{Lung93}} at large $q$.
} }  \end{center}
\end{figure}

The importance of \gen\ for nuclear form factors  can be appreciated by 
considering the charge form
factor for an $N=Z$ nucleus. Nucleon finite size leads basically to a factor
 $(G_{en}(q)+G_{ep}(q))$ multiplying the body form factor calculated from the 
ground state wave function. At $q=4 fm^{-1}$ the ratio \gen / \gep \ amounts to 
0.22 already \cite{Hoehler76},  at larger $q$ it could be much larger if 
$G_{en}(q)$ continues to fall as slowly as present data for $q < 4 fm^{-1}$ 
seem to imply. Parameterizations such as the one of Gari and Kr\"umpelmann
\cite{Gari92} have predicted large ratios at large $q$, {\em e.g.}
 $G_{en}/G_{ep}$=1.4 at 
$q=9 fm^{-1}$. Experiment \cite{Lung93} only provides an upper limit of 
$\sim 0.8$ (see fig.~\ref{gen}). The uncertainty in \gen ~ is of course 
particularly detrimental for \hyt\ \cite{Brandenburg75a} where the neutron
charge form factor comes in with the double weight.

The determination of $G_{en}$ is difficult as there is no free neutron target; 
in addition, the measured electron-neutron cross section is largely 
dominated by the contribution from the magnetic form factor $G_{mn}$.  
In the past, $G_{en}$ therefore
was mainly determined from electron-deuteron elastic charge scattering which
eliminates the magnetic contribution. In this approach, the
removal of the deuteron structure (deuteron body form factor), the subtraction
of the proton contribution and 
non-IA contributions to the reaction mechanism lead to large systematic 
uncertainties, however. The most reliable determinations were probably those of Platchkov 
\et\ \cite{Platchkov90} and Galster \et\ \cite{Galster71} which, using various
calculations of the deuteron wave function and non-IA contributions, provided
values of \gen\ up to 4 $fm^{-1}$. The resulting parameterizations have considerable
systematic uncertainties due to both the deuteron input and the different 
theoretical results for the non-IA contributions 
\cite{Plessas00,Buchmann96,Carlson98,Arenhoevel00}.

The neutron electric form factor can, in 
principle, also be determined via quasi-elastic electron-deuteron scattering,
subtraction of the proton contribution and longitudinal/transverse (L/T)
 separation. The resulting 
uncertainties are substantial, however; at low $q$ the \gen\ contribution is comparable
to the experimental systematic uncertainties, at high $q$ (where it seems that
the neutron and proton contributions are more similar in size) the overlap of
quasi-elastic and Delta peaks makes difficulties.  In fig.~\ref{gen} we only 
show the results of Lung \et\ \cite{Lung93} at very large $q$.  

Today, $G_{en}$ can be determined via polarization observables in $(e,e'n)$ 
experiments. The detection of the knocked out neutron --- feasible at  
CW electron beam facilities --- largely removes the contribution of the proton. 
The use of polarized electrons and polarized target (or, alternatively, 
measurement of the polarization of the recoil neutron)
allows one to measure an interference term $G_{en} \cdot G_{mn}$ which,
once $G_{mn}$ is known, provides a measurement of $G_{en}$. 

Given the experimental difficulties of polarized targets or recoil 
polarimeters, only a few results from $\vec{d}(\vec{e},e'n)$, 
$\vec{^3 \hspace*{-0.5mm} He}(\vec{e},e'n)$ and $d(\vec{e},e'\vec{n})$ 
are available 
\cite{Meyerhoff94a}-
\nocite{Becker97,Becker99,Rohe99,Herberg99,Eden94,Ostrick99,Passchier99}
\cite{Ahmidouch01}  and
the data still have fairly large uncertainties.  Fortunately, however, these
uncertainties are mainly statistical in nature, and thus can be improved upon
with the progress in polarized target and polarimeter technology. 
The presently available set of $G_{en}$ data from double-polarization
experiments is displayed in fig. \ref{gen}. Clearly, the extent of the data 
still is far less than what is desirable as an input to the calculation of nuclear
form factors of light nuclei. 
\subsection{N-N potentials \label{NN}}
An important ingredient to the calculations of the few-body wave functions
is the nucleon-nucleon (N-N) potential employed. As compared to other 
interactions occurring in physics, the N-N interaction has an unusually rich and 
complicated structure. It depends on the spin and isospin of the nucleons and
 their relative  momentum. Only the long-range part, dominated by one-pion exchange,
can easily be calculated. The construction of N-N 
potentials is a field that has a long tradition and covers many aspects which we 
here cannot possibly do justice to; so we refer the reader to the 
original literature (for a review see {\em e.g.} papers by Machleidt 
\cite{Machleidt89,Machleidt94}). Here, we only can mention a few aspects 
that are relevant for the specific few-body calculations discussed.

The various N-N potentials are all fit to the N-N (p-p and p-n) scattering 
data and the deuteron binding energy. Typically, the scattering data up to 
energies of $\sim$350$MeV$ are employed. At energies higher than 350$MeV$ 
inelasticities due to $\pi$-production {\em etc.} lead to a much more involved
description of N-N scattering.   
While earlier potentials \cite{Reid68,Cottingham73,Nagels78,Wiringa84} 
were generally  fitted to the full set of data available, 
more recent fits \cite{Wiringa95,Stoks94,Machleidt96} often have used a 
restricted data set \cite{Stoks93} (see below).

Most often, the nuclear force is parameterized in terms of a sum of One Boson
 Exchanges 
(OBE). These OBE potentials for the longest range  component mostly use the local version
of the one-pion exchange potential. For the intermediate and short range the 
potentials are parameterized in different ways. The Argonne V18 potential 
developed by Wiringa \et\ 
\cite{Wiringa95} has a one-pion exchange piece for the longest range, 
but uses essentially phenomenological parameterizations for the shorter ranges.
The Nijmegen 
potential of Stoks \et\ \cite{Stoks94} uses empirical masses of existing mesons and 
meson distributions. All potentials are regularized at short distances by 
{\em e.g.}
exponential cut-off factors. These cut-off form factors simulate the effect
of the exchange of heavier mesons which have not been taken into account.
Modern versions of these OBE potentials are
 'charge dependent', {\em i.e.} they provide different parameterizations 
for p-p, n-p and n-n.

The CD-Bonn OBE potential differs mainly in the use of a non-local one-pion 
exchange term, which is derived from a relativistic Lagrangian for 
meson-nucleon coupling. This leads to a non-local force with a tensor 
component which is substantially weaker {\em off-shell}. As a consequence, 
this potential predicts a deuteron D-state probability which is $\sim$0.8\% 
smaller than the one predicted by other potentials (the other deuteron
static properties are very similar). Also, the A=3 binding energy for the Bonn
potential is 0.4--0.5$MeV$ larger. 

Several potentials include more than OBE. The Paris 
\cite{Lacombe80} and full-Bonn \cite{Machleidt87} potentials include
 2$\pi$-exchange (or phenomenological fits to it), and
thus largely can do away with the fictitious $\sigma$-meson exchange. The two-pion
contribution is calculated using time-ordered perturbation theory, or via a dispersion-relation 
analysis of $\pi-N$ scattering. Terms like 
the $\pi\rho$ and $\pi\omega$ exchange have been included as well 
\cite{Machleidt87}. 
The full-Bonn potential is more difficult to use in nuclear structure 
calculations due to its energy dependence, and can not consistently be used 
for cases of $>$2 nucleons; as an alternative, the 
energy-independent (but still nonlocal) Bonn-B potential has been 
derived by Machleidt \et\ \cite{Machleidt89}.   

The modern OBE potentials fitted to the cleansed N-N scattering data base
 (see below), the V18, CD-Bonn and 
Nijm-II potentials, achieve a similarly good fit. The major difference 
between them shows up in the calculations of nuclear properties where 
off--shell properties of the force do matter. In particular,  a force 
with a weaker off-shell tensor component 
leads to a stronger binding of the A$\geq$3 systems.

In the past, the potentials were usually fit to the full data base on N-N 
scattering. More recently, the Nijmegen group has performed a fit to the 
data employing different 
energy-dependent potentials for different partial waves. The $\chi^2$ of 
individual data points was then used as a criterion to keep the data, or 
remove them from the data set. A large fraction ($\sim$30\%) was removed. 
The remaining, cleansed,  data base  can be fit with very good $\chi^2$, which makes it 
attractive to  those groups constructing N-N potentials and therefore has been
used to fit the modern potentials mentioned above.

This procedure of removing data is fine as long as the parameterized energy 
dependence used in the data elimination procedure is correct, and as long as 
enough degrees of freedom have been allowed for. That this is the case is by 
no means obvious. It can be
checked by comparing the resulting phases to the ones obtained from 
single-energy phase shift analyses of the full data set. 
If there are significant differences
(as judged by the error bar resulting from the single-energy analyses) then
one must assume that the multi-energy analysis used to reject data was not 
general enough and introduced a bias. When looking at the comparison of 
\begin{figure}[bht]
%Figur mit topp/sideways hergestellt, modif boundingBox: 66 206 516 556
\centerline{\mbox{\epsfysize=7cm \epsffile{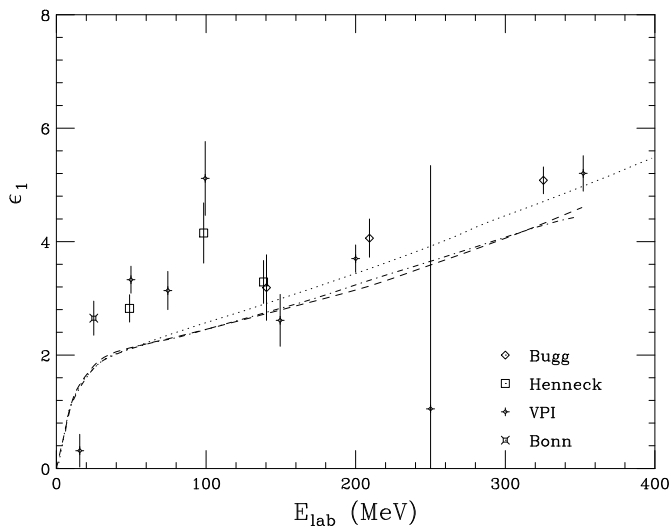}}}
\begin{center} \parbox{14cm}{\vspace*{-0.5cm} \caption{
\label{epsilon} 
S-D mixing parameter $\epsilon_1$ resulting from single energy phase-shift analyses (data 
points, \protect{\cite{Bugg92,Henneck93,Arndt94,Ockenfels91}}), 
the multi-energy fit used to cleanse the data set (dashed, 
\protect{\cite{Stoks93}}), the phases from the CD-Bonn potential (dotdash
 \protect{\cite{Machleidt96}}) and the $V18$ 
potential (dotted, \protect{\cite{Wiringa95}}) fit to this set. 
} }  \end{center}
\vspace*{-0.4cm} \end{figure}
phases resulting from single- and multi-energy analyses,
 one finds worrisome systematic differences in some phases. This is true in 
particular for the $\epsilon_1$ phase (see fig.~\ref{epsilon}) \cite{Wiringa95}
 which affects the strength 
of the tensor force and is responsible for the S-D 
transitions that play an important role for few-body systems 
(deuteron electrodisintegration, A=3 magnetic form factors).

When using OBE potentials in calculations of the wave function of light 
nuclei with A$>$2, one generally finds an underbinding, which amounts to 
0.5--1 $MeV$ for A=3 and several $MeV$ for A=4. This indicates that it is also 
necessary to consider the role of the three-body force (3BF). Such a 3BF results in
particular from the internal degrees of freedom  of the nucleon which have been 
frozen out when describing nuclei in terms of interacting nucleons only. 

The longest-range 3BF  involves the intermediate excitation of the 
nucleon to the $\Delta$, and two pions  exchanged with two other nucleons. 
This interaction, originally  given by Fujita and Miyazawa \cite{Fujita57}, 
is attractive in light nuclei. When introducing the $\Delta$ degrees of freedom
explicitly in the wave function, as done {\em e.g.} by P. Sauer and collaborators
\cite{Sauer92}, one finds that the attractive contribution is significantly 
reduced by dispersive effects (energy dependence of effective N-N interaction). 
Other sources for a 3BF need to be considered as well.

Often used are variants of the Tucson-Melbourne force of Coon and 
collaborators \cite{Coon79,Coon93,Stadler95} 
which are  based on 2$\pi$ and $\pi \rho$ exchange. At short range these forces are 
phenomenologically adjusted to produce {\em e.g.} the correct binding energy 
of the A=3 systems and nuclear matter density. Alternatively, the Brazil 3BF is
sometimes used \cite{Robilotta85}.

A somewhat different approach is followed by the Argonne/Urbana group 
\cite{Carlson83,Pudliner95} who also
uses at long range a two-pion exchange shape, but at short range uses a 
purely phenomenological spin- and isospin-independent repulsive term, with
parameters that are again adjusted to fit the binding energies of light 
nuclei.

For the time being, these three body forces all contain adjustable parameters. The calculation 
of the 3BF from first principles is still too difficult, as this force not only
has to account for non-nucleonic degrees of freedom integrated out when describing 
nuclei in terms of nucleonic constituents; it to some degree has to 
account as well for
relativistic aspects which are suppressed in the standard, non-relativistic, 
calculations. 
\subsection{Wave functions \label{wf} }
In order to predict the various charge- and magnetic form factors as measured
by electron scattering experiments, accurate wave functions of the nuclear 
ground states are
the basic ingredient. These wave functions can be calculated employing a 
number of different techniques, some of which we will address in other
chapters of this review. In this section, we only want to discuss in somewhat 
more detail the approaches that are used in  the ''standard model of 
nuclear physics''. Often these approaches are non-relativistic, but for the 
deuteron relativistic calculations are available as well. The non-relativistic
approaches calculate the nuclear  wave function 
as a solution of the Schr\"odinger equation for nucleons
interacting via one of
the modern N-N potentials which reproduce the N-N scattering data with a good 
$\chi^2$. Calculations performed within this framework have
reached a high state of perfection, and they have been applied to all of the 
light nuclei.

In this picture of the nucleus, one has already made significant 
simplifications. Rather than allowing for all sorts of constituents --- 
nucleons, excited nucleons, mesons, ... --- one has projected onto a purely
nucleonic basis.   

For the \de\ the wave function depends only on the 
n -- p relative distance (relative momentum). The solution of the Schr\"odinger equation 
for the corresponding one-body  problem is fairly straightforward. Technical complications 
arise with the 
momentum-dependent N-N  interactions such as the full Bonn N-N potential for
calculations in coordinate space.

For the three-body system, a number of different approaches to compute the
wave function within the above-mentioned frame work have been developed. Over 
the years, some of these approaches  to calculate ground-state 
wave functions have been perfected to a high degree of 
reliability \cite{Harper72}-\nocite{Brandenburg74,Laverne73}\cite{Gloeckle82}.

In the Faddeev approach the Schr\"odinger equation is reformulated in terms of
three coupled equations separated according to the pair of nucleons that have
interacted last. 
The sum of the three equations reproduces the Schr\"odinger equation.
While the separation may appear to complicate the description, it actually 
corresponds to a simplification as, in the limit of three identical particles,
the 3 equations differ only by a permutation of indices. The variables typically
used (for the case of coordinate space calculations) are the relative distance
vector {\bf x} of two nucleons and the relative radius vector {\bf y} of
the third nucleon relative to the center of mass of the pair (Jacobi 
coordinates), or the corresponding momenta. 

The spin- and isospin degrees of freedom of the three nucleons have to be 
treated explicitly such to satisfy the 
Pauli principle. Often the wave function components are decomposed according
to the relative angular momentum of pairs of 
particles, or one particle relative to the CM of the other two. Truncations
of the space in terms of the angular momenta allowed for often are necessary 
(and can be shown to be justified).

Accurate calculations today typically keep all channels where the interacting
N-N pair is in partial waves with $j \leq 4$, which requires 34 channels.
For scattering states at high nucleon momenta, more channels are included 
in accordance with the larger total angular momenta of interest.
These Faddeev calculations can be done both in coordinate- or momentum space. 
The former has the advantage that the (for many observables not so important) 
Coulomb interaction can be included more easily.

In realistic calculations of the A=3 ground state wave function, one wants to 
include the effect of a three body force (3BF). This force is more
short-ranged than the $\pi$-exchange force, and gives typically a 2\% contribution 
to the potential energy. 

 For the momentum-space approaches, it has been shown by the group of 
Gl\"ockle to be 
practical  also to solve for the lower energies ($<$100$MeV$, say) the Schr\"odinger equation for continuum states
\cite{Gloeckle96}, a 
great asset when one wants to access the rich physics information contained in
the two- and three-body break-up channels.

The three-body Faddeev equations actually can be generalized to four
nucleons. The corresponding Faddeev-Yakubovsky equations \cite{Yakubovsky67}
still can be
solved although the number of channels required for an accurate description of the
observables grows very quickly with the number of
nucleons.

During the last years, the Correlated Hyperspherical Harmonics (CHH) approach 
also has successfully been used to describe the A=3,4 bound states as well as
the two-body continuum channels. The hyperspherical approach involves 
rewriting the Faddeev equations in terms of new coordinates which use
(for the case of coordinate space representation) the hyperspherical 
radius which corresponds to the quadratic sum of the usual Faddeev radial 
variables  {\bf x} and  {\bf y}, and the angle between {\bf x} and {\bf y}. This hyperspherical
approach has long been used, without too much success. Only with the 
introduction in the wave function of explicit N-N correlation operators, with 
parameters  determined in a variational calculation, has this approach become 
a successful one. These correlation operators account for 
the strong 
state-dependent N-N correlations induced by the N-N interaction at small N-N 
distances. Explicitly allowing for these correlations, as done {\em e.g.} by
Kievsky \et\ \cite{Kievsky94},   greatly accelerates
 the convergence of the calculation in terms 
of the number of hyperspherical harmonics that need to be included.

The most recent addition to the arsenal of methods  are Variational 
Monte-Carlo (VMC) calculations. 
The VMC calculations \cite{Lomnitz81,Carlson83,Schiavilla86}
start with a trial wave function with a Jastrow core that
 includes single-particle orbitals LS-coupled to the desired J,M values. Pair 
and triplet spatial correlations are introduced from the very beginning. 
The Jastrow core is then acted on by products of two-body spin-, isospin- tensor-
and spin-orbit correlation operators. Three-body correlation operators are 
included as well according to the 3BF used. The correlation operators 
allowed for thus contain the same operators as occur in the N-N interaction 
employed. The wave function is diagonalized 
in a basis of different Jastrow spatial symmetry components to project
out the lowest state. 

The VMC trial wave functions have typically 20-30 parameters. Using the
Metropolis algorithm \cite{Metropolis53}, the parameters are determined by 
minimizing the energy.
Such calculations today are feasible up to 
A=10, and produce ground state energies that are already quite accurate 
($\sim$ 1 $MeV$).

In the Greens Function Monte Carlo (GFMC) approach used by
Carlson \et\ and Pudliner \et\ \cite{Carlson87,Pudliner95},
 one starts with a trial wave function (often the VMC wave function) 
and operates on it with the imaginary time operator $e^{-(H'-E_0)\tau}$, where 
$H'$ is a simplified Hamiltonian and $E_0$ is an estimate of the eigen value, 
with $\tau$ being the imaginary time. The excited-state components of the 
trial wave function  will then be damped out for large $\tau$, leaving the
 exact lowest eigen state with the desired quantum numbers. The expectation 
value of $H$ is computed for the sequence of different 'times' to verify the 
convergence. Differences between the simplified $H'$ and the true $H$ 
are treated perturbatively.

These GFMC calculations converge to the essentially exact wave function. 
Difficulties due to the ''sign problem'', which limited the times $\tau$
one could go to,  recently have largely been overcome (see {\em e.g.} 
\cite{Pudliner97}).

\subsection{Two-body terms \label{MEC}}
The work on electromagnetic observables of few-body systems has shown that two-body
terms of the nuclear electromagnetic operators give important contributions to the 
observables. The explanation of the longstanding discrepancy between 
theory and experiment of the $n+p$ radiative capture cross section 
\cite{Riska72,Cox65} and the deviation of the A=3 magnetic moments from 
theory \cite{Villars47}  
provided the first concrete evidence. These exchange terms influence both the 
electromagnetic {\em current} and {\em charge} operators. The theoretical treatment of these
two-body terms has made great progress during the past 20 years, as documented
in several review papers \cite{Riska89,Carlson98}.

The discussion below will start from the assumption that the
nuclear system is basically non-relativistic, and that relativistic effects can be 
taken care of by the lowest-order expansion in $v/c$ (for calculations in a relativistic
framework see sect. \ref{deuttheo}).  Calculations of light-nucleus form
factors in a fully relativistic frame work at present are still limited to the A=2 
system; for the other nuclei we still have to rely on the non-relativistic approaches
involving as complete as practical a set of relativistic corrections.   

We begin by discussing the non-relativistic limit only.
The electromagnetic current operator {\bf j}  in a nuclear many body system, which 
we address first,  is related
to the Hamiltonian  $H$ of the system via the continuity equation
%\begin{eqnarray*}
${\bf \bigtriangledown \cdot j} + i [ H,\rho ]  = 0 $
%\end{eqnarray*}
where $\rho$ is the  charge operator of the system.  In non-relativistic nuclear 
physics one usually assumes that the Hamiltonian can be separated into a sum of 
single-nucleon kinetic energy terms $T$ and two-nucleon interactions $V$. As a 
consequence, the electromagnetic current operator {\bf j} contains a sum of both 
single-nucleon and two-body operators. Two-body, or exchange, operators thus
have to be present whenever the commutator between the N-N potential and the single-nucleon
charge operator is non-zero.  This is the case in particular for the isospin-dependent 
and velocity-dependent components of the N-N interaction. Such components
occur prominently in the N-N interaction, as the long-range one-pion exchange interaction
is isospin dependent, and thus leads to an important two-body operator.
The non-locality of the N-N interaction also leads to a non-vanishing commutator, 
hence an exchange current operator  needs to be taken into account
\cite{Friar77}.
\begin{figure}[htb]
%Figur mit topp/sideways hergestellt, modif boundingBox: 66 206 516 556
\centerline{\mbox{\epsfysize=3cm \epsffile{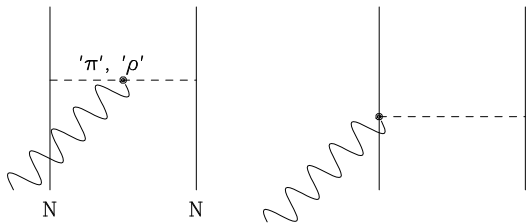}}}
\begin{center} \parbox{14cm}{ \caption{
\label{MEC1} 
Contributions to the two-body currents which are model-independent in the 
leading non-relativistic order. The exchange 
particles '$\pi$' and '$\rho$' are $\pi$- and $\rho$-like objects.
} }  \end{center}
\vspace*{-0.4cm} \end{figure}

Further two- or many-body terms of the electromagnetic interaction would have to be
considered once many-body forces between nucleons are allowed for. While such forces
are indeed introduced to account for the binding energies of light nuclei (and 
nuclear matter) the corresponding many-body terms in the currents are still poorly 
understood.

The isospin dependent N-N interaction is associated with the exchange of 
charged particles. Within the simplest models, it involves the exchange of $\pi$'s 
and $\rho$'s. The corresponding interactions and the associated current operators
which satisfy current conservation are known. This allows one to derive the 
two-body operators from the N-N interaction employed 
\cite{Ohta89a,Ohta98b,Riska85,Adam89}. The corresponding terms have been
classified as ''model independent'' by Riska \cite{Riska85}. In terms of a diagrammatic
representation, these terms, accounting for the exchange of $\pi$'s and $\rho$'s,
are displayed in fig.~\ref{MEC1}. By construction these terms respect
gauge invariance.  When dealing with realistic N-N interactions, 
these terms are replaced by exchange terms involving particles that are 
$\pi$-like (pseudoscalar
exchange with quantum number $0^-$) and $\rho$-like (vector-meson type exchange with 
quantum numbers $1^\pm$), 
in accordance with the underlying structure of the N-N interaction
\cite{Riska85,Buchmann85}. The identification of the $\pi$-like and $\rho$-like 
contributions is not necessarily dictated by the potential \cite{Adam89}, so 
the qualification of ''model independent'' has to be understood with the 
quotes. The additional two-body terms then are not necessarily constrained
by current conservation.

These terms {\em a priori} do not account for the finite size of the nucleon. 
A simple (but not necessarily unique) approach to include the finite size replaces
 the coupling constants by vertex form factors. 
When using  vertex form factors known from N-N scattering, additional 
vertex currents have to be introduced in order to respect current conservation. 
Fortunately, the numerical consequence of these form factors for few-body
electromagnetic form factors is not great, as their effects in the $\pi$- and 
$\rho$-like terms largely cancel.

Other two-body terms related to other components of the N-N interaction, such a 
{\em e.g.} a velocity dependence, can be and have been included 
\cite{Carlson90,Schiavilla96}, but are 
typically of lesser numerical importance. 

While the above exchange current operators can be derived directly from the
N-N interaction, and thus can be regarded as largely model independent, there are others 
that are not constrained by the continuity equation. These terms, addressed in the 
following, then are  more uncertain.
\begin{figure}[htb]
%Figur mit topp/sideways hergestellt, modif boundingBox: 66 206 516 556
\centerline{\mbox{\epsfysize=3cm \epsffile{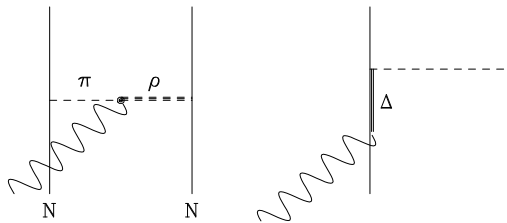}}}
\begin{center} \parbox{14cm}{ \caption{
\label{MEC2} 
Two-body contributions involving the Delta, and coupling to two different mesons. 
} }  \end{center}
\vspace*{-0.4cm} \end{figure}
These ''model dependent'' terms (in the classification of Riska \cite{Riska85}) 
are associated with the excitation of virtual intermediate nucleon resonances and 
exchange operators involving more than one type of meson.  

The most important model dependent transition-type exchange current operator involves 
excitation of an intermediary $\Delta_{33}$ resonance (see fig.~\ref{MEC2}), with the 
exchanged particle being $\pi$- or $\rho$-like. Processes involving higher nucleon 
resonances are suppressed by larger energy denominators. While some of the relevant
coupling constants are known from experiment, others have to be derived via model
calculations. For electron scattering at large $q$, the $q$-dependence of the couplings
becomes a source of uncertainty. In calculations of the nuclear ground state wave
function that allow for {\em both} nucleons and delta's as done {\em e.g.}
by Henning \et\,  this term is, of course,
included (in a non-perturbative way) in the wave function and does not appear 
as  an exchange current contribution \cite{Henning92a}.  

The second type of ''model dependent'' two-body contribution accounts for  terms of the
type shown in fig.~\ref{MEC2}a. Such terms, which involve {\em e.g.} the 
$\rho \pi \gamma$   transverse isoscalar exchange current or the 
$\omega \pi \gamma$ current, are more uncertain due to the poorly known
coupling constants involved.  The $ \rho \pi \gamma$  term can, depending on the
 observable, represent an important source of uncertainty at large $q$, the 
$\omega \pi \gamma$  and other terms are numerically not so significant. In most 
cases, though, the model-independent  terms dominate, and these are well under control.
 
Coming to the two-body {\em charge} operators, the situation  is less favourable.
In the non-relativistic limit, there are no pion exchange contributions to the
nuclear charge operator. The (longest range) pion exchange charge operator may be
looked at as the contact term of the type shown in fig.~\ref{MEC1}b. This term, however,
is already of relativistic order and cannot directly be linked to the N-N interaction
employed to compute the nuclear wave  function. 
The form of this two-body term (for the exchange of both $\pi$'s and heavier mesons)
still is somewhat uncertain. To derive it one has to
systematically reduce both the interaction and the electromagnetic current operator
from a fully relativistic to a non-relativistic description.
These contact-type contributions to the two-body charge operator
 are of considerable numerical importance when calculating the 
few-body C0 charge form factors. 

When deriving the two-body charge operators by reduction from a relativistic 
description of the coupled system of meson and nucleon fields, such as 
{\em e.g.} done by Arenh\"ovel \et\ 
 \cite{Arenhoevel00,Adam89}, one also has to deal with the boost term which,
at very high $q$, has a considerable effect.    

The diagrams involving $\pi$- and $\rho$-exchange again need to be modified relative to 
the propagation of bare particles by vertex form factors in order to account
 for the finite size
 of the particles involved. These form factors can be taken as to be the same ones that 
occurred in the current operators, where they were constrained by current 
conservation and the N-N interaction employed.
 
As for the two-body currents, there are contributions to the two-body charge operator
involving  vector mesons, such as the $\rho \pi \gamma$ term (see 
fig.~\ref{MEC2}). 
These processes, as for the current operators, are of fairly short range and contribute 
only at very large momentum transfer. 
 
The situation concerning exchange currents  is different when considering the intrinsically
relativistic approaches used by VanOrden \et\ and Hummel and Tjon 
\cite{VanOrden95,Hummel94} which solve the Bethe-Salpeter
equation. If care is taken to insure that the truncations necessary in 
practical calculations do maintain current conservation, then (in PS-coupling)
no two-body terms appear
other than the pion-in-flight diagram (which usually gives a small contribution)
and the ones due to the type of diagrams shown in fig.~\ref{MEC2}. The 
$\rho \pi \gamma$ diagram represents the dominant term, and introduces 
considerable uncertainty.  Depending on whether a rather soft vertex form
factor or a rather hard one (from {\em e.g.} Vector Dominance)
 is used, the
deuteron $A(q)$ and $B(q)$ at large $q$ are close to the data, or too high.  
%\newpage
\section{Deuteron}
\subsection{Introduction \label{deutintro}}
The \de, the only bound two-nucleon system, is one of the fundamental systems of
nuclear physics. Accordingly, many studies, both experimental and theoretical,
have been devoted to it. Of particular interest today is the degree to which
the \de\ can be understood as a system of two nucleons
interacting via the  known nucleon-nucleon interaction. 

When addressing, more specifically, the electromagnetic properties of the \de,
 the corresponding question concerns the ability to predict the three \de\ form
factors starting from the calculated \de\ wave function and the nucleon form 
factors
known from electron-nucleon scattering. At low momentum transfers predictions
and data agree quite well when accounting for one-body terms only, at the 
higher momentum transfers, two-body contributions 
are known to be important. Whether quark degrees of freedom do need to be
allowed for is still a matter of debate.

The approach traditionally taken for the two-nucleon system has been to 
derive the nucleon-nucleon (N-N) interaction from N-N scattering data, without
inclusion of the information from the deuteron (other than the binding energy).
The data for elastic electron-deuteron scattering then are used as a ''check''
of our understanding of the N-N system. 
 
The study of the electromagnetic form factors of the \de\ is complicated by
the fact that the deuteron has spin one. As a consequence there are {\em three} 
form factors corresponding to the multipolarities $C0$, $M1$ and $C2$,  
needed to
fully describe the electromagnetic structure. Without data on  polarization
observables --- the measurement of which up to recently presented too many
difficulties ---
only a combination of $C0$ and $C2$ could experimentally be determined, a 
fact that limited greatly our ability to fully exploit the information
contained in the deuteron structure functions. This limitation fortunately now
has largely been overcome.

The \de\ charge form factor is particularly interesting for the understanding
of the role of exchange currents. To leading order, (non-relativistic)
two body currents are absent due to their isovector character.
The dominant two-body currents in the isoscalar charge form factor then are
of relativistic, $v^2/c^2$, order. These currents still are less than perfectly 
understood. Therefore the  $C0$ form factor is a good place to make comparisons
between the non-relativistic approaches --- supplemented by $v^2/c^2$
relativistic corrections --- and the theories based on a fully relativistic
approach.

The deuteron electromagnetic form factors most often are studied in order to
check our understanding of the two-nucleon system. In parallel, however,
the deuteron form factors are also exploited to get a better handle on the
neutron form factors. In the past,  much of our knowledge on the 
neutron charge form factor $G_{en}(q)$ came from precision studies of the deuteron 
structure function $A(q)$. Only very recently, experiments involving both
polarized electrons and polarized target/recoil-nuclei have allowed us to get 
access to $G_{en}$ via other observables (see sect.~\ref{nucleon}). 
At large $q$, however, $G_{en}$ is 
still largely unknown, a fact that represents a serious handicap for the
quantitative understanding of the deuteron charge form factors.

The deuteron is a very loosely bound system (probably the still best example 
for a ''halo nucleus''). Accordingly, many of the deuteron properties of 
relevance to strong-interaction physics concern the large-radius behaviour of 
the deuteron wave function which is well under control.
 With the electromagnetic form factors at the higher 
momentum transfers $q$, one does have the chance to access the 
{\em shorter-range} 
properties of the two-nucleon system. This makes the study of the deuteron 
form factors --- especially the $C0$ form factor --- particularly interesting.

\begin{figure}[htb]
%Figur mit topp/sideways hergestellt, modif boundingBox: 66 206 516 556
\centerline{\mbox{\epsfysize=9cm \epsffile{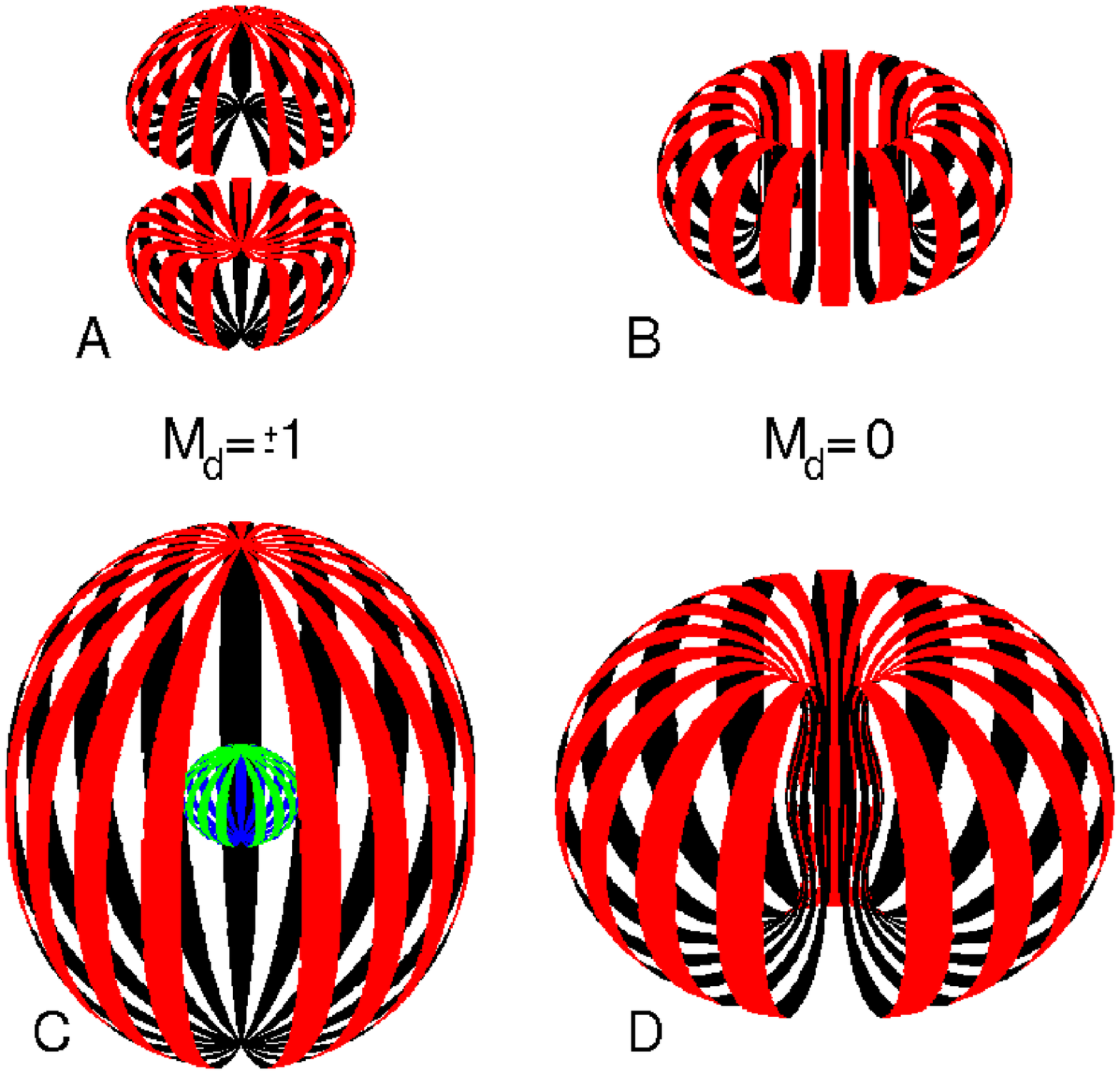}}}
\vspace*{2.cm} 
\begin{center} \parbox{14cm}{\vspace*{-3.5cm} \caption{
\label{dd} Equidensity surfaces of the deuteron for $m_z$ = 1,0 for 
high (A,B) and medium (C,D) values of the density \protect{\cite{Forest96}}.
Fig. A shows that for high density the deuteron is dumbbell like for 
m=$\pm$1, while B,C show a torus-like shape for m=0.
} }  \end{center}
\vspace*{-2.cm} 
\end{figure}
\vspace*{1.cm} 

At  small inter-nucleon distances,  the \de\ actually has quite a bit more 
structure than appreciated when looking at the usual plots of the radial
wave functions $u(r)$ and
$w(r)$ of the $S$- and $D$-state, respectively. The strong tensor force
at small $r$ leads to an interesting behaviour of the \de\ density: for 
spin-projection $m_z = \pm 1$ the equi-density contours have the shape of a 
dumbbell, for $m_z$=0 they have the shape of a donut, as displayed in 
fig.~\ref{dd} from Forest \et\ \cite{Forest96}.

Traditionally,  the calculations of the \de\ electromagnetic structure are
compared to the structure functions \aq\ and \bq\ and, more recently, to
the data on the polarization observable $T_{20}$ that are becoming available.
In the present review, we mainly compare to the $C0$, $C2$  (and $M1$) form 
factors
rather than \aq\ and $T_{20}$. This is done for two reasons. First, the 
individual
$C0$ and $C2$ form factors are much more sensitive to the physics ingredients 
than \aq .
Second, the use of $C0$ and $C2$ allows us to more directly compare to
the A=3 (T=0) and A=4 charge form factors.
 
We also want to deemphasize the discussion of the observable $T_{20}$ for another reason: this
quantity  refers to a specific scattering angle (although the dependence
on the angle is weak as compared to the one on momentum transfer). It thus
is better to avoid this dependence on angle altogether and concentrate on
 $C0$ and $C2$. Moreover, the physical significance of $T_{20}$ is rather
convoluted (even if expressed in terms of the form factors $F_{00}$
and $F_{11}$, see ref.~\cite{Forest96} and below). We therefore rather use 
the $C0$ and $C2$
form factors, which today  can be determined with adequate precision from
the experimental data.
\subsection{Electron scattering data \label{deutdata}}

Elastic electron-\de\ scattering has been investigated in many experiments, and
the cross section data 
\cite{Abbott99} \nocite{Alexa99,Akimov79}
\nocite{Arnold75,Arnold87,Auffret85a,Benaksas66,Berard73a,Bosted90}
\nocite{Bumiller70,Buchanan65,Cramer85,Drickey62}
\nocite{Elias69,Friedman60,Galster71,Ganichot72,Goldemberg64}
\nocite{Honegger97,Grossetete66b}
\nocite{Martin77,Platchkov90,Rand67,Simon81,Stein66}-
\cite{Voitsekhovskii86}
today cover a large range of momentum transfers. Some of that data obviously 
is not very precise, other data, mainly of more recent origin, has reached
accuracies down to the 1\% level. In the analysis to be described below, we will
employ the full set of {\em world} data today available. We discuss  below in
more detail only some of the sets.
 
%Early data came from McIntyre and collaborators 
%\cite{McIntyre55} \nocite{McIntyre56,McIntyre57} - \cite{McIntyre58}.
The earliest measurements useable in todays determination
of deuteron form factors, were performed at Stanford by
Friedman {\em et al.} \cite{Friedman60}. These measurements actually were aimed 
mainly at the neutron magnetic form factor, which was studied via magnetic
scattering from the deuteron. 
%The experiment of Goldemberg {\em et al.} 
%\cite{Goldemberg64} 
%was the first one to measure cross sections at 180$^\circ$ in order to 
%separate at low $q$ the small magnetic contribution from the dominating charge contribution.
%The measurements of Benaksas {\em et al.} \cite{Benaksas66} continued in the 
%same direction, the main difference being the detection of the recoil 
%deuteron at 0$^\circ$. 
%
%Subsequent experiments extended the momentum transfer to significantly 
%larger values. To reach them, Buchanan \et\ \cite{Buchanan65} used
%the e-d coincidence technique to cleanly identify elastic scattering despite
%the insufficient energy resolution.
%
The experiment of Rand \et ~\cite{Rand67} provided the first more extensive 
set of data on the magnetic form factor measured at 180$^\circ$. The main
result of this experiment actually concerned threshold electro-disintegration, 
the process that later on \cite{Bernheim81} was recognized to be one of the 
showpieces for the contribution of meson exchange currents (see 
sect.~\ref{deutdis}).

The experiment of Elias \et\ \cite{Elias69}, performed at the Cambridge 
electron accelerator, provided data up to 6$fm^{-1}$. In this experiment, the 
 electron beam was scattered from a target placed inside the synchrotron. Elastic
scattering was identified by detecting the recoil deuterons using a quadrupole
spectrometer. These data cover an important range in $q$, where few other
data were available up to very recently. 
Some care is advisable in using this data: the deuteron spectra, measured with 
a combination of scintillators, spark chambers and range counters, appeared to 
show an unidentified constant ''background'' of 20 $\div$ 50\% under the
deuteron peak. This ''background'' was 
subtracted by the authors. After closer inspection of the deuteron spectra, 
and with the benefit of hindsight, it would appear that
there actually was no significant background; rather, the deuteron peak (measured
over too narrow a momentum region to allow one to identify a ''constant 
background'') was 
simply a bit  wider than expected.      

The Bonn synchrotron also contributed to the world supply of electron-deuteron
data \cite{Cramer85}. Using two magnetic spectrometers, the reaction products were
detected in coincidence at both forward and backward angles, in order to 
perform a selfconsistent Rosenbluth separation. (Note that the last entries 
in table 2  of \cite{Cramer85}  seem to be inverted). 

Data at the highest momentum transfers were measured by Arnold \et\ at 
SLAC \cite{Arnold75,Arnold87,Bosted90,Martin77} both at forward and backward angles. 
In both experiments scattered electron and recoil deuteron were detected in 
coincidence. For the case of the backward-angle scattering, the reaction
products were detected at zero and 180$^\circ$, respectively. 
This experiment identified for the first time the diffraction minimum occurring 
in $B(q)$ at $q \sim 7 fm^{-1}$.  

Two recent experiments at Jefferson Laboratory by Abbott \et\ and Alexa \et\
provided additional data  \cite{Abbott99,Alexa99}.
These experiments could reach very large momentum  transfers due to the 
large beam intensities available at JLAB and the large acceptance of the spectrometers.
While the hall-A experiment was carried out using the two high-resolution 
spectrometers, the hall-C experiment employed for the recoil-detection the
magnetic channel built for the $T_{20}$ polarimeter (see below). In this latter
experiment the spectrometer acceptance and target length were  reduced on 
purpose in order to achieve a well known solid angle and spectrometer 
acceptance.  At the time of
the writing of this review, there still is a significant discrepancy between 
the two data sets; there is the suspicion that this is due to discrepancies
in the determination of the beam energy.

Of particular interest to the precise determination of the deuteron form 
factors are the sets of data which have reached the highest accuracy.
At the low and medium momentum transfers,  these are the  
data of Galster \et, Berard \et, Simon \et~ and Platchkov \et\ 
\cite{Galster71,Berard73a,Simon81,Platchkov90}. 
Precision experiments on the deuteron face extra problems as compared to 
experiments involving heavier nuclei, for two reasons: \\
$\bullet$  Achieving a reasonably large target
thickness involves in general the use of long liquid-deuterium targets.  
The determination of the acceptance of spectrometers for long targets 
requires a special effort. This poses particular problems for modern 
multi-element spectrometers which have a complicated
acceptance function. \\
$\bullet$  Due to the low deuteron mass, the energy loss of the electron due to deuteron 
recoil is large. At the larger momentum transfers the deuteron elastic peak 
then overlaps with the quasi-elastic peak of the target windows. This in general
requires to move the windows outside the spectrometer acceptance, or to 
detect electron and recoil deuteron in coincidence.

The experiment on e-d scattering carried out by Galster 
{\em et al.} \cite{Galster71} was performed at the DESY accelerator using the extracted beam. 
In this experiment the scattered 
electron and recoil deuteron were detected in coincidence in order to 
identify elastic scattering at the incident electron energy of 2.5 GeV.
Galster {\em et al.} did a very precise comparison between e-p and e-d 
scattering, with the aim to determine from the deuteron cross section the
small contribution of charge scattering from the neutron. This determination
of $G_{en}$, performed up to momentum transfers of 4$fm^{-1}$,  actually has 
withstood quite well the test of time.    

The low-$q$ data of Berard \et~\cite{Berard73a} were taken using
cooled H$_2$ and D$_2$ gas targets, in the range of momentum transfer 
$q = 0.2 \div 0.7 fm^{-1}$. 
The experiment essentially measured ratios of cross sections relative to 
Hydrogen; to get cross sections for the \de, normalization to absolute 
data on the proton is needed (for small corrections, see \cite{Sick98}). 

The data of Simon \et~\cite{Simon81} covers the range 
0.2 $\div~  2 fm^{-1}$. The experiment used both a low-temperature gas- and 
a liquid target, for both Hydrogen and Deuterium. The Hydrogen data taken with 
the gas target and a special small-angle spectrometer served as the absolute 
cross section standard, the data with 
the liquid targets were measured relative to that standard. 

The experiment of Platchkov \et~\cite{Platchkov90}  provided absolute data in the range $q = 0.7 
\div 4.5 fm^{-1}$ measured with a liquid Deuterium target. This 
experiment reached a very high accuracy, $\sim$1\%. This became 
possible as the spectrometer acceptance was restricted to the purely 
geometric one, which is easily measured. The acceptance of the spectrometer 
for finite target length had been studied in detail using solid targets 
displaceable along the beam direction.  The absolute efficiency of the detection
system in the focal plane could be measured precisely by exploiting the 
redundancy of the various detector elements. Data taken with 
a liquid Hydrogen target were used to confirm the accuracy of the absolute 
cross section measurement.
The Saclay group also provided data on magnetic scattering up to
momentum transfers of $5.3 fm^{-1}$ 
\cite{Auffret85a}, measured at a scattering angle of 155$^\circ$.  

%Other low-q data have much larger uncertainties 
%\cite{Drickey62,Ganichot72,Grossetete66b} or have been superseded by later work of the 
%same group \cite{Bumiller70}. The data of Akimov {\em et al} \cite{Akimov79}
%  have a floating  normalization as they have been taken without the benefit
%of the charge measurement, with a normalization referred to the impurity of 
%Hydrogen in the CD$_2$-target used; this makes the data much less useful.  

Some of the  \de\ data \cite{Abbott99}--\cite{Voitsekhovskii86}
listed above have been taken by determining the overall normalization 
using electron-proton scattering. 
The cross sections given in the publications were obtained by normalizing the 
measured ratios using the best information on the parameterization of the 
proton form factors  available at the time. Today, we have a set of more 
accurate and more complete proton data, and a comprehensive fit to all 
the data. It is therefore advisable to renormalize the corresponding \de\
data using the fit to todays world data on the proton \cite{Mergel96} 
in order to 
obtain the most accurate absolute \de\ data.

% Care is also necessary as in some cases the published low-$q$ data incorporate
%corrections for relativistic
%effects and M1, C2 contributions. For these cases one needs to add
%the corresponding corrections back in  to get the purely experimental 
%cross sections.  

For a detailed investigation of the form factors and their uncertainties, it 
is very
important to account for the systematic errors of the data. Their 
effect in general is much larger than the uncertainty due to the statistical 
errors. In the publications cited, the systematic uncertainties often are 
not adequately discussed, and several times important sources of systematic
errors {\em other} than normalization ({\em e.g.} electron beam energy, beam
halos) are not 
even mentioned.   For the data sets  that have a particular weight in
the determination of the form factors, one can consult the corresponding 
thesis works 
\cite{Simon77}-- \nocite{Schmitt77,Auffret85c,Berard73b,Topping72}
\cite{Hacene86} in order to 
extract a  reasonable estimate for the relevant systematic uncertainties.

%For the case of the data by Berard \et, the energy spectra of scattered 
%electrons show the presence of a heavier impurity in the hydrogen gas 
%\cite{Topping72}. The contribution of this impurity was removed in the scattered 
%electron spectra, but no
%corresponding correction had been made for the contribution of this impurity to
%the target pressure. Given the temperature of the gas (liquid nitrogen 
%temperature) and the source of hydrogen, the most likely contaminant is air
%(other likely impurities would condense). 
%The contribution of air to the target pressure can be calculated from the 
%energy spectra shown, and amounts to 0.28\%. We have added the corresponding 
%correction to the cross section ratios. 

%Fig.~\ref{et} gives a qualitative impression of the data today available for
%momentum transfers below $8 fm^{-1}$. 
%%
%\begin{figure}[htb]    
%%Figur mit topp/sideways hergestellt, modif boundingBox: 66 206 516 556
%\centerline{\mbox{\epsfysize=7cm \epsffile{et.pss}}}
%\begin{center} \parbox{14cm}{\vspace*{-0.5cm} \caption{ 
%\label{et} World data for electron-deuteron scattering as function of 
%momentum transfer and virtual photon polarization $\epsilon=(1+2 \vec{q}
%\hspace*{0.5mm}^2/Q^2 \cdot tg^2 \theta /2)^{-1}$. 
%% Only data for $q < 4 fm^{-1}$ are shown.
%} }  \end{center}
%\vspace*{-0.4cm} \end{figure}
%%

During the last years, it has increasingly become possible to measure not only
 cross sections, but also {\em spin observables}. The knowledge of these spin 
observables is imperative if one wants to separate the two contributions
of the $C0$ and $C2$ multipolarities to the $A(q)$ structure function. 
The separation of $C0$ and $C2$ is of particular interest in the region 
of the predicted diffraction zero of the $C0$ form factor, near $ q = 4 fm^{-1}$.
To separate $C0$ and $C2$ one needs data involving {\em tensor} 
spin observables. 

Two techniques basically are  available to measure such spin observables: 
\begin{itemize}
\item
At storage rings, one can use polarized, internal  deuteron gas targets 
from an atomic beam source. The
high intensity of the circulating electron beam allows one to achieve acceptable
luminosities despite the very low thickness of the gas target. 
\item
At facilities with external beams, one can use polarimeters to measure 
the polarization of the recoil deuterons. High beam intensities are a prerequisite
as the polarization measurement, which requires a second reaction of the 
deuteron, involves a loss of a few orders of magnitude in count rate.
\end{itemize}
A further potential approach, the use of an external polarized target (such as 
$ND_3$ at very low temperature \cite{Averett99}) has not yet been employed, as the luminosities
reachable with these targets were not  sufficient. This in part is also
related to the difficulties of making, instead of the usual vector-polarized
targets, a {\em tensor}-polarized target as
needed for the separation of $C0$ and $C2$.

The pioneering experiment on e-d scattering involving polarization observables
was performed by Schulze \et\ \cite{Schulze84} at Bates. 
The polarimeter employed made use
of the $^3 \hspace*{-0.5mm}He(\vec{d},p)^4 \hspace*{-0.5mm}He$ reaction, 
which represents an efficient analyzer  of the tensor
polarization for  deuterons with very low momentum. This experiment provided data near 
$q \sim 2 fm^{-1}$. 

\begin{figure}[htb]    
%Figur mit topp/sideways hergestellt, modif boundingBox: 66 206 516 556
\centerline{\mbox{\epsfysize=9cm\epsffile{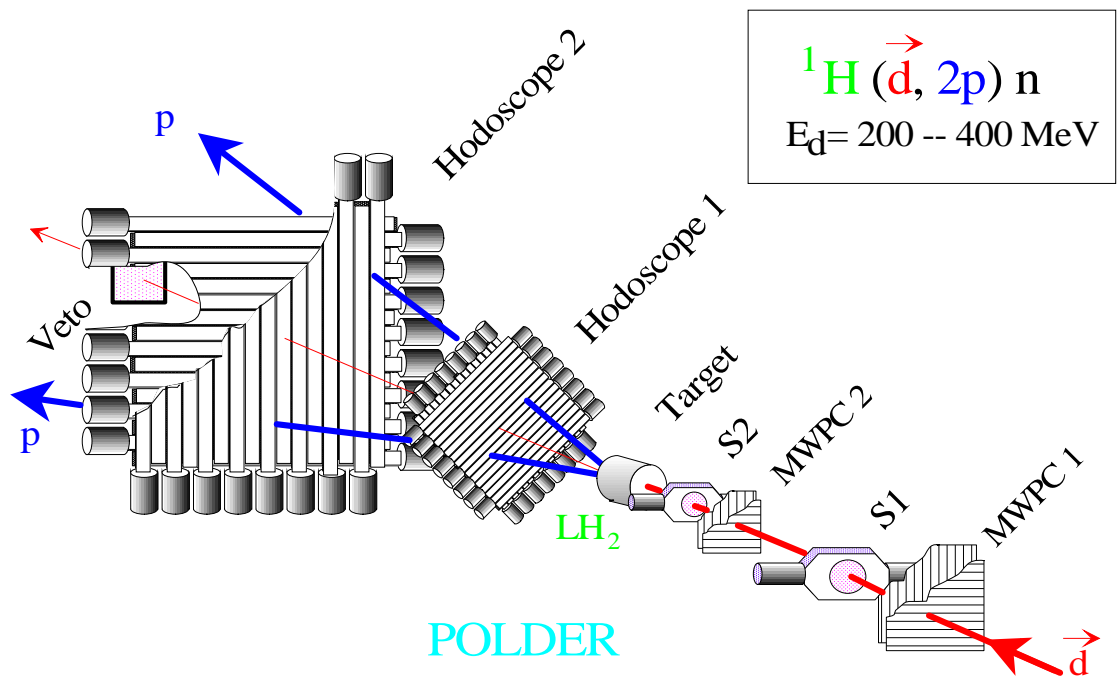}}}
\begin{center} \parbox{14cm}{\vspace*{-0.5cm} \caption{ 
\label{pol} POLDER deuteron polarimeter used in the experiment of Abbott 
{\em et al.} \protect{\cite{Abbott99}}. The deuteron track is measured before the 
$\vec{d} + p \rightarrow p + p + n $ reaction using multi-wire 
proportional chambers. The two protons are detected 
using the hodoscope counters. 
} }  \end{center}
\vspace*{-0.4cm} \end{figure}

The first experiment with an internal electron beam and a gas-jet polarized 
target was performed at the VEPP-storage ring in Novosibirsk \cite{Dmitriev85}.
 The data cover the region near 1$fm^{-1}$ momentum transfer. The follow-up 
experiment at VEPP \cite{Gilman90} reached 3$fm^{-1}$, but had
rather large statistical uncertainties.

 These early experiments did not
yet cover the region where the separation of $C0$ and $C2$ is expected to 
provide new information, but they showed that --- with a suitable
increase in luminosity and polarimeter efficiency --- the determination
of polarization observables would be  practicable.

Higher momentum transfers could be reached using external beams and different
types of polarimeters. The experiment of The \et~ \cite{The91}, performed at the 
Bates accelerator, used a polarimeter based on $\vec{d}-p$ elastic scattering. With 
this setup it became possible to reach the region of the zero crossing of the 
$C0$ form factor at 4.4$fm^{-1}$.   

Further measurements  using an internal tensor-polarized gas target were 
carried out at the AMPS storage ring \cite{Ferro96,Bouwhuis99}. The deuterons
were produced using an atomic-beam source which injected the deuterons into a storage
cell traversed by the electron beam. Scattered electron and recoil deuteron were measured in 
coincidence. This experiment provided $T_{20}$-values up to a momentum 
transfer of 3.2$fm^{-1}$.

The highest momentum transfers were reached in the recent experiment performed at 
JLAB by Abbott \et\ \cite{Abbott99}. With this experiment the data could be extended beyond
the diffraction maximum of the $C0$ form factor, reaching 6.7$fm^{-1}$.
This became possible with the construction of a polarimeter optimized for the 
high deuteron momenta, a polarimeter based on the 
$\vec{d}+p \rightarrow p+p+n$ break-up reaction depicted in fig.~\ref{pol}.
This experiment also resolved a worrying discrepancy between the data and theory
in the 4--5$fm^{-1}$ region where the normally most successful calculations 
seemed to lie significantly above the previous data.      

\begin{figure}[htb]    
%Figur mit topp/sideways hergestellt, modif boundingBox: 66 206 516 556
\centerline{\mbox{\epsfysize=7cm \epsffile{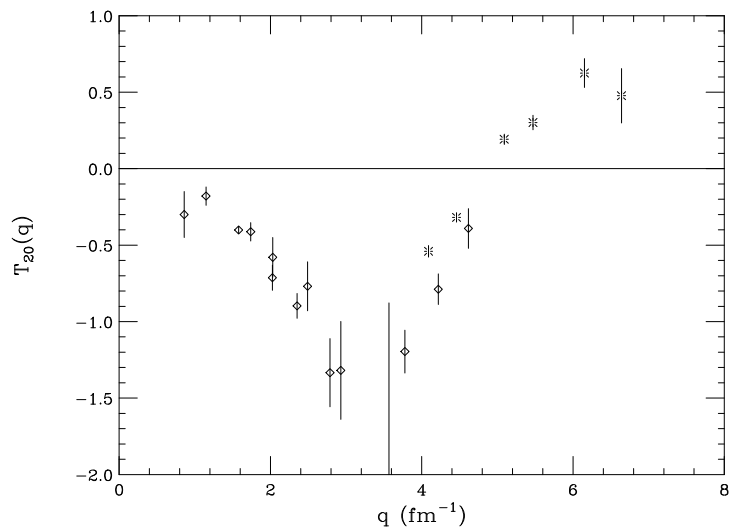}}}
\begin{center} \parbox{14cm}{\vspace*{-0.5cm} \caption{ 
\label{t20d} World data for the tensor analyzing power $T_{20}$ for
electron-deuteron scattering as function of momentum transfer (the slight
dependence on scattering angle has not been removed).
} }  \end{center}
\vspace*{-0.4cm} \end{figure}

In fig.~\ref{t20d} we display the $T_{20}$ data which today are available. 
Additional data have been measured for $T_{21}$ and $T_{22}$. These quantities,
however, are not very constraining in separating the multipolarities $C0$ and
$C2$; rather, they serve as consistency checks. 

\subsection{Experimental form factors \label{deutfit}}
Here we use the world data on electron-\de\ scattering up to high momentum
transfer to determine the various form factors.
Some 340 data points on e-d scattering are available for momentum transfers
below 8$fm^{-1}$ (the higher transfers are ignored for the moment 
as the experimental information is more fragmentary).

As discussed in more detail in section \ref{deutrms}, the 
experimental cross sections are first converted to effective PWIA cross 
sections by removing the Coulomb distortion 
effects. This is done by multiplying the experimental cross section 
$\sigma(E,\theta)$ 
with the calculated ratio $\sigma_{1.Born}(E,\theta)/\sigma_{2.Born}(E,\theta)$.
These $\sigma^{PWIA}$ then can be interpreted in terms of PWIA.

The deuteron cross sections measured in the various experiments 
\cite{Akimov79}-\cite{Voitsekhovskii86} contain contributions from the 
form factors of multipolarity $C0$ (charge monopole), $M1$ (magnetic dipole) 
and $C2$ (charge
quadrupole), which in turn determine the two structure functions \aq\ and \bq: 
\begin{eqnarray*}
\sigma(E,\theta)^{PWIA} = \sigma_{Mott}(E,\theta)
[ A(q) + B(q)~ tg^2(\theta/2)]
 \end{eqnarray*}
where $\sigma_{Mott}$ includes the recoil factor, with 
\begin{eqnarray*}
A(q)=F_{C0}^2(q) + (M_d^2 Q)^2 \frac{8}{9} \tau^2 F_{C2}^2(q) +
(\frac{M_d}{M_p}\mu)^2\frac{2}{3} \tau (1+\tau) F_{M1}^2 (q)
\end{eqnarray*}
\begin{eqnarray*}
B(q)=(\frac{M_d}{M_p}\mu)^2 \frac{4}{3} \tau (1 + \tau ) F_{M1}^2 (q)
\hspace*{0.5cm} with \hspace*{0.5cm} \tau = q^2/(4 M_d^2)
\end{eqnarray*}
In the above equations, $M$ are the masses, $\mu$ the
deuteron magnetic  moment in units of magnetons, $Q$ the quadrupole moment
 and $q$ is the four momentum transfer (>0). All the $F$-form factors are 
normalized to $F(0)=1$. 
                    
Up to recently the study of the form factors in general
was limited to the $A(q)$ and $B(q)$ structure functions. 
With the polarization data that became available during the last years
\cite{Schulze84,Dmitriev85,Gilman90,Voitsekhovskii86,Ferro96}, 
one also can perform an analysis of the world data in terms of the form 
factors \fcz,  \fmo, \fct.
The tensor polarization observables give a handle to separate the contributions
of $C0$ and $C2$. In particular 
\begin{eqnarray*}
T_{20} = -(\frac{8}{3} \tau F_{C0} F_{C2} QM_d^2 + \frac{8}{9} \tau^2
F_{C2}^2  Q^2 M_d^4 + \tau (\frac{1}{3} + \frac{2}{3}( 1 + \tau )~ 
tg^2 \frac{\theta}{2})
 F_{M1}^2 (\frac{M_d}{M_p}\mu)^2) / \sqrt{2}(A + B tg^2 \frac{\theta}{2})
\end{eqnarray*}
contains the interesting interference term $C0 \cdot C2$.
The other tensor observables, $T_{21}$ and $T_{22}$, are not so useful.

The  separation of  $A(q)$ and $B(q)$ (or \fcz,  \fmo, \fct ) in the past 
often has been done by the 
experimental groups providing the cross section data. In most cases, the 
individual experiments 
measured  cross sections that were dominated by either $A(q)$ or $B(q)$, so that 
the ''Rosenbluth separation'' could simply be done by subtracting the 
non-dominating structure function using the best values  known at the time. 

Here, we rather concentrate on a more reliable approach which involves a fit 
of the unseparated 
world cross sections with a parameterization of \aq\ and \bq\ 
{\em simultaneously}. This allows one to use the full set of data today 
available to determine {\em all} form factors. The ''Rosenbluth separation''
thus is done implicitly during the fit of the data. As the parameterizations 
of \aq\ and \bq\ provide  continuous functions, this approach also takes
care of the problem that rarely the forward-angle and backward-angle data 
are available at exactly  the same $q$. For the standard Rosenbluth 
separation this usually implies somewhat obscure interpolations or 
extrapolations. The same reasoning applies when determining the three form 
factors $C0$,  $M1$ and $C2$ by adding the polarization observables.

In order to analyze the \de\ data, the various form factors  are  parameterized
using a flexible functional form. From these form factors, one calculates cross
 sections in PWIA, and fits the parameters to the data (already corrected
for Coulomb distortion). 

We here discuss the results obtained with the 
Sum-of-Gaussians (SOG) parameterizations \cite{Sick74} for the $F$'s or, 
alternatively, \aq\ and \bq. 
The form factors are written as
\begin{eqnarray*} \label{sog}
F(q) = e^{-\frac{1}{4}q^{2}\gamma^{2}} \sum _{i=1}^{n} \frac{Q_{i}}{1 + 2 
R_{i}^{2}/\gamma ^{2}} \left(cos (qR_{i}) + \frac{2 R_{i}^{2}}{\gamma ^{2}} 
\frac{sin (qR_{i})}{qR_{i}}\right )
\end{eqnarray*}
For the comparison to theory mainly the $q$-space version is of interest 
{\em a priori}. In coordinate space, 
this parameterization for \fcz\ corresponds to the Fourier transform of a 
density $\rho(r)$ written as a sum of (symmetrized) gaussians that are placed at 
arbitrary radii $R_{i}$, with amplitudes $Q_{i}$ that are fitted to the 
data, and a fixed width $\gamma $.

The choice of this parameterization is governed by the following 
considerations. To extract form factors not biased by the choice of the 
analytical form, one needs a parameterization that in principle is totally 
general, with restrictions that can be justified on physical grounds. In 
the limit $\gamma \rightarrow 0, R_{n} \rightarrow \infty$ (hence $n 
\rightarrow \infty$) the SOG is a perfectly general basis. For a fit of 
the data, one introduces two restrictions: 1) The gaussians are given a finite 
width $\gamma $. In both $\rho_{ch}(r)$ 
and $\rho_{m}(r)$ we do not expect any structure that is significantly 
smaller than allowed for by the proton size. Accordingly, $\gamma $ is defined  via the   
$rms$-radius  $\gamma$ $\sqrt{3/2} = 0.6 fm$. It has also been verified that
with this value the form factors  from several theoretical calculations can be 
accurately represented up to the largest $q$. This finite 
width essentially imposes that the form factors on average fall no slower 
than the proton form factor. 2) The gaussians are placed at radii 
$R_{i} \leq R_{max} = 10 fm$. This 
is justified given the fact that, from independent knowledge on the 
behavior of wave functions at large radii, one can easily specify the radius at 
which the tails of densities give no significant $(< 10^{-3})$ contribution 
to $F(q)$. A finite $R_{max}$ imposes that the form factor cannot oscillate 
in $q$-space more quickly than with a minimal wave length $\Delta q \propto  
1/R_{max} $.

The data have been fitted with the SOG parameterization,
 and  
10 free parameters (for $r < 4 fm$) for each structure function (for the 
treatment of the region $4<r<10$ fm see sect. \ref{deutrms}).
 The data set for 
the \de\ is thus fitted  with a total of 30 free  parameters (20 for the fit
 of \aq , \bq ). 

The statistical error of the fit is easily calculated using the error matrix.
%This quantity can also be used to calculate the error of any other quantity 
%G(Q) that also depends on the $Q_{i}$, via 
%\begin{eqnarray*} \label{err}
%\delta G^{2} = \sum_{ij} \frac{\partial G}{\partial Q_{i}} \frac{ 
%\partial G}{\partial Q_{j}} E_{ij}. 
%\end{eqnarray*}
%
The systematic errors, which in general are the dominating ones, have been 
evaluated by changing 
each individual data set by the quoted error, and refitting the 
complete data set. The dominant
systematic errors are the ones due to overall normalization, and knowledge of
the electron energy. 
The changes due to systematic errors  of the different, independent,
sets of data are evaluated separately, and added quadratically. 
 The total error then is the quadratic  sum of statistical and systematic  errors. 

With the SOG parametrization both the form factors $F_{C0}, ~F_{M1},~ F_{C2}$ and 
the structure functions $A(q),~ B(q)$ have been extracted from the data. 
(Numerical values are available upon request).
For the former quantities, the separation is limited to a maximum momentum
transfer of $\sim 7 fm^{-1}$, due to the limited set of $T_{20}$ data 
available.  

The $\chi^2$ of the fit is quite satisfactory, given the very diverse origin
of the data. When fitting the three form factors to the data where the uncertainties
include the statistical errors only, the $\chi^2$ amounts to $\sim$610 for
373 data points \footnote{Due to the presently still unresolved discrepancy 
between the data
sets of \cite{Abbott99,Alexa99} mentioned in sect.\ref{deutdata} we have included only
the set \cite{Abbott99}.} (In this fit the large-r behaviour was constrained as 
described in sect. \ref{deutrms}). When adding quadratically the systematic
error, as is often done, the $\chi^2$ of this fit amounts to 485. A large 
contribution to the
$\chi^2$ ($\sim$110) comes from the data of ref.\cite{Elias69} which, as discussed
in sect. \ref{deutdata}, suffer from a probably  incorrect background 
subtraction.

%%
%\begin{figure}[htb]
%%Figur mit topp/sideways hergestellt, modif boundingBox: 66 206 516 556
%\centerline{\mbox{\epsfysize=7cm \epsffile{ratal1.ps}}}
%\begin{center} \parbox{14cm}{\vspace*{-0.5cm} \caption{
%\label{ratal} Ratio of world data to fit. Only data of dominantly longitudinal
%character are included. For clarity, the data of 
%\protect{\cite{Elias69}} are not shown.
%} }  \end{center}
%\vspace*{-0.4cm} \end{figure}
%%

The form factors resulting from the fit, together with their total error 
(statistical and systematic) will be used in sect. \ref{deuttheo} when 
comparing to the theoretical calculations.

\begin{figure}[htb]
%Figur mit topp/sideways hergestellt, modif boundingBox: 66 206 516 556
\centerline{\mbox{\epsfysize=7cm \epsffile{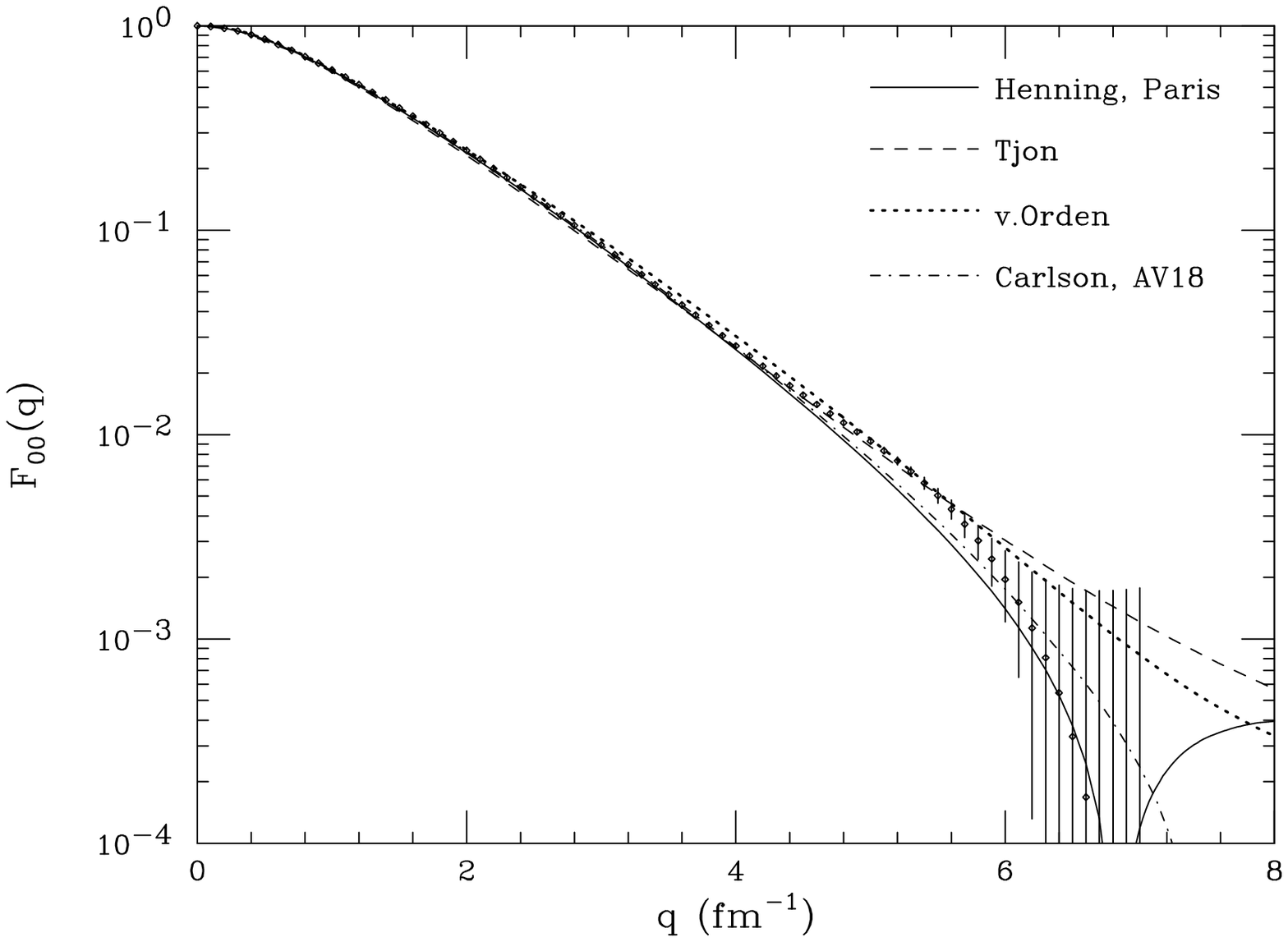}}}
\centerline{\mbox{\epsfysize=7cm \epsffile{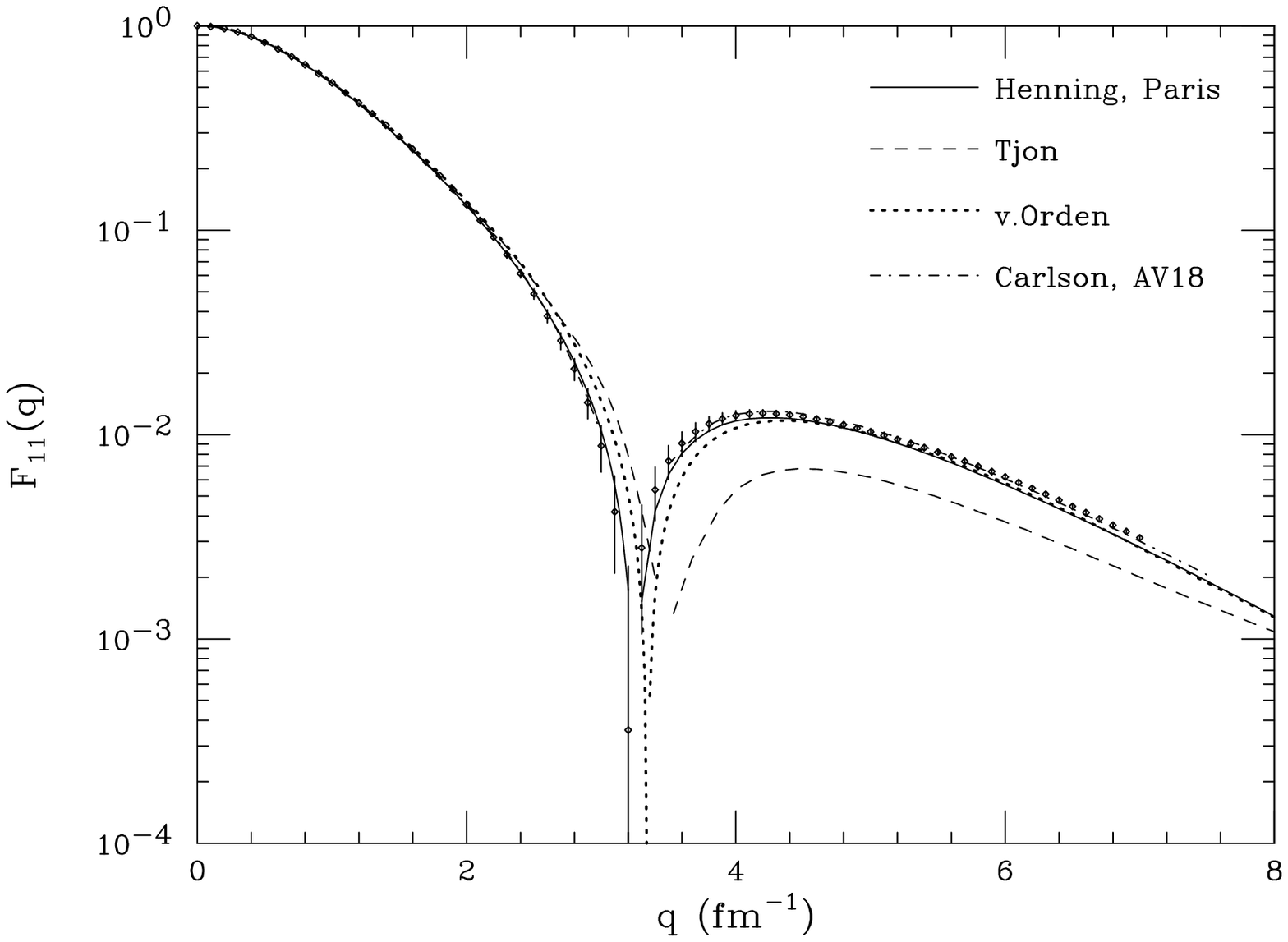}}}
\begin{center} \parbox{14cm}{\vspace*{-0.5cm} \caption{
\label{vijay} Charge form factors of the deuteron for spin-projection 
$m_z = 0$ and $m_z = 1$. The error bars include statistical and systematic 
uncertainties. The experiment is compared to selected theoretical calculations
(for discussion of the calculations see sect.~\protect{\ref{deuttheo}}).
} }  \end{center}
\vspace*{-0.4cm} \end{figure}

Traditionally, the deuteron electromagnetic structure has been discussed
in terms of the $A(q), B(q)$ structure functions or the $F_{C0}, F_{M1}, F_{C2}$
form factors. As pointed out by Forest \et~\cite{Forest96}, it may be more
elucidating to employ, for the charge form factors, the quantities which (in IA) 
correspond to the Fourier transforms
of the deuteron density in welldefined states of the projection $m_z = 0,\pm 1$
of the deuteron spin, $F_{00},~F_{11}$. (To distinguish these form factors from
the usual ones, we use subscripts to indicate the initial and final 
spin projection 0, $\pm$1). These form factors are linear combinations of 
the usual $C0$, $C2$ ones  (the corresponding magnetic form factor $F_{01}$, 
which describes the transition from the $m_z=0$ to the $m_z=\pm1$ state
  is  equivalent to $F_{M1}$): 
\begin{eqnarray*}
%c fc0 = c000*fa00+c011*fa11
%c fc2 =(c200*fa00+c211*fa11)/q**2
%c fa00= b000*fc0+b002*fc2*q**2
%c fa11= b110*fc0+b112*fc2*q**2
%      c000= 1./3.
%      c011= 2./3.
%      c200= 2./dqfm
%      c211=-2./dqfm
F_{C0} = 1/3~F_{00} + 2/3~F_{11} \hspace*{1cm} 
F_{C2} = 2~ F_{00}/Qq^2 -2~F_{11}/Qq^2
\end{eqnarray*} 
We give in fig.~\ref{vijay} plots for these quantities as the uncertainties
of the $F_{00},~F_{11}$ are not easily computed from the ones of 
$F_{C0}, F_{C2}$ shown in other figures of this review.

At large $q$, the form factor  $F_{11}$ gets its largest contribution from 
the two peaks of $\rho _1(r)$ occurring at distance $r=d$ from the deuteron 
center of mass CM (see fig.~\ref{dd}).
 If these peaks would have very narrow width, the form factor would have a 
$cos(qd)$-dependence, with the first sign change occurring at $qd = \pi$. 
The position of
the first zero of $F_{11}$ thus gives information related to the radius r 
where the maximum density occurs.

When approximating the deuteron density in the $m_z=0$ state as a disc
of thickness $t$, the corresponding form factor would have a 
$sin(qt/2)/(qt/2)$ dependence. The first diffraction zero thus gives information 
on the thickness $t$; the presently available form factors do not yet allow 
to determine it directly.

When ignoring the (rather small) magnetic contribution to $T_{20}$, this 
observable is in a fairly simple way related to the $m_z=0,\pm 1$ form factors:
\begin{eqnarray*}
T_{20}(q) \sim - \sqrt{2} \frac{F_{00}^2 (q) - F_{11}^2 (q)}
{F_{00}^2 (q) + 2 F_{11}^2 (q)}    
\end{eqnarray*}
The minimum of $T_{20}$ thus occurs when $F_{11}$=0, while for the maxima 
$F_{00}$=0. The minima and maxima  thus occur at those values of $q$ where 
the recoiling deuterons are only in the  $m_z=0$ and
$m_z = \pm 1$ states, respectively. 

Similar considerations apply to the magnetic form factor \cite{Forest96}. The 
zero of the magnetic form factor, which experimentally is located near 7$fm^{-1}$, 
gives information on the thickness of the torus, which amounts to about
0.9$fm$.

In fig. \ref{vijay}, we give the experimental results from our analysis of 
the world data  and compare to some theoretical 
form factors for the cases $m_z = 0,~ \pm 1$; a more detailed discussion 
of the theoretical calculations is given in  section \ref{deuttheo}.  

In the above figures, as in most of the other figures in this review, we 
plot the form factors as a function of $q$, and not, as is often done, $q^2$. 
The use of $q^2$ leads to an overemphasis on the highest $q$'s where the 
uncertainties of the data generally are largest. To the degree that two-body 
effects can be ignored, the densities corresponding to the form factors 
are Fourier transforms involving $q$ as a variable, and not $q^2$.  
%\newpage
%\input{deutfit.tex}   
%\input{deutrms1.tex}   
\subsection{Radius \label{deutrms}}
The electromagnetic properties of the \de\  at low 
momentum transfer could be hoped to be {\em accurately}  predicted
as non-nucleonic degrees of freedom or relativistic effects --- aspects that are not yet under perfect 
control  --- should be unimportant. The form factors at very low  $q$ are
 dominated by the 
parts of the \de\ wave function where the two nucleons are far apart, and the
properties of the \de\ should be determined
by the known N-N interaction and the known nucleon
form factors. 

For these reasons, the \de\ rms-radius has been a favourite observable to compare 
experiment  and calculation. 
%The interpretation of the experimental data at low momentum
%transfer in terms of the rms-radius appears   to be simple and clean.
 The theoretical calculations of the rms-radius are particularly 
reliable as the calculation is largely independent of the particular 
nucleon-nucleon potential used; for a very broad class of nucleon-nucleon 
potentials the radius depends essentially on the binding energy and the well known
n-p scattering lengths or, alternatively, the known asymptotic norm.
 A comparison between calculation and experiment therefore
promises to be particularly constraining.

Over the years, many theoretical calculations aiming at the \de\ rms-radius 
have been performed (for a review see \cite{Wong94}).  Much of 
the insight from theory, and an analysis of the experimental data, have been
put together by Klarsfeld {\em et al.} \cite{Klarsfeld86}, who
discovered a disturbing discrepancy: the rms-radius derived using nucleon-nucleon
potentials was 0.019$\pm$0.003 $fm$ higher than the one obtained from electron 
scattering data. The detailed analysis of Wong \cite{Wong94} confirmed
this finding. Given the failure to find a plausible mechanism to explain this 
discrepancy, this led to a rather confused situation.

This discrepancy has triggered several authors to  look more closely into 
potential corrections. In particular, Buchmann \et\ and Herrmann \et\ have
studied the 
effects of meson exchange currents \cite{Buchmann96} and dispersive corrections 
\cite{Herrmann97} in much greater detail than had been done 
previously. The effects found, however, were quite minor in terms of a change 
of the rms-radius, and go in the direction of increasing the discrepancy. 
The modification of the \de\ wave function due to
other aspects that are not well under control, such as the energy-dependence 
of the N-N interaction off-shell \cite{Desplanques88,Friar97}, 
 have been studied, 
also with little success in explaining  the discrepancy. 

It was finally found \cite{Sick96b} that 
much of the discrepancy originated from the fact that the \de\ data always 
were analyzed in Plane-Wave Born Approximation (PWBA), {\em i.e.}
 by neglecting the
Coulomb distortion. Although Coulomb distortion is indeed small as Z$\alpha 
\sim$ 0.01, the distortion effects are significant at the level of precision
the comparison of radii from various sources has reached today. They should be taken 
into account not only for the deuteron. Also for the proton it turns out
that, when considering the rms-radius, Coulomb distortion makes a non-negligible
difference, as pointed out by Rosenfelder \cite{Rosenfelder00}.

In order to calculate the Coulomb distortion, ref.~\cite{Sick96b} used
the second order Born approximation. The series
in $Z\alpha$ is expected to be very accurate for the $Z\alpha \sim$0.01 of 
interest for Hydrogen and, as shown in ref.~\cite{Herrmann97}, there 
indeed is no significant difference between this approach and the 
exact partial-wave analysis cross section.

The cross section in second order 
Born approximation \cite{Lewis56} can be written,
for scattering angles not very close to $\theta=180^\circ $, as 
\begin{eqnarray*}
\sigma_{2.Born} \simeq \sigma_{Mott} |F(q)|^2(1+R)
\end{eqnarray*}
where $F(q)$ is the elastic form factor,  $\vec{q}=\vec{k_i}-\vec{k_f}$ 
is the momentum transfer, $\vec{k_i}$ is the incident electron momentum and
$\vec{k_f}$ is the final electron momentum. $R$ is the Coulomb correction 
factor and is given by a principal value integral as:
\begin{eqnarray*}
R=\frac{2Z\alpha}{\pi^2}\frac{k \tilde\Delta^2}{F(q)} PV
\int d^3k' \frac{\vec p \vec{k'} + \frac{1}{2} p^2 }{{k'}^2-k^2}
\frac{F\left(|\vec{k'}-\vec{k_f}| \right)
      F\left(|\vec{k'}-\vec{k_i}| \right)}
{\left| \vec{k'}-\vec{k_f} \right|^2 \left|\vec{k'}-\vec{k_i} \right|^2}
\end{eqnarray*}
where  $k=|\vec{k_i}|=|\vec{k_f}|$ and where we have used the 
abbreviation $\tilde{\triangle}=q/p=\tan\frac{\theta}2$, with 
$\vec{p}=\vec{k_i}+\vec{k_f}$.
This integral, which in its original form is rather difficult to calculate 
numerically, has been  simplified  \cite{Sick96b}
by regularizing the principal value integral by subtracting the
singular part of the integral. 
The resulting integrals are numerically well behaved and can be 
calculated easily for any form factor given in analytic or numeric form.
This calculation \cite{Sick98} is used throughout this paper to convert 
the experimental data to PWIA form factors.  

At the low $q$ of main interest for a radius-determination, the dominant effect 
of Coulomb distortion actually is 
{\em not} the one due to the familiar difference 
$q \leftrightarrow q_{eff}$ (which has little effect indeed). Rather, it is
the change in cross section given, for a point nucleus, by $e.g.$ McKinley
\cite{Kinley48}. For a point nucleus, the Mott cross section then reads
\begin{eqnarray*}
\tilde{\sigma}_{Mott}=\left(\frac{Ze^2}{2E}\right)^2 \frac{cos^2(\theta/2)}{sin^4(\theta/2)} 
\left[1+\frac{\pi Z}{137}\frac{sin(\theta/2)(1-sin(\theta/2))}{cos^2(\theta/2)}
\right] 
\end{eqnarray*}

The additive term proportional to Z/137 is the one that is mainly responsible
for the Coulomb distortion at the low $q$'s as it influences the finite size 
effect which is given by the difference
between the experimental cross section and the point nucleus value. 

%The most accurate radius can be obtained by performing a model-independent 
%analysis of the cross sections. This type of analysis has been described in 
%sect.\ref{deutanal}. 

%The structure function $A(q)$, which can be determined from the data without 
%the input from polarization observables (which are not very accurate at low $q$)
%contains contributions from the C0 and C2 multipolarities.
%The contribution of C2 to $A(q)$ is very small at low $q$, and can be 
%removed using theory; for the determination of the charge rms-radius, it 
%actually does not come in at all given the $\eta^2$-factor. 

The  \de\ is special in the sense that the density
extends to rather large radii, given the low binding energy. The long tail 
has caused considerable difficulties in the past, as it influences the deuteron
form factor at extremely low $q$, where accurate values for the difference of 
the form factor to the point nucleus value  
1 are hard to measure. Fortunately, at these large distances, the shape of the 
\de\ density is easily calculable. Outside
the range on the N-N force, the wave functions $u(r)$ and $w(r)$ have an analytic
form that is well known (see e.g. \cite{Ballot80}) 
that  depends only on the \de\ binding energy. 
For the determination of the \de\ radius, this shape for radii $r > 4fm$, where the distance 
$r$ refers to the distance of the nucleons to the \de\ center of mass, can be 
imposed. 

The fit of the world data for $q < 8 fm^{-1}$ has been performed as described 
in section \ref{deutfit}. 
For the rms-radius of the \de, ref. \cite{Sick96b} finds 2.130 $fm$, with a 
random uncertainty 
of 0.003 $fm$ and a systematic uncertainty of 0.009 $fm$. 
The results of the fit is reported in table \ref{radii}, together with 
results from optical isotope shifts and \de\ wave functions 
to be discussed below.

The  comparison to other results is made on a basis of the quantity that comes 
closest to  the charge rms-radius of the non-relativistic two-nucleon system. 
We remove from the quantity directly measured in $(e,e)$ the effect related to 
non-nucleonic degrees of freedom (two body currents) as far as 
presently possible. We also remove the effects due to the (virtual)
excitation of internal degrees of freedom of the deuteron by the 
electron (dispersion corrections). In the same spirit, we remove from the 
charge radius measured in optical transitions the contributions of meson exchange
currents and nuclear polarization. To compare to the nuclear size calculated
by solving the Schr\"odinger equation for given N--N potentials, we add
to the ''matter rms-radius'' (the expectation value of $r^2$ of the wave 
function)
%$<\Psi r^4>/<\psi r^2>$ 
the contribution of the  
proton and neutron charge radii, and the Darwin-Foldy term. This makes the
various radii comparable; the radius we discuss is closest to what could be 
considered as ''charge radius'' in $(e,e)$ in Impulse Approximation.

 The dispersive effects --- corresponding to a 
two-step scattering process with excitation of the deuteron in the intermediate
state --- have recently been studied by Herrmann and Rosenfelder \cite{Herrmann97} who 
take into account the Coulomb excitation only and use an S-wave separable 
potential (Yamaguchi) to calculate the deuteron wave function. When analyzing the data
of Simon \et~\cite{Simon81}, they find a change of the rms-radius of --0.003 $fm$
when correcting the data for the dispersive effects. This calculation gives a
significantly smaller effect than a previous estimate \cite{Bottino72}, 
but is much more reliable. The contribution
of non-nucleonic degrees of freedom have been studied in great detail by 
Buchmann \et\ \cite{Buchmann96}. When
correcting the experimental data for the two-body effects, these authors find
a change of the rms-radius of --0.001$fm$, with a fluctuation of .001$fm$ 
depending on the approach used. The estimate for the contribution of 6-quark 
components is much smaller, and presumably also much more uncertain.

For the optical isotope shift, the accuracy has recently greatly 
improved, as a consequence of progress in the area of two-photon spectroscopy 
on hydrogen and deuterium. The atomic transition energies are now 
known with much higher precision, and a number of additional higher-order 
QCD corrections have been calculated by Pachucki \et\ \cite{Pachucki96,Beauvoir97}. 
We here quote the
analysis of these results by Friar \et\ \cite{Friar97}. Using 
for the proton charge rms-radius the standard value of 0.862$fm$ 
\cite{Hoehler76},  correcting
for the small nuclear polarization effect and using, for consistency,  for the 
two-body effects,  the value as calculated in  \cite{Buchmann96}, we find 
2.1316 $\pm$ 0.001 $fm$
for the deuteron  rms-charge radius.

The radius of the \de\ as determined from N--N scattering has remained
relatively stable since the work of Klarsfeld \et\ \cite{Klarsfeld86}. 
 Friar \et\ have also updated the determination of the deuteron radius starting
from the nucleon-nucleon potentials and the known asymptotic normalization
 of the 
S-state wave function. This yields, when adding the remaining electromagnetic 
contributions due to proton and neutron intrinsic
charge distribution and the Darwin-Foldy term, 2.1286 $\pm$ 0.002$fm$. With 
a small further correction resulting from the usually neglected
proton--neutron mass difference, we get the value listed in table \ref{radii}. 
\begin{table}[htb]
\begin{center}
\begin{tabular}{l r r}
Source & rms-radius ($fm$) & reference \\
\hline
\rule[0mm]{0mm}{5mm}
\hspace*{-1.5mm}$(e,e)$ world data & $2.130~~ \pm 0.010 $ & \cite{Sick98} \vspace*{-1.5mm}\\
\multicolumn{3}{l}{\dotfill} \\
 $(e,e)$ --disp. --MEC & $2.123~~ \pm 0.010$ &  \cite{Sick98,Herrmann97,Buchmann96}\\
 isotope shift --pol. --MEC &  $2.1316 \hspace{0.5mm} \pm 0.001$ &
\cite{Buchmann96,Friar97,Pachucki96} \\
 N--N scattering data & $2.130~~ \pm 0.002 $ & \cite{Friar97} \vspace*{1mm} \\
\hline 
\end{tabular}
\end{center}
\begin{center}
\parbox{12cm}{\caption{\label{radii} Charge rms-radii from different sources.
Corrections for dis\-persive effects (disp), two-body currents (MEC) and 
nuclear polarization (pol) have been applied to make the three last entries 
comparable.}}
\end{center}
\end{table}

The comparison of the second and fourth entry of table 
\ref{radii} shows that there is good agreement between the 
radius determined from $(e,e)$ and N-N scattering. The radius coming from the
optical isotope shifts (third entry) is within errors compatible with the one 
from electron scattering, but lies slightly above the value deduced from 
N--N potentials and the \de\ asymptotic norm. 
Overall, we find quite a consistent picture, as the three sources on the \de\ 
size give results which are quite close. 

Occasionally, the equivalent of the charge rms-radius for the M1 part is of
interest. This quantity is not really a ''radius'', but it does describe
the $q^2$-dependent term of the $M1$ form factor at $q = 0$. This 
''magnetic radius'' amounts to 2.072 $\pm$0.018 $fm$.  

\subsection{Comparison to theory \label{deuttheo}}
The theoretical understanding of the \de\ form factors involves a number
of issues. We first list the main questions before discussing in more
detail selected calculations and their comparison to the experimental
results.

The impulse approximation (IA) calculation of the form factors in a non-relativistic
frame work is straightforward as the Schr\"odinger equation can easily be 
solved for a given nucleon-nucleon (N-N) potential. Uncertainties originate
from the N-N potential employed (mainly its off-shell properties related to
the non-locality of the potential) and the
nucleon form factors used (mainly the neutron electric
form factor $G_{en}$). For the deuteron $C0$ form factor, the sum of
neutron and proton \gen\ and \gep\ comes in as a multiplicative factor; uncertainties
in \gen\ then directly propagate in the charge form factor. 

It may be instructive to consider the IA-relation between the S- and D-state 
radial wave functions $u(r)$ and $w(r)$ and the form factors. The form factors
are given by the integrals
\ba
F_{C0}(q)=2G_e^s (q) \int_0^\infty \left[u(r)^2+w(r)^2 \right]~j_0(qr/r)~dr
\ea
\ba
F_{C2}(q)= \frac{12 \sqrt{2} G_e^s (q)}{q^2 Q} \int_0^\infty w(r) \left[
u(r)-w(r) / 2 \sqrt{2} \right]~j_2(qr/2)~dr
\ea
\ba
F_{M1}(q) = \frac{1}{\mu} \left[ G_e^s (q) F_l(q) + 2 G_m^s (q) F_s(q) \right]
\ea
where the $M1$ form factor is split up into the contributions from the
intrinsic magnetization and the spin-orbit term
\ba
F_l (q) = \frac{3}{2} \int_0^\infty w(r)^2 \left[ j_0(qr/2)+j_2(qr/2) \right] ~dr
\ea
\ba
F_s (q) = \int_0^\infty \left( \left[u(r)^2-w(r)^2/2 \right] j_0(qr/2) + w(r) 
/ \sqrt{2} \left[u(r) + w(r) / \sqrt{2} \right] j_2(qr/2) \right)  dr ~~,
\ea
where the isoscalar nucleon form factors are given by
\ba
G_e^s (q) = \left[G_{ep}(q) + G_{en}(q) \right] /2, \hspace*{1cm} 
G_m^s (q) = \left[G_{mp}(q) + G_{mn}(q) \right] /2.
\ea
and $r$ here refers to the $n-p$ distance.
The D-state contribution affects, besides the $C2$ form factor, also significantly
the $M1$ form factor due to the S-D transition, by shifting the IA diffraction minimum by about $2 fm^{-1}$ 
to larger $q$ and decreases the height of the diffraction maximum. 
The contribution of $w$ in $F_{C0}$ is a modest increase of the height of the
diffraction maximum,  the effect of 
$F_l$ in $F_{M1}$ is small. 

In order to quantitatively understand the \de\ electromagnetic structure, the 
contribution of two-body terms needs to be accounted for. In
order to maintain current conservation, the two-body currents should be consistent with the 
N-N interaction employed. The calculation of such two-body terms has become possible during the 
last years. Uncertainties arise mainly from those two-body  terms 
(such as the $\pi \rho \gamma$-diagram) that are not constrained by current 
conservation. The \de\ is special in the sense that the leading-order
nonrelativistic MEC from $\pi$-exchange are absent due to their isovector
character; thus relativistic contributions are expected to be more
important.

\begin{figure}[htb]
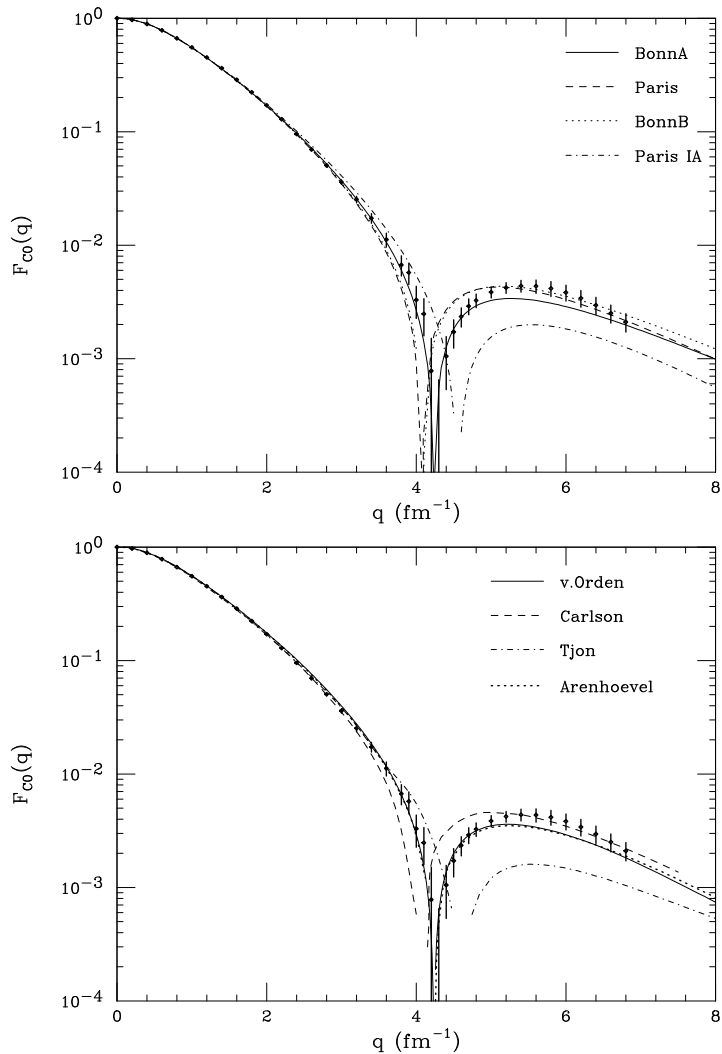

%Figur mit topp/sideways hergestellt, modif boundingBox: 66 206 516 556
\centerline{\mbox{\epsfysize=7cm \epsffile{fc01.pss}}}
\centerline{\mbox{\epsfysize=7cm \epsffile{fc02.pss}}}
\begin{center} \parbox{14cm}{\vspace*{-0.5cm} \caption{
\label{fc0} Comparison of experimental $C0$ form factor and calculations of
Henning \et\ (top),  and vanOrden \et , Carlson \et,  Hummel and Tjon and 
Arenh{\"o}vel \et\ (bottom). For references see text.
} }  \end{center}
\vspace*{-0.4cm} \end{figure}
\begin{figure}[htb]
%Figur mit topp/sideways hergestellt, modif boundingBox: 66 206 516 556
\centerline{\mbox{\epsfysize=7cm \epsffile{fm11.pss}}}
\centerline{\mbox{\epsfysize=7cm \epsffile{fm12.pss}}}
\begin{center} \parbox{14cm}{\vspace*{-0.5cm} \caption{
\label{fm1} Comparison of experimental $M1$ form factor and calculations of
Henning \et\ (top), and vanOrden \et , Carlson \et , Hummel and Tjon and 
Arenh{\"o}vel \et\ 
(bottom). For references see text. } }  \end{center}
\vspace*{-0.4cm} \end{figure}
 
In a relativistic frame work --- which is needed if one wants to understand
the \de\ electromagnetic structure up to the largest momentum transfers --- 
the calculation of the form factors gets more involved and different 
approximations are used by different authors.

Below, we compare the experimental results to a number of theoretical
calculations.  
We first address calculations that assume that 
the basic dynamical content of the \de\ is non-relativistic in nature, and 
that relativistic effects can be described as corrections in a $v/c$-expansion
\cite{Friar75}. In these calculations, the N-N interaction is one of the  
standard  potentials
with parameters fitted to the N-N scattering data. The two-body terms are 
then constructed to
be, as far as possible, consistent with the meson exchanges of the potential.
We subsequently address calculations which start from the Bethe-Salpeter equation
representing a summation of multiple-scattering series written in terms of Feynman 
diagrams. This approach is covariant, but the solution of the Bethe-Salpeter
equation usually requires truncations which can lead to problems in maintaining Lorentz covariance 
and current conservation. 

A large number of calculations are available 
\cite{Buchmann96,Chemtob74,Gari76,Plessas95,Schiavilla91,Dymarz90,Leidemann87,Wiringa95}, 
so we here discuss only 
a few representative ones which have appeared more recently.

The Hannover group \cite{Buchmann96,Henning92,Sauer94,Henning95a,Henning97} 
has performed calculations for
various N-N potentials (Paris, Bonn-A,B,C,..). For the nucleonic form 
factors the authors use the Hoehler parameterization. For the calculation of
the two-body terms, they use the approach developed by Adam \et\ \cite{Adam89} who
employ a modified S-matrix method to derive the two-body contribution 
 including the
leading order relativistic corrections. The electromagnetic nuclear currents are
derived from the exchange of all mesons which the Bonn potentials involve. 
Thus, the two-body contact-, pair- and retardation terms are consistent with the
underlying N-N potential. Boost corrections in the M1 are not included. 
The $\rho \pi \gamma$ term is added separately. 

\begin{figure}[htb]
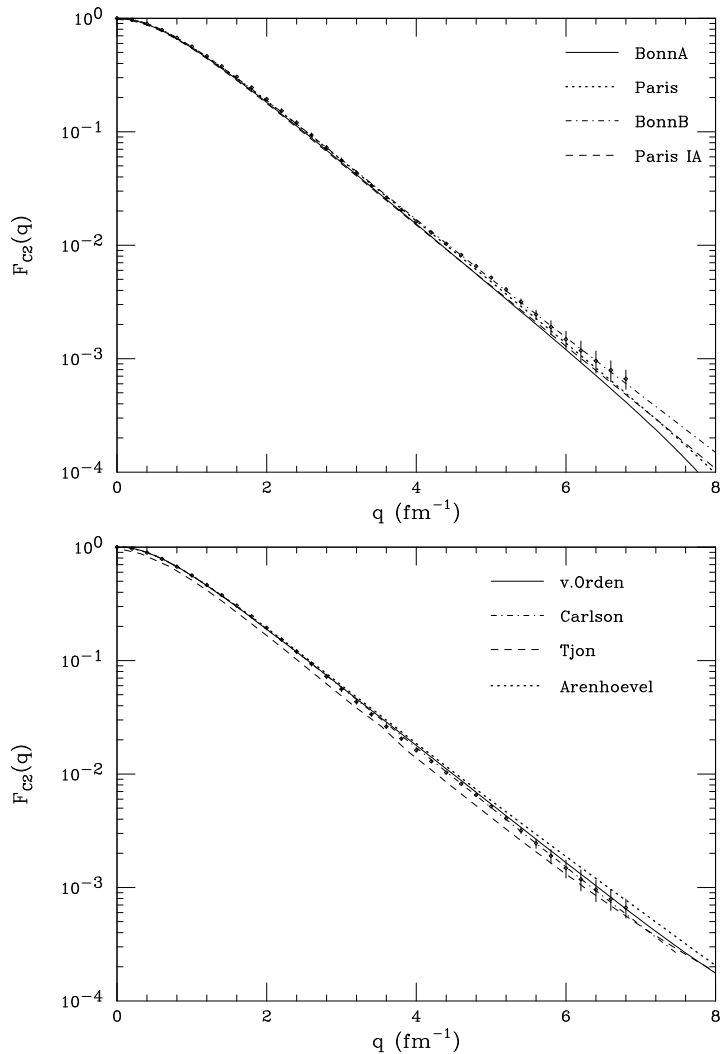

%Figur mit topp/sideways hergestellt, modif boundingBox: 66 206 516 556
\centerline{\mbox{\epsfysize=7cm \epsffile{fc21.pss}}}
\centerline{\mbox{\epsfysize=7cm \epsffile{fc22.pss}}}
\begin{center} \parbox{14cm}{\vspace*{-0.5cm} \caption{
\label{fc2} Comparison of experimental $C2$ form factor and calculations of
Henning \et\ (top),  and vanOrden \et , Carlson \et ,  Hummel and Tjon and 
Arenh{\"o}vel \et\ (bottom) For references see text.
} }  \end{center}
\vspace*{-0.4cm} \end{figure}

The results of these calculations are reported in the upper parts of 
figs.~\ref{fc0},\ref{fm1},\ref{fc2}.
In order to illustrate the two-body contribution, the IA result is also shown
for one particular N-N interaction. The comparison to the full calculation shows
that for the $C0$ and $M1$ form factors the contribution of two-body 
terms is very important.

The calculations of Carlson and Schiavilla \cite{Carlson98} are based on the 
Argonne V18 N-N potential. The one-body operator uses the standard Hoehler
form factors. The two-body current has '''model-independent'' pieces, related 
to the $\pi$-like and $\rho$-like exchange terms in the N-N interaction; these 
are largely 
derived from the N-N interaction and are constructed to satisfy current
conservation. The model-dependent part of the two-body current is mainly 
associated with the  $\Delta$-isobar. By means of transition correlation operators
  $\Delta \Delta$  (and, for $A>2$, N$\Delta$) components in the wave function are created.
%The two-body charge operators also has pieces related to the exchange of $\pi$-like
%and $\rho$-like mesons that can be derived consistently with the N-N interaction
%used. 
The contribution of the $\rho \pi \gamma$ and $\omega \pi \gamma$ 
exchange charge operators is more ambiguous as it requires the introduction of
phenomenological cut-off form factors.
The results of this calculation are also reported in figs. \ref{fc0}-\ref{fc2}.

In the figures we also show the results obtained recently by the Mainz
group \cite{Arenhoevel00}. This calculation  starts from a system of coupled
nucleon and meson fields, and, by means of the Foldy-Wouthuysen transformation,
 derives the non-relativistic limit including all the leading order 
relativistic contributions. For the exchange currents the terms 
consistent with the Bonn OBEPQ-B potential, which is used to calculate the
wave function, are used. The lowest order $\rho \pi \gamma$ contributions are 
also included when calculating  the M1 form factor. The dipole nucleon 
form factors (which do not well reproduce $G_{ep}$ at the larger $q$'s, see
above) together with the Galster $G_{en}$ are employed.

This calculation finds, for the $C0$ form factor,  a sizeable effect of the
boost corrections (not included in \cite{Wiringa95}), which are however 
largely canceled by   the $\pi$-two body term. The boost contribution for the
$M1$ multipolarity, together with the $\pi$-two body-term, leads to a significantly too large 
$M1$ form factor at the larger $q$'s.

\begin{figure}[htb]
%Figur mit topp/sideways hergestellt, modif boundingBox: 66 206 516 556
\centerline{\mbox{\epsfysize=7cm \epsffile{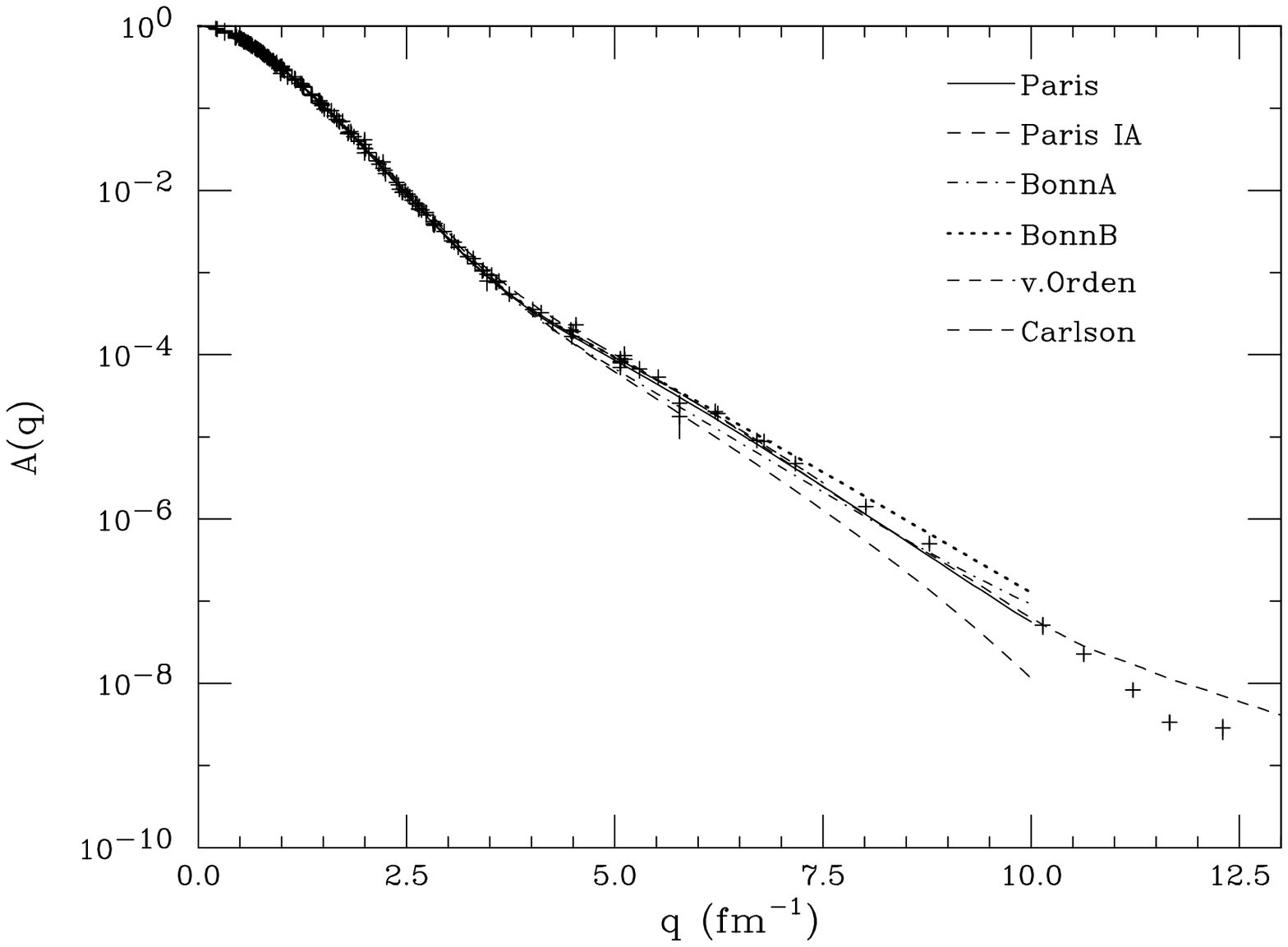}}}
\centerline{\mbox{\epsfysize=7cm \epsffile{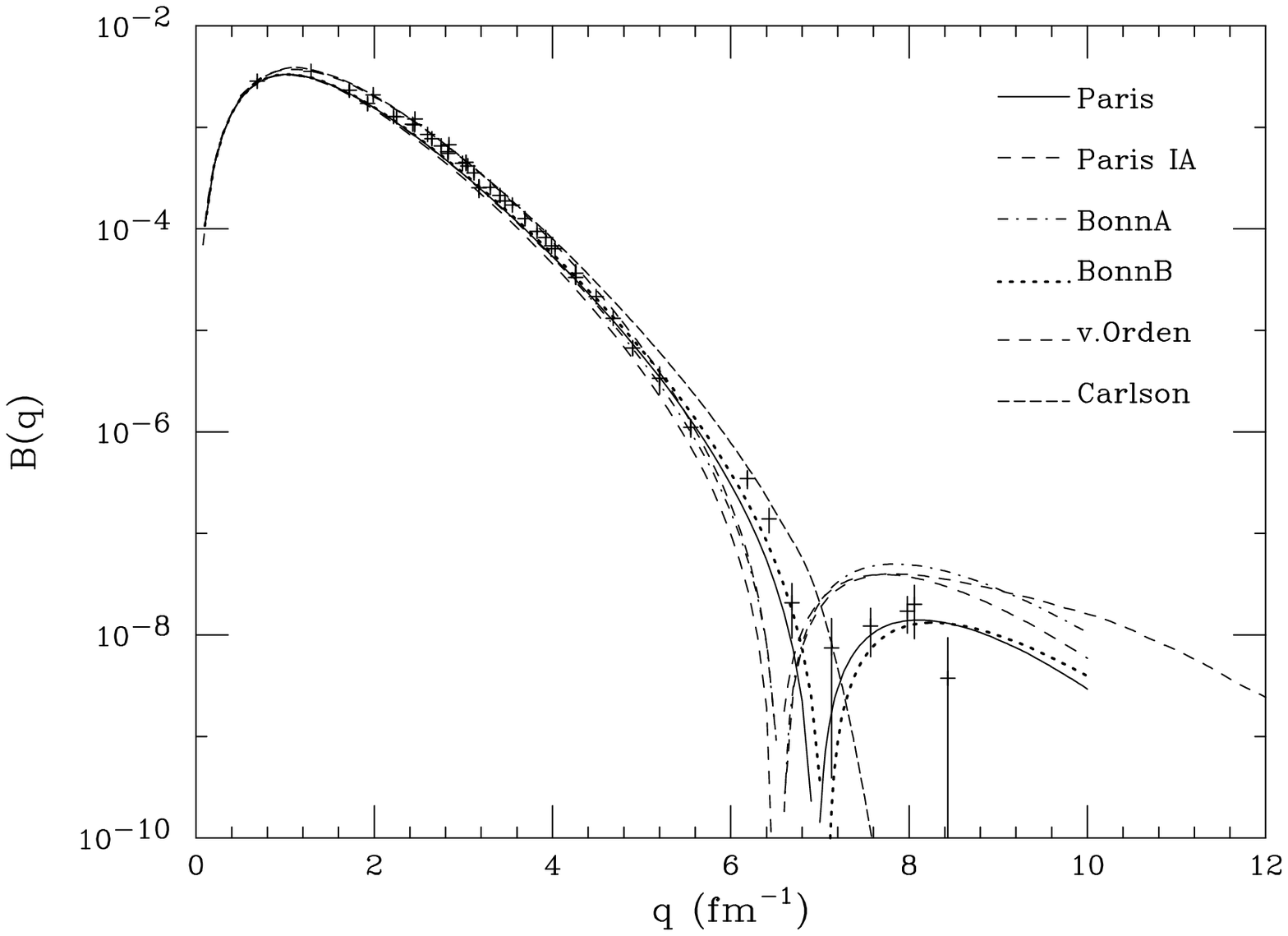}}}
\begin{center} \parbox{14cm}{\vspace*{-0.5cm} \caption{
\label{ab} Comparison of experimental $A(q)$ and $B(q)$ structure functions 
and calculations of
Henning \et ,  vanOrden \et , Carlson \et \ For references see text.
} }  \end{center}
\vspace*{-0.4cm} \end{figure}

The calculated two-body contributions obviously go a long way in closing the 
gap between the IA results and experiment. Some differences do remain, however. 
This is also seen in the $q=0$ observables. For instance, the IA \de\ magnetic 
moment for the V18 potential is 0.847 magnetons, with the two-body terms 
added Wiringa \et\ \cite{Wiringa95} find 0.871, while experimentally the 
magnetic moment is 0.857. Similarly, for the quadrupole moment, the IA value
of 0.270 $fm^2$ is increased by two-body contributions to 0.275 only, while 
experiment finds 0.286. For a discussion of the $rms$-radius, see 
sect. \ref{deutrms}.  Other types of calculations find similar 
differences, the origins  of which are not understood.
 
We next address calculations that belong  to the second class of approaches 
mentioned at the beginning of this section. Relativity is imposed from the 
very beginning by requiring that currents and interactions are consistent with 
Poincar\'e invariance. Several groups have published results belonging to this
category
\cite{Gross74}\nocite{Arnold80,Zuilhof81,Hummel90}-\cite{Devine93},\cite{VanOrden95}.
We here can discuss only some selected ones.

The calculation of Hummel and Tjon \cite{Hummel94}  employs a quasi-potential 
approximation to the Blankenbecler-Sugar equation. The description
of the nucleonic and mesonic structure is performed within the 
one-boson-exchange (OBE) model, with the various coupling  constants fitted to 
the experimental  phase shifts of N-N scattering with the same relativistic
frame work. 
For the results shown here, the Hoehler nucleon form factors have been 
employed. To the form factors resulting from this relativistic IA (RIA) 
calculation, the contributions of the $\rho \pi \gamma$ and 
$\omega \epsilon \gamma$ MEC diagrams have been added, with form factors taken 
from the vector dominance model (VDM).  

In figs. \ref{fc0}-\ref{fc2} we also display the results of the calculations
by vanOrden, Gross and Devine \cite{VanOrden95}. These authors also start
from the Bethe-Salpeter equation, which has been reduced to a quasi-potential 
equation by assuming that one of the nucleons is on mass shell. This calculation again is
Lorentz covariant and gauge invariant. The parameters of the OBE interaction 
have directly been fit to the N-N scattering data. The calculation includes the contribution
from the $\rho \pi \gamma$ exchange current, with vertex form factors coming
from model calculations.  Contrary to most other calculations cited above, 
here the dipole nucleon form factors have been used, together with a particular 
prescription for the off-shell nucleon form factors.

It is found that wave function components which occur only in the covariant
description, {\em e.g.} the p-state contribution, give a rather small contribution,
even at large $q$. The main effect of the fully relativistic description 
then must be assumed to come from the use of a fully relativistic electromagnetic
 operator, as has been found in refs. \cite{Jeschonnek00,Amaro98}.

When comparing the results of the different calculations, it becomes clear that
the $C2$ form factor is not very sensitive to the dynamical ingredients of
the calculations. The same is true for $A(q)$ which, over a significant part of the 
$q$-range, is dominated by the $C2$ multipole form factor. 
The $C0$ and the $M1$ multipolarities 
are the ones that allow for  the most sensitive test of our understanding of 
deuteron structure and two-body effects. 

When considering all three form factors, the calculations using the Bonn-B, 
Paris and V18-potentials yield the best overall agreement with the data. The 
deviations of the Bonn-A curves (C2 form factor falling too quickly, M1 
diffraction zero at too low a $q$) are characteristic for a wave function
with too large a ''defect'' at small $r$ in the D-state wave function (leading 
to a low  D-state probability).

The role played by the neutron electric form factor has been investigated
several times, most recently by Plessas \et\ \cite{Plessas00}. As pointed out
in sect.\ref{nucleon} the deuteron charge observables $F_{C0}^2$, $F_{C2}^2$
 and $A(q)$ basically contain a factor $(G_{ep}(q)+G_{en}(q))^2$. Large 
values of $G_{en}(q)$ at large $q$, as given {\em e.g.} by the parameterization of 
Gari and Kr\"umpelmann \cite{Gari92} can lead to a large factor. For this 
parameterization the ratio   of the predicted \gen\ to the experimental \gep\ at
$q = 8 fm^{-1}$ (the upper limit of the charge form factors shown in figs. 
\ref{fc0},\ref{fc2}) is \gen / \gep $\sim 1.2$, while the Hoehler 
parameterization yields 0.3; the data only provide an upper limit of $\sim$0.8. 

In order to judge the agreement between theory and data at the higher momentum
transfers where no separation of $C0$ and $C2$ is possible, we show in 
fig.~\ref{ab} the structure functions $A(q)$ and $B(q)$. At the highest momentum
transfers, the non-relativistic calculations with relativistic corrections
may increasingly get doubtful, while the covariant calculations still
could be valid. The comparison between theory and experiment, however, is not
indicative of a real breakdown of the calculations which start from a non-relativistic
frame work. The  two-body terms that are poorly controlled --- 
such as the $\rho \pi \gamma$ term --- are mostly responsible for problems at
large $q$. The calculations that use a soft $\rho \pi \gamma$ 
form factor do better than the ones using a hard one such as given by the VDM.

While most of the calculations use the Hoehler nucleon form factors,  the
calculation of \cite{VanOrden95} uses the dipole parameterization. The Hoehler
$G_{ep}$, the most important nucleon form factor entering the calculation
of the deuteron charge form factors, actually does quite well up to 
$q\sim 9 fm^{-1}$ in comparison to the precise $G_{ep}/G_{mp}$ ratios 
recently measured at JLAB by Jones \et\ \cite{Jones00}; the comparison between data and theory 
shown in this section thus should not be affected by shortcomings in
the proton form factors employed. The exception might be the calculation of 
$A(q)$ of ref. \cite{VanOrden95} at the highest $q$'s, where the use of the dipole
parameterization does lead to too large an $A(q)$.   

The success up to large $q$ of the calculations that start from a basically 
non-relativistic frame work --- supplemented by $v^2/c^2$ relativistic 
corrections --- may at first sight come as a surprise. Upon reflection, 
this may not be so astonishing, however, as the relation between $q$ and 
the typical nucleon momenta probed is somewhat special for the A=2 system.
For a heavy nucleus the typical momenta probed are of order $q/2$: when
describing the form factor using a momentum space wave function, the biggest
overlap between initial and final state occurs for $\Psi (-\vec{q}/2) \Psi (
-\vec{q}/2 + \vec{q})$. For the deuteron, the center-of-mass 
motion leads to the fact that this overlap is maximal at initial nucleon
momenta $\sim q/4$; for the momentum transfers reached by experiment, $q/4$ 
does not yet correspond to genuinely relativistic momenta.  
  
A source of uncertainty in the prediction of the deuteron form factors 
resides in the contribution of $\Delta \Delta$ components. Various groups 
have studied wave function components involving the $\Delta$ or higher nucleon
resonances 
\cite{Plessas94}-\nocite{Fabian74,Gari76b,Weber78,Amghar99,Blunden89,Dyrmarz90}\cite{Sauer86}
 using $N-\Delta$ transition 
potentials derived within OBE  or boundary condition models. The calculation
of the $\Delta$ components turn out to be difficult, and partly ambiguous, 
as the  $NN \rightarrow \Delta \Delta$ transition is of very short
range; the corresponding cut-off and vertex form factors are poorly known. As a 
consequence,
there is little agreement between the calculations. While several approaches
give $\Delta \Delta$ probabilities in the 0.5-1\% range, some give much larger 
probabilities. There is no agreement either on  whether the contribution to   
{\em e.g.} $A(q)$ is positive or negative at low $q$. It seems clear that the 
$\Delta \Delta$ components have a larger effect on $B(q)$ (near the diffraction
minimum). It also emerges that the effects of $\Delta \Delta$ components 
are significantly smaller than the ones of the standard two-body currents. 
Given the uncertainties of the latter, it is very difficult to learn from the
comparison of theory and experiment something on the $\Delta \Delta$ contribution.

The calculations compared above to the experimental results obviously only
represent a small fraction of the presently available theoretical predictions.
Calculations of the \de\ form factors  within different theoretical 
frame works are available: for more detailed consideration of  alternative 
relativistic approaches such as the light 
front \cite{Carbonell98} and  equal-time 
approximation to the solution of Bethe-Salpeter equation \cite{Phillips98}, 
we refer the reader to the original literature. 

A rather different framework has been employed by Buchmann, Yamauchi and 
Faessler \cite{Buchmann89}. These authors use the quark cluster model and
the technique of the resonating-group method. The \de\ 
wave function is derived from a microscopic six-quark Hamiltonian which, in 
addition to the quark confining potential, includes a one-pion and one-gluon 
exchange potential between quarks. In this calculation the electromagnetic 
current operators are constructed directly on the quark level. 

\begin{figure}[htb]
%Figur mit topp/sideways hergestellt, modif boundingBox: 66 206 516 556
\centerline{\mbox{\epsfysize=7cm \epsffile{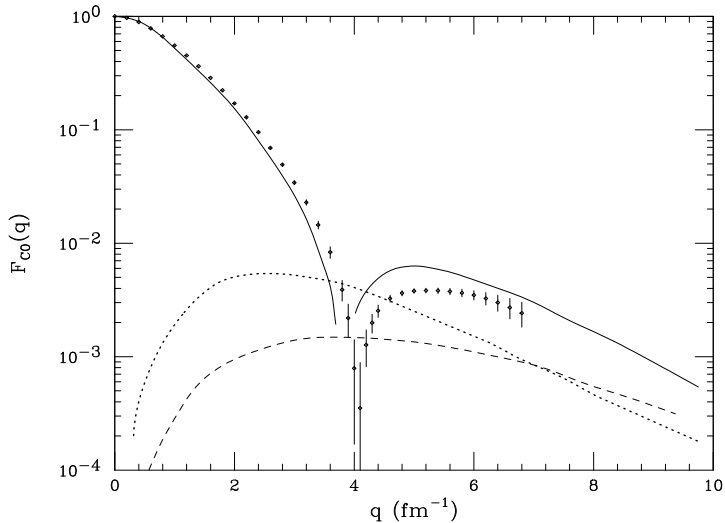}}}
\begin{center} \parbox{14cm}{\vspace*{-0.5cm} \caption{
\label{fc03} 
Data for the deuteron monopole form factor compared to the quark cluster model 
predictions of Buchmann {\em et al.} \protect{\cite{Buchmann89}}. Dotted curve: conventional
pion exchange current, dashed curve: additional quark exchange current, solid:
full calculation. 
} }  \end{center}
\vspace*{-0.4cm} \end{figure}

The particular interest in this approach lies in the fact that, due to the 
application of the Pauli principle on the quark level, there are new quark 
interchange terms in both the one- and the two-body current matrix elements 
that are not present in a nucleon-level calculation. The occurrence of these
terms can perhaps best be used to gauge the relevance of a six-quark (rather
than two-nucleon) description of the \de. Fig.~\ref{fc03} shows some of the
results of this calculation. While it is probably not appropriate to 
quantitatively compare the one-body results  
to the data, the dashed curve shows the new  exchange terms that appear in 
addition to the standard ones (dotted line).  

These additional terms clearly become important at the very large momentum 
transfers. Similar results have been obtained by Buchmann \et\ for the M1 and 
C2 form factors, where the contribution of the quark exchange currents is 
smaller, however.  

The \de\ form factors in the past also have been discussed in terms of a 
very different approach. The large momentum transfers achieved experimentally,
 and the smooth
fall-off of $A(q)$, made it tempting to look at the asymptotic behaviour of 
the \de\ form  factors in terms of the quark model \cite{Brodsky76}
(the asymptotic behaviour in terms of nucleonic constituents 
\cite{Alabiso74,Alabiso75} curiously received much less attention).

 These predictions, starting
from quark counting rules, were applied despite the fact that the condition 
for the applicability, $q/6 \gg 3fm^{-1}$, where $3fm^{-1}$ is the typical 
momentum of quarks and $6$ is the number of (valence quark) constituents in the deuteron, 
is far from true ($12/6 \not\gg 3$). These considerations on the
asymptotic behaviour also continued to be used despite the fact that the estimates
for the 6-quark-contribution at the $q$'s of interest were orders of magnitude
below the data \cite{Isgur89}. Today, with the experimental proof that the form
factors display diffraction features (not $q^{-n}$ behaviour) and are 
{\em negative} at the largest $q$  (while the counting-rule predictions
are positive as the normalizing constant is a probability), these quark counting
guesstimates have fallen somewhat out of fashion.
%\newpage
%\input{deutdis.tex}
\subsection{Electrodisintegration \label{deutdis}}
In addition to the elastic form factors, there is one transition form
factor to a nearly discrete state of interest when discussing the 
form factors of few-body 
nuclei --- the M1 transition to the singlet-S state of the deuteron right above 
disintegration threshold. This transition form factor has gained particular 
prominence as it has become one of the showcases for the role of two-body 
currents. We therefore include it in the discussion here.

The first data for the M1 transition were measured by Rand \et\ at 180$^\circ$
in conjunction with a measurement of the \de\ magnetic form factor
\cite{Rand67}. These 
measurements reached a momentum transfer $q$ of 3.1 $fm^{-1}$ and showed, 
despite the limited resolution, the enhancement due to the singlet-S 
state. These  data, however,
did not attract particular attention at the time, as they appeared to
agree with the then accepted IA-theory of Jankus. 

Only later on it was  realized by Hockert \et\ \cite{Hockert73}
that, similar to the A=3 magnetic form factor, the transition form factor 
in IA should display a pronounced minimum near $3.5 fm^{-1}$ due to the destructive 
interference of the $^3$S-$^1$S and $^3$D-$^1$S transitions. This D-S 
transition had not been 
included in the Jankus calculation, as the deuteron ground state was assumed
to be a pure S-state.  The occurrence of this minimum leads to a 
pronounced disagreement of IA-calculations with the data, by a factor of 10. 
Only when allowing for two-body currents agreement with the data can be restored. 
As for A=3, the two-body currents give a  basically opposite contribution to the S-D 
transition. The large effect of the two-body terms is essentially a consequence of the fact that 
electrodisintegration corresponds to an M1 isovector transition. 

\begin{figure}[htb]    
%Figur mit topp/sideways hergestellt, modif boundingBox: 66 206 516 556
\centerline{\mbox{\epsfysize=6cm \epsffile{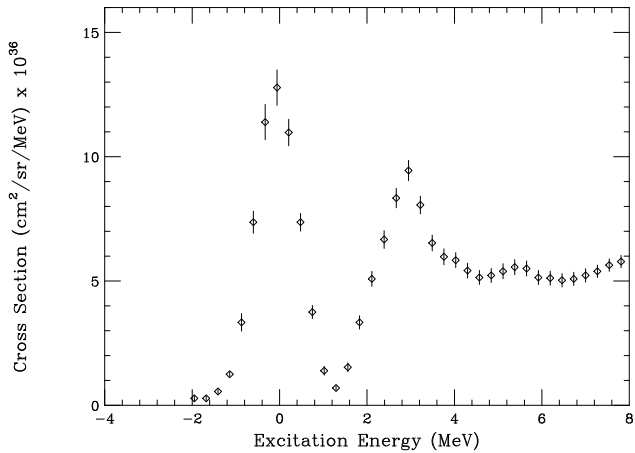}}}
\begin{center} \parbox{14cm}{\vspace*{-0.5cm} \caption{ 
\label{cusp}  Inclusive cross section for $d(e,e')$, displaying the elastic 
peak ($E_x = 0$) and the enhancement at low E$_{x}$ due to the transition to the 
singlet-S state \protect{\cite{Bernheim81}}.
} }  \end{center}
\vspace*{-0.4cm} \end{figure}
Detailed measurements of deuteron electrodisintegration then were performed
at low $q$ at Mainz \cite{Simon76,Simon79} who made an L/T-separation 
up to 2 $fm^{-1}$. Measurements up to high $q$ were done at 
Saclay by Bernheim \et\ and Auffret \et\ \cite{Bernheim81,Auffret85} who 
measured the cross section at
a scattering angle of 155$^\circ$. This experiment provided precise data 
up to momentum transfers of $5.2fm^{-1}$, thus covering fully the region
where the IA-minimum and maximum  of the cross section would occur. 

\begin{figure}[htb]    
%Figur mit topp/sideways hergestellt, modif boundingBox: 66 206 516 556
\centerline{\mbox{\epsfysize=7cm \epsffile{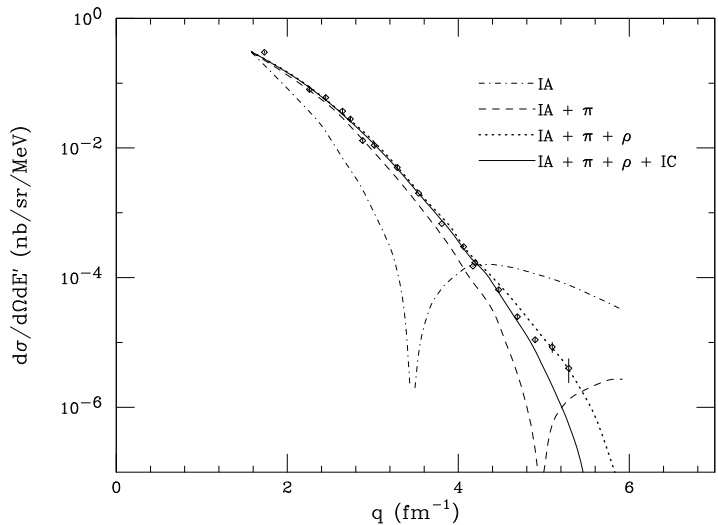}}}
\begin{center} \parbox{14cm}{\vspace*{-0.5cm} \caption{ 
\label{deenpsac}
Cross section for deuteron electrodisintegration in the region of the IA 
minimum, averaged over the range 
E$_{np}$ = 0--3 $MeV$ \protect{\cite{Bernheim81}}, compared to the calculation of 
Mathiot \protect{\cite{Mathiot84}} (for E$_{np}$ = 1.5 $MeV$).   
} }  \end{center}
\vspace*{-0.4cm} 
\end{figure}

Fig.~\ref{cusp}
shows the enhancement of the cross section right at threshold due to the 
$^1$S-state. 
In fig.~\ref{deenpsac} the data are compared to an early calculation of 
Mathiot \cite{Mathiot84}. The IA calculation, accounting for the nucleonic
contribution only, shows a pronounced interference minimum which does not 
occur in the data. The pionic two-body term gives the main contribution that brings theory
closer to experiment. The $\rho$-exchange currents and isobar components 
produce non-negligible contributions as well, but the quantitative 
contribution was very sensitive to the cut-off masses assumed for the 
vertex form factors. 

When comparing data and calculations in general, some caution is advisable. 
While the data in most cases are averaged over the region of
energy $E_{np}$=0--3 $MeV$ above disintegration threshold (0--10 $MeV$ for the SLAC data),
 the calculations often consider
a fixed  energy of 1.5 $MeV$ above threshold. Given the cusp-shaped 
behaviour of the cross 
section near threshold (see fig.~\ref{cusp}), this is not entirely equivalent 
and can, depending on
momentum transfer, lead to appreciable differences. In particular, near the minimum
at $q = 5 fm^{-1}$, the minimum for E$_{np}$ = 1.5 $MeV$ is much sharper than
the one in the excitation-energy averaged data which includes contributions from multipolarities
other than M1 which dominate at the larger $E_{np}$.

\begin{figure}[htb]    
%Figur mit topp/sideways hergestellt, modif boundingBox: 66 206 516 556
\centerline{\mbox{\epsfysize=7cm \epsffile{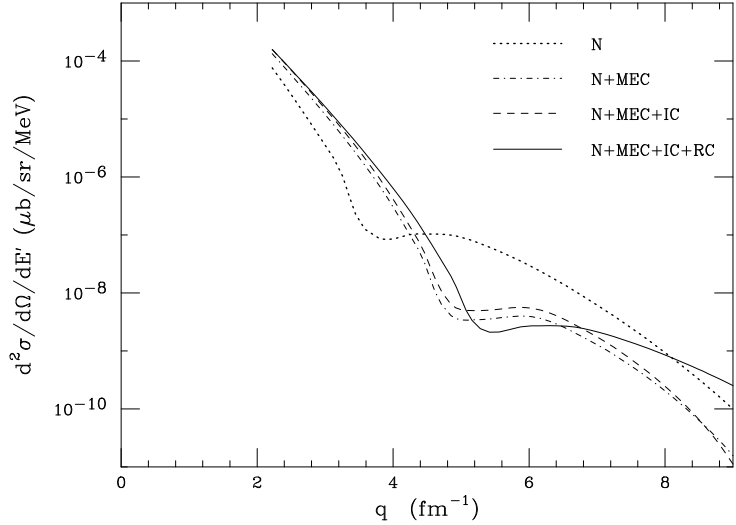}}}
\begin{center} \parbox{14cm}{\vspace*{-0.5cm} \caption{ 
\label{deenp3}
Prediction for  d(e,e') cross sections up to large momentum transfer by
Ritz \et\ \protect{\cite{Ritz97}}.   
} }  \end{center}
\vspace*{-0.4cm} \end{figure}

\begin{figure}[htb]    
%Figur mit topp/sideways hergestellt, modif boundingBox: 66 206 516 556
\centerline{\mbox{\epsfysize=7cm \epsffile{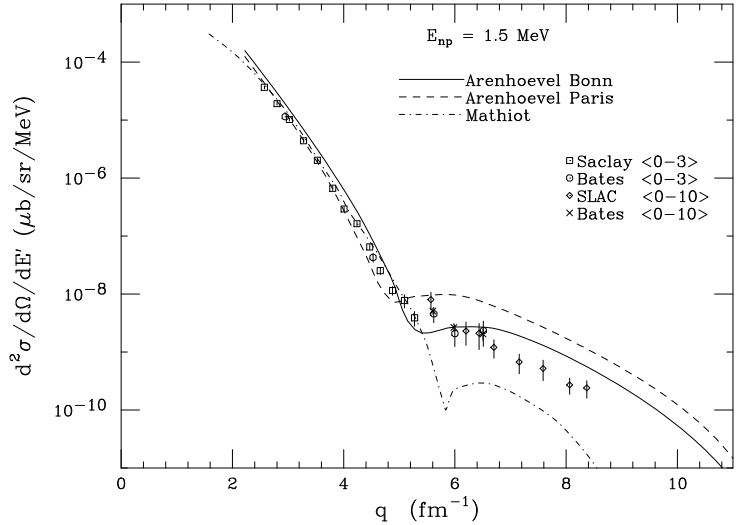}}}
\begin{center} \parbox{14cm}{\vspace*{-0.5cm} \caption{ 
\label{deenp1}
World data for d(e,e') integrated over E$_{np}$-regions as indicated. The cross
sections are compared to calculations including MEC of Mathiot 
\protect{\cite{Mathiot84}} and the group of H. Arenh\"ovel
\protect{\cite{Ritz97}}, for two different N-N potentials.   
} }  \end{center}
\vspace*{-0.4cm} \end{figure}

\begin{figure}[hbt]    
%Figur mit topp/sideways hergestellt, modif boundingBox: 66 206 516 556
\centerline{\mbox{\epsfysize=7cm \epsffile{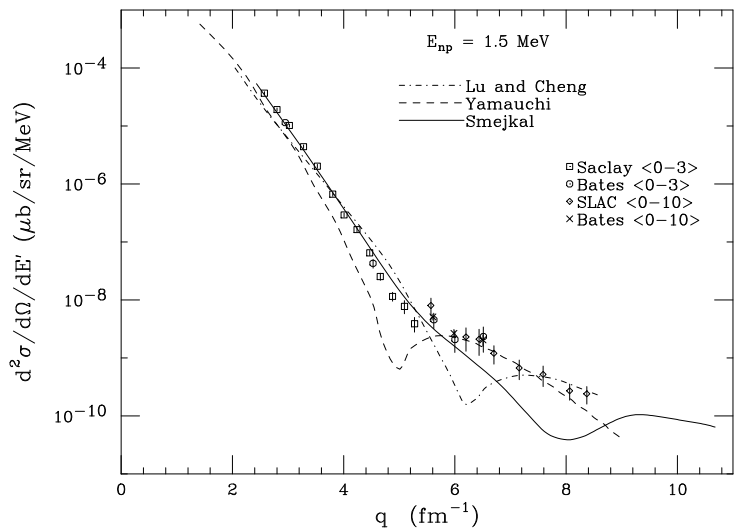}}}
\begin{center} \parbox{14cm}{\vspace*{-0.5cm} \caption{ 
\label{deenp2}
Cross sections for d(e,e') predicted using quark-based models 
\protect{\cite{Yamauchi84,Lu96}}, or a chiral
Lagrangian \protect{\cite{Smejkal97}}.  
} }  \end{center}
\vspace*{-0.4cm} \end{figure}

Additional data have been provided by an experiment performed at Bates \cite{Schmitt97}
and the NE4-experiment at SLAC \cite{Frodyma93}. This latter experiment, which
reached the highest momentum transfers, was optimized to measure 
at $\theta$ = 180$^\circ$ the elastic
magnetic form factor, and had a fairly poor energy resolution (as much as 20 $MeV$). 
After unfolding of this resolution the cross section over the interval
0--10 $MeV$ could be extracted. The complete set of data referred to above 
will be used in figs.~\ref{deenp3}-\ref{deenp2} when comparing to theory.

The calculation of the two-body contributions has made considerable progress since the early 
calculations cited above. In particular, it is now possible to derive 
the $\pi$ and the $\rho$-like exchange currents consistently with the 
nucleon-nucleon potential employed for the calculation of the nucleonic
part. Also, the 
relativistic effects of order v$^2$/c$^2$ can be fully included, thus removing
the earlier ambiguity of whether to use $G_e$ or $F_1$ as vertex form factors.
The isobar contribution continues to be somewhat uncertain, as it can not
be fixed via the N-N potential and current conservation.

Fig.~\ref{deenp3} shows, for a modern calculation \cite{Ritz97},
the different contributions to the cross section, averaged over the region 
E$_{np}$= 0--3 $MeV$. Clearly, the two-body terms lead to the biggest change relative to 
IA in the region 2--6 $fm^{-1}$.  Isobar currents (IC) somewhat increase the cross section
without changing much the overall $q$-dependence. Relativistic effects (RC) become
very important at the largest momentum transfers.

As shown by fig.~\ref{deenp1}  the calculated cross sections are reasonably
sensitive to the N-N potential employed. This is  mainly due to the 
different predictions for the D-state, where the energy-dependent Bonn potential
produces quite different results as compared to {\em e.g.} the Paris 
potential. One also should note that the calculated cross sections are 
not too sensitive to the nucleon form factors employed \cite{Schiavilla91} as
 disintegration at threshold is a magnetic transition, so \gen\ is not
important.

The large momentum transfers reached, together with the ambiguities of the 
calculations of the two-body terms occurring when dealing with inter-nucleon 
distances that are comparable to the nucleon  size, 
have led to alternative attempts to calculate the disintegration 
cross section. The results of some of these approaches are displayed in 
fig.~\ref{deenp2}. Lu and Cheng \cite{Lu96} use a hybrid model 
\cite{Cheng86} where the deuteron is  divided into two distinct regions: an 
exterior one described in terms of baryons, and an interior one described in
terms of six-quark configurations. The matching radius chosen was near 1$fm$. 
This model predicts a second maximum of the cross section at 
$q \sim 7 fm^{-1}$, a maximum taken as a signature for a six-quark 
contribution. This maximum, however, was not seen in the data.

Quark-based calculations have been published by Yamauchi \et\ \cite{Yamauchi84}
and Chemtob \et\ \cite{Chemtob86}. In the calculation of Yamauchi \et\ the 
two-nucleon system is described using the resonating-group method. The 
quark-quark interaction includes quark interchange effects, 
with gluon-exchange corrections, and pion-exchange effects.
While the calculation is quite successful at large momentum transfer, 
it predicts too pronounced a minimum at $q=5fm^{-1}$, which would only 
partly be filled in when averaging over the experimental E$_{np}$-range. 

Smejkal \et\ \cite{Smejkal97} construct a 
Lagrangian in terms of N, $\pi$, $\rho$ and A$_1$. The exchange terms include 
$\pi$ and $\rho$ exchange and $A_1-\pi$ exchange. The resulting cross 
sections are also shown in fig. \ref{deenp2}.    
  
The comparison to the data shows that the calculations in terms of nucleons
with added two-body terms consistently derived from the N-N interaction 
are quite successful 
in explaining the data. The calculations in terms of quark constituents 
for the time being do not correctly reproduce the data. Part of the success
of the conventional approach may again be due to the special nature of the 
deuteron mentioned already above. The typical internal momenta probed in the
deuteron wave function are of order $q$/4, {\em i.e.} less than 2 $fm^{-1}$ for 
electrodisintegration; at these momenta, the conventional description in 
terms of nucleons and mesons still
can be expected to work well, while quark-model based approaches could be 
expected to be valid at much higher $q$'s only.
\section{3-body nuclei}
\subsection{Introduction}
Three-nucleon elastic form factors are particularly useful for assessing our understanding
of the few-nucleon systems and our ability to quantitatively predict nuclear 
properties based on  the knowledge of the nucleon-nucleon interaction.
The wave functions and form factors  can reliably 
be calculated starting from modern nucleon-nucleon interactions known from
the two-nucleon system; two-body 
exchange currents can be taken into account in a way that is largely 
consistent with the nucleon-nucleon interaction. 
Apart from a much higher nucleon density, the main difference between the two- and 
three-nucleon  
systems is related to the presence of the third nucleon, which can lead to the 
appearance of three-nucleon forces. 
The most sensitive observables to test these  theoretical 
predictions are the electromagnetic elastic form factors. The beak-up cross section at 
low energy loss, which in analogy with the deuteron potentially could also
provide interesting information, has hardly been exploited up to now 
\cite{Viviani99}.

 For momentum transfers below
$5 fm^{-1}$ the one-body contribution (impulse approximation) 
is now considered to be quantitatively  under control. It is a well known fact, however, 
that the impulse approximation alone can not explain  the magnetic 
or the charge form factors. Two-body currents as well as relativistic effects 
have to be added in order to get satisfactory agreement with the data. 

All trinucleon form factors are
 accurately known up to momentum transfers $q$ of the order of 5.5
$fm^{-1}$, transfers large enough to cover the region of the secondary diffraction 
maxima. These data are now adequate for an extensive comparison with the 
present theoretical calculations performed within the framework of nucleonic 
constituents and two-body currents. Up to the momentum transfers where
 the \het\ and \hyt\ form 
factors reach  similar values of $q$, this comparison can be extended to 
the combinations of isospin T=0 and T=1, thus  allowing a direct confrontation 
with form factors for the 2- and 4-body systems.
These include  the deuteron charge and magnetic form factors
which are both pure $\Delta$T~=~0 transitions, and deuteron electrodisintegration
at threshold, which is a $\Delta$T~=~1 transition. 

The \het\ and \hyt\ magnetic form factors are of particular interest due to 
the destructive interference that occurs at large $q$, where the S-S and the S-D
terms partly cancel \cite{Brandenburg74}. This leads to a pronounced shift in $q$ 
of the predicted interference minima. The A=3 magnetic form factors are also
of special interest when studying the effects of two-body currents. Already 
at $q=0$ their contribution is sizeable: the magnetic moments in IA for 
\hyt(\het) are 2.57 (--1.76) magnetons \cite{Carlson98}, while the moments
including two-body effects amount to 2.98 (--2.09).

\subsection{Electron scattering data \label{hhedata}}
The \hyt\ and \het\ nuclei present particular challenges for the experimentalist.
For both nuclei targets of adequate thickness are difficult to produce and use,
particularly of course for the radioactive \hyt, which we discuss first

The pioneering experiment on \hyt\ was performed by Collard \et\ at Stanford
\cite{Collard65}. For both 
\hyt\ and \het\ high-pressure (100bar) sealed gas targets in the form
of cylinders of $2cm$ diameter and $20cm$ length were used. The \hyt\ target 
contained as much as 25'000 Curies.  The \hyt\ and \het\ data could be 
measured quasi-simultaneously, thus ensuring low systematic errors in the 
relative cross sections. Data were taken up to a momentum transfer of 
2.8$fm^{-1}$ 
for both the charge and the magnetic form factors, so the data did not yet
reach the diffraction feature located at larger $q$. 

An experiment using a somewhat different  technology for the gas-target
was carried out more than 20 years later   by Beck \et\ \cite{Beck87} using the Bates 
accelerator.  Vertical cylinders of $\sim$10$cm$ diameter were used as target.
The \hyt\ was stored in a Uranium oven, and transferred to the target which was kept
at 45K temperature and a pressure of 15 bar, yielding  a target thickness 
of $\sim 50 mg/cm^2$.  
This target contained a rather large quantity of \hyt, 140 kC. Contrary to the 
target of ref.~\cite{Collard65} the target-system did require handling 
 of the radioactive gas in the experimental hall.   
This experiment, which also reached momentum transfers of $\sim$ 2.8 $fm^{-1}$, 
 did achieve higher accuracy as the luminosity was considerably higher than 
the one utilized in ref.~\cite{Collard65}. The use of \hyt\ not too far from the 
point of liquification did, however, introduce some uncertainty on the
target thickness as the thermodynamical properties of \hyt\ have to be 
extrapolated from Hydrogen.
Again, \hyt\ and \het\ data were measured in the same experiment. This 
experiment also measured quasi-elastic scattering data \cite{Dow88}.

\begin{figure}[htb]
%Figur mit topp/sideways hergestellt, modif boundingBox: 66 206 516 556
\centerline{\mbox{\epsfysize=10cm \epsffile{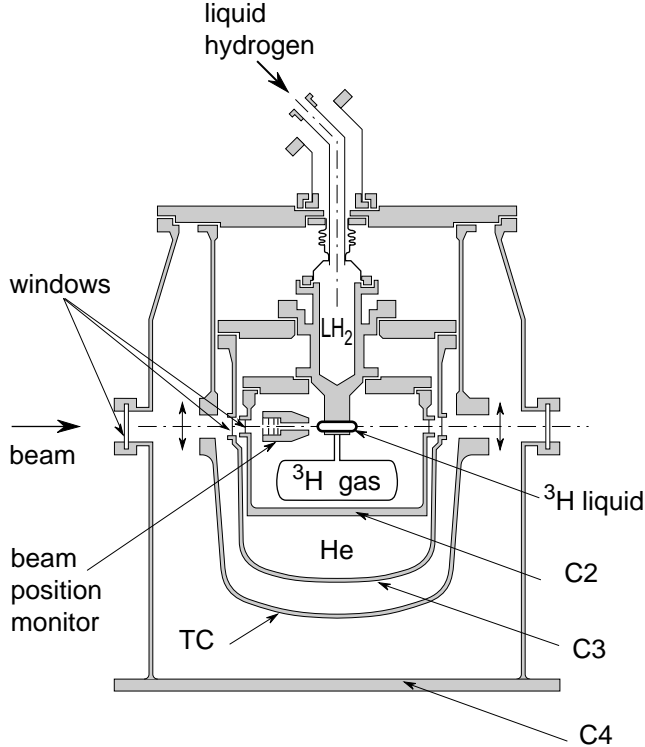}}}
\begin{center} \parbox{14cm}{\vspace*{-0.5cm} \caption{
\label{target} Saclay tritium target. The volume between containers C3 and C4
is filled with gaseous $^4 He$ for a continuous check of integrity  of C3 and
C4. The volume C4 could retain the tritium at pressures below 1 bar. 
} }  \end{center}
\vspace*{-0.4cm} \end{figure}

For the Saclay experiment \cite{Juster85,Amroun94} a very different technology 
was utilized.
Amroun \et\ used a  target containing {\em liquid} tritium, as only this 
option provided the luminosity required to reach momentum transfers beyond the 
expected diffraction features in $F_{ch}$ and $F_{m}$. For reasons of 
safety,  a {\em sealed} cryogenic system was employed that 
allowed to liquefy \hyt\ without any radioactive-gas handling, except for the 
initial fill and final gas recuperation carried out in a specialized laboratory.
 Again for safety 
reasons, the tritium target was enclosed in three further sealed containers with 
thin windows for the electrons, and a massive container closed during 
transportation of the target. The general layout of the target is 
shown in figure \ref{target}.

The \hyt-target proper consisted of a stainless steel cylinder, 40$mm$ 
long, 10$mm$ in diameter with hemispherical end caps located outside the 
acceptance of the spectrometer. 
%The wall thickness of the 
%cylinder was 0.5~mm. The thickness of the end caps varied between 0.5 and 
%0.05~mm from the edge to the axis of the target. 
%At our extreme angle of  
%155$^{o}$, they were shielded by  an additional collimator (see below). 
%The calculated bursting pressure of the target was 110 bars. Destructive tests
%of similar cells lead to bursting pressures that were in agreement with the
%calculation.
The target was in permanent connection with an expansion-vessel of 
180$cm^{3}$ volume. This vessel was equipped with a heater, in order to 
keep it at T $\sim$ 300K while the target was cold. When the entire system 
was at room temperature, the tritium pressure was 23 bars.

This combination of cold target and warm storage vessel had a number of
outstanding benefits, such as high target density (liquid \hyt), low 
operating pressure  while the  target was placed in the 
high-intensity electron beam (3 bars), very high efficiency in the use of the 
\hyt\ 
(98\% in the cold target cell), well known density of liquid \hyt, and thin 
target windows. Only 10kC of tritium  were needed to reach the high
target thickness.

Use of a liquid target with minimized amounts of \hyt\ involves a large amount 
of heavy nuclei (Cu, Fe) near the beam. A uranium collimator 
(diameter 8~$mm$) 
upstream of the target and outside the spectrometer acceptance eliminated 
potential stray electrons. A secondary emission monitor upstream of the 
target verified that no beam-halo was present. A  split secondary 
emission monitor ensured stabilization of the beam position to  0.1  $mm$ 
accuracy.

For the extreme forward- and backward-angles 
(25$^{o}$ and 155$^{o}$) the end caps of the target were not entirely 
outside the acceptance of the spectrometer. Insisting on this condition 
would have led to 
an undue increase of target length ({\em i.e.} quantity of \hyt). For the extreme 
angle of 155$^{o}$ --- which is of great interest to the separation of 
$F_{ch}$ and $F_{m}$ --- a special collimator  placed close to the 
\hyt\ target eliminated all contributions from the end caps. 
%It also led to a region of inaccessible 
%angles between 110$^{o}$ and 150$^{o}$, which did not impair the quality of
% the experimental results.

%For the tritium experiment, we needed to  accurately determine the target 
%density, as we intended to measure {\em absolute} cross sections. 
%The temperature and pressure of \hyt\ were measured with calibrated 
%sensors, and continuously registered by the data acquisition computer 
%during all runs. The liquid tritium pressure and temperature, LH$_{2}$ 
%pressure and temperature, provided (under thermal equilibrium conditions) 
%four independent measurements of the  \hyt\ density, allowing to identify 
%potential problems with time-dependent changes of calibrations, or 
%inaccuracies during the initial calibration. The quantities required to 
%determine the 
%absolute density of the target at zero beam intensity are given in 
%the reports of Roders {\em et al} \cite{Roder73}, Souers {\em et al}
% \cite{Souers73} and Guizouarn  
%\cite{Guizouarn65}. 
At the nominal temperature of 21.7K the target density 
was 0.2710$\pm$0.0014~$g/cm^2$. This led to a thickness of $\simeq$ 1 $g/cm^2$
 at \mbox{$\theta$ = 155$^\circ$}. With this large target thickness, data could
be taken at momentum transfers where the cross section is very small. 
As a consequence, the region of the expected diffraction minimum and 
secondary maximum could be covered, and the experiment reached a maximum $q$ of
5.7$fm^{-1}$.

For \het\ the life of the experimentalist is somewhat easier, as no 
precautions due to radioactive gas are required; achieving an adequate target
thickness  still is a challenge.

Besides the experiments mentioned already above \cite{Collard65,Beck87} several 
groups have contributed to the \het\ data base. An experiment performed at 
Mainz by Ottermann \et\ \cite{Ottermann85} provided data up to $q=1.9 fm^{-1}$,
lower-$q$ data were measured by  Szalata \et\ \cite{Szalata77} and Dunn \et\
\cite{Dunn83}. 

The region of the diffraction minimum and secondary maximum
of the charge form factor was first reached by the experiment of 
McCarthy \et\ \cite{McCarthy70,McCarthy77} who, contrary to all other 
experiments, used a {\em liquid} target cooled to 1K using superfluid Helium.
The diffraction feature in the 
magnetic form factor was reached in the experiment of Cavedon \et\ 
\cite{Cavedon82}.    

The data on the charge form factor going to the largest $q$ 
were taken by Arnold \et\ \cite{Arnold78} in a coincidence experiment 
performed at SLAC. Due to the cryogenic target (42$cm$ of \het\ at 20K and 50 bar)
a large target thickness was achieved, $\sim 5 g/cm^2$. In combination with
the high energy of $E_e$=15 GeV used --- the cross section at a given $q$ is 
roughly proportional to $E_e^2$ ---  very small form factors 
(F $\sim 3 \cdot 10^{-5}$) 
could be measured. This allowed Arnold \et\ to reach a maximum momentum 
transfer of 10$fm^{-1}$, and cover the region of the second diffraction zero. 
 
\begin{figure}[htb]
%Figur mit topp/sideways hergestellt, modif boundingBox: 66 206 516 556
\centerline{\mbox{\epsfysize=9cm \epsffile{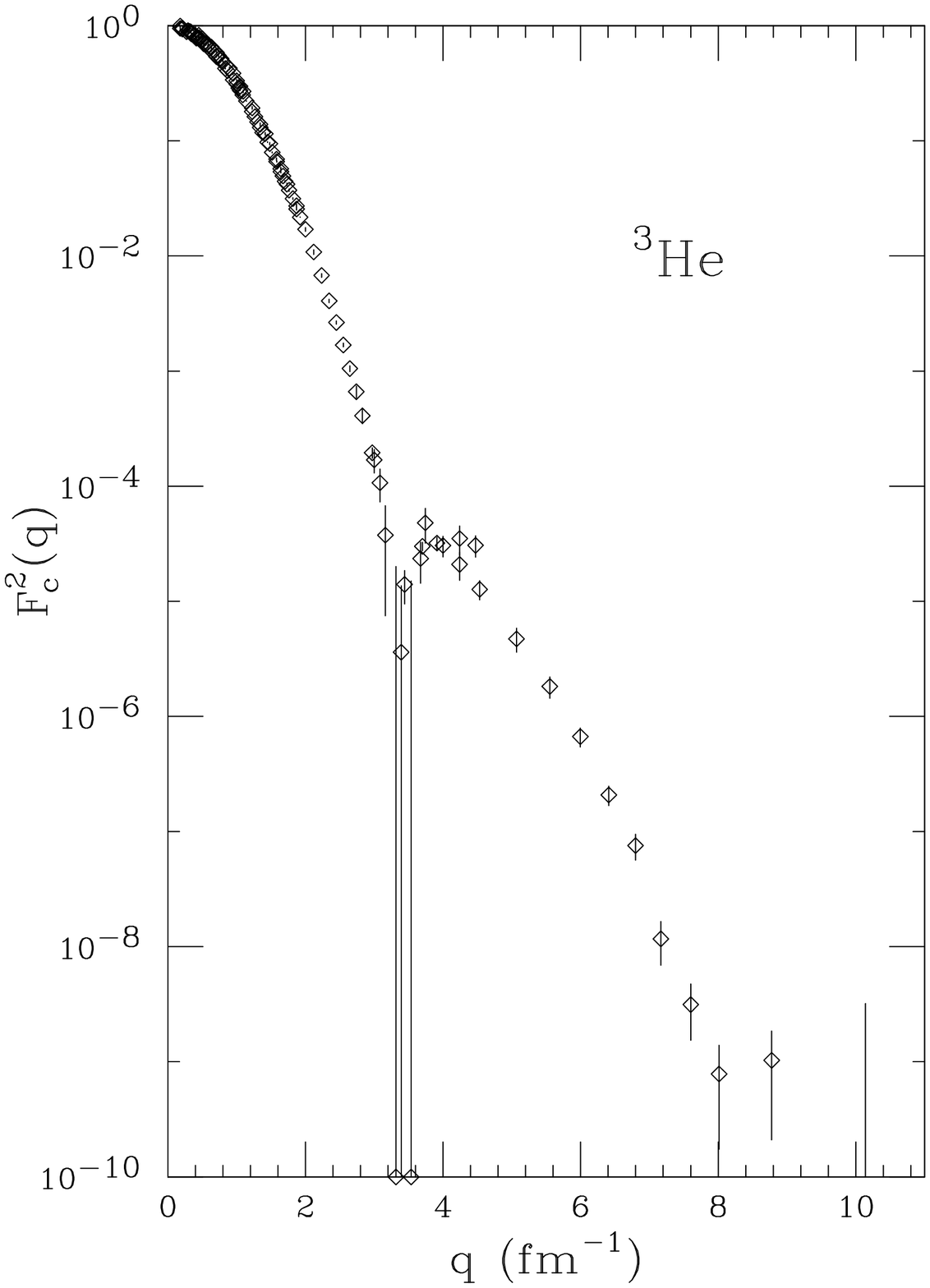}
\hspace*{5mm}
\epsfysize=9cm \epsffile{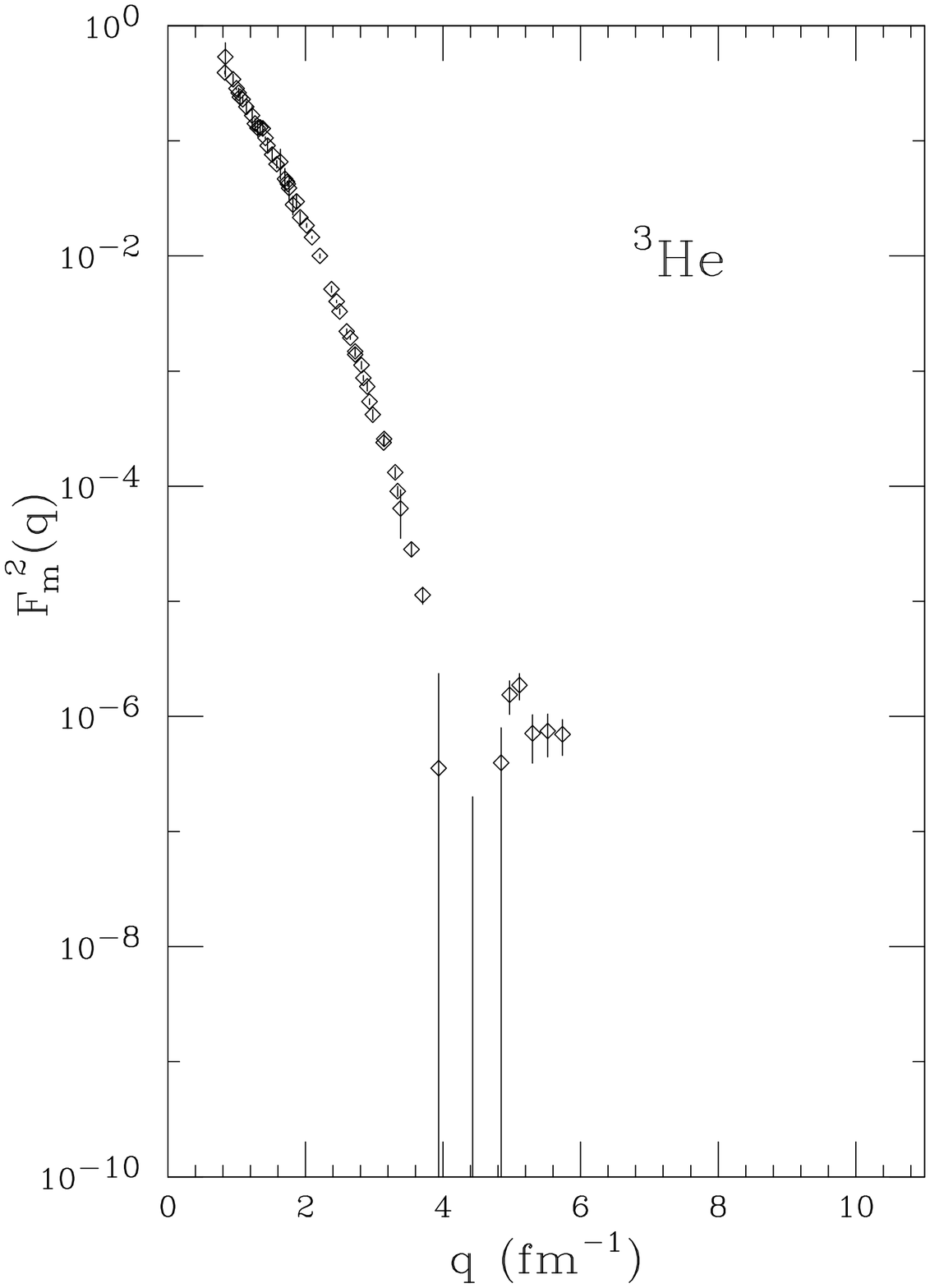}}}
\begin{center} \parbox{14cm}{\vspace*{-0.5cm} \caption{
\label{fcmhe3} 
Charge and magnetic form factors of \het.
} }  \end{center}
\vspace*{-0.4cm} \end{figure}

The form factors  presently available, as extracted directly from the various 
experiments, are shown in fig. \ref{fcmhe3}. 
%
%\cite{Collard65,McCarthy77,Szalata77,Arnold78,Dunn83,Cavedon82,Juster85}
%\cite{Ottermann85,Beck87,Amroun92}. 

An experiment performed at Bates \cite{Nakagawa99} has produced two points
at higher $q$ for magnetic scattering, the data have, however, not yet
been published.

\subsection{Experimental form factors \label{hhefits}}
The {\em world} set of cross sections measured for \het\ and \hyt\ have been
analyzed by Amroun \et\ \cite{Amroun94} in the same way as described in 
sect.~\ref{deutfit} for the deuteron. The various form factors  were parameterized 
using the SOG parameterization. From these form factors, the cross sections 
were calculated and the parameters fitted to the data. 
The separation of longitudinal and transverse form factors thereby was 
automatically achieved.   

In order to analyze the electron scattering data it was assumed that
the cross section can be expressed in the plane wave Born approximation
(PWBA) as:
\begin{eqnarray*} 
\sigma 
   =   \sigma_{Mott} \frac{1}{\eta} 
     \left[ \frac {q^2}{\vec{q}^2} F_{ch}^{2}(q)
     + \frac { \mu^{2}q^2}{2 M^{2}} \left( \frac {1}{2} \frac {q^{2}}{\vec{q}^{2}}
     +tg^{2} \frac {\theta}{2}\right) F_{m}^{2}(q) \right]
\end{eqnarray*}
%where $f_{rec} = (1+2E/M\sin^2{\frac{\theta}{2}})^{-1}$ is the usual 
%recoil factor, $\eta=q^2/4M^2$, $\mu$ is the trinucleon magnetic moment and 
%$M$ the nucleon mass.
where the Mott cross section contains the usual recoil and $Z^2$ factors,
 $\mu$ is the 
\hyt (\het) magnetic moment,  $\eta$ is given by 
$(1+ q^{2} / 4 M_T^2)$. $M$ ($M_T$) is the nucleon (three-nucleon)
 mass and $q$ ($\vec{q}$) is the four (three)-momentum transfer.

As the incident electron wave is distorted by the Coulomb field of 
the nucleus, the PWBA is not entirely valid.
The DWBA form factors $F^2_{ch,DW}(E,\theta)$ and 
$F^2_{m,DW}(E,\theta)$ are related to the Born form factors $F^2_{ch}(q)$ and 
$F^2_m(q)$ by:
\begin{eqnarray*} 
 F^2_{ch}(q)=F^2_{ch,DW}(E,\theta)f_c(E,\theta), \hspace{2cm} 
    F^2_m(q)=F^2_{m,DW}(E,\theta)f_m(E,\theta)
\end{eqnarray*} 
where $f_{c}$ and $f_m$ are Coulomb correction factors. Unfolding of these 
corrections allows to restore the PWBA formula and to infer the PWBA
form factors that can be compared to theoretical predictions and will be 
used below.

\begin{figure}[htb]
%Figur mit topp/sideways hergestellt, modif boundingBox: 66 206 516 556
\centerline{\mbox{\epsfysize=7cm \epsffile{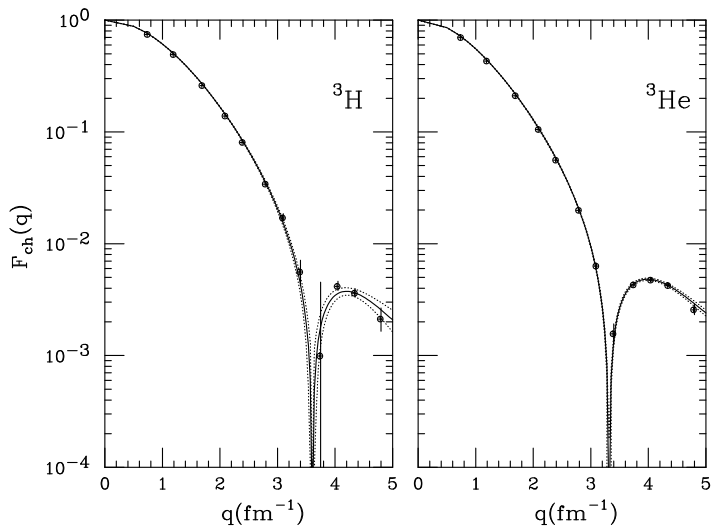}}}
\begin{center} \parbox{14cm}{\vspace*{-0.5cm} \caption{
\label{hhecros} 
Best fit charge form factors (curves with error band) and form factors
derived using standard Rosenbluth plots in individual regions of $q$.
} }  \end{center}
\vspace*{-0.4cm} \end{figure}
%
%
%\begin{figure}[htb]
%%Figur mit topp/sideways hergestellt, modif boundingBox: 66 206 516 556
%\centerline{\mbox{\epsfysize=7cm \epsffile{hhemros1.ps}}}
%\begin{center} \parbox{14cm}{\vspace*{-0.5cm} \caption{
%\label{hhemros} 
%Best fit magnetic form factors (curves with error band) and form factors
%derived using standard Rosenbluth plots in individual regions of $q$.
%} }  \end{center}
%\vspace*{-0.4cm} \end{figure}
%%

In order to unfold the Coulomb corrections from the experimental cross 
sections approximate charge and magnetization densities were used to compute
the  cross sections as a function of the actual energies 
and angles. This was done using a phase shift 
code for charge scattering \cite{Heisenberg70}, and the DWBA code HADES \cite{Andresen94} 
 for magnetic scattering. 
The Coulomb correction factors $f_c, f_m$, obtained from the ratio of the
exact to the PWBA cross sections, were found to be insensitive
to the exact densities employed. 

Amroun \et\  have also performed standard Rosenbluth separations of the data. 
As forward- and backward-angle data are not exactly at the same momentum 
transfer (although close as by now data at {\em many} angles are available), 
they have used the fit to shift the data to the $q$-value desired. At the 
$q$-value of interest the ratio $\sigma _{exp} / \sigma _{fit}$ of the 
neighbouring experimental $q$-value was used. The resulting $F_{ch}$ and $F_{m}$ 
agree with the ones from the global fit. The error bars are less precise, 
however, as for a determination at a given $q$ the global fit {\em de facto} 
uses  the values of all data in the neighbouring $q$-interval;  the width of 
this interval is given by $\sim 1/R_{max} \simeq 0.2 fm^{-1}$.

In the paper of Amroun \et\ \cite{Amroun94} the best fit parameters for
 \hyt\ and \het\  are given together with the 
uncertainties of the form factors,  so we dont need to repeat them
here. (Note that eq.(1) contains a misprint, the numerical factor in the 
exponent should be 1/4 and not 1/2).
 
In fig. \ref{hhecros}
%,\ref{hhemros} 
the comparison between the global fit 
(curves with error
band) and the form factors determined from Rosenbluth separations in 
individual intervals of $q$ is shown. 

\begin{table}[htb]
\begin{center}
\begin{tabular}{l|l|l}
nucleus & type & $rms$-radius \\
\hline
\hyt & $C0$ & 1.755 $\pm$ 0.087 $fm$ \\
\hyt & $M1$ & 1.840 $\pm$ 0.182 $fm$ \\
\het & $C0$ & 1.959 $\pm$ 0.034 $fm$ \\
\het & $M1$ & 1.965 $\pm$ 0.154 $fm$ \\
\hline
\end{tabular}
\label{rms3}
\end{center}
\end{table} 

On occasion, it is useful to know the $rms$-radii of the A=3 nuclei.
They are listed in table \ref{rms3}, where the error bars given include both
statistical and systematic uncertainties. For the magnetic form factors, the 
''radii'' given do not really correspond to a genuine radius; rather, they 
simply reflect the coefficient that occurs in the $q=0$ expansion of the
magnetic form factor $F(q) = 1 - q^2 r_{rms}^2/6 + ..$

\subsection{Comparison to theory \label{hhetheo}}
As mentioned in sect. \ref{wf}  a number of techniques are available to calculate the
wave function starting from the (non-relativistic) Schr\"odinger
equation and the known N-N interactions: Faddeev calculations in both 
coordinate- and momentum space 
\cite{Harper72,Brandenburg74,Laverne73,Chen85b,Witala91,Brandenburg75,Hadjimichael83,Maize84}, 
the Correlated Hyperspherical Harmonics  approach 
\cite{Kievsky94,Viviani95} and Greens-function Monte Carlo methods  \cite{Carlson95}.
%These different methods now give results that agree quite closely; for
%the binding energy, for instance, the results agree within $\sim$10keV for
%the same N-N interaction.
%
%These non-relativistic approaches underbind the A=3 nuclei by 0.5--1 $MeV$, 
%depending on the N-N interaction used. The calculations therefore in general 
%include a three-body interaction (see sect. \ref{NN}).
%
Relativistic calculations are also becoming available
\cite{Carlson93,Forest95,Machleidt96,Stadler97}. Calculations of the 
A=3 nuclei have been performed using the Blankenbecler-Sugar formalism 
\cite{Machleidt96}  and the relativistic spectator (Gross) equation 
\cite{Stadler97}. 
%These calculations tend to give a stronger binding, as a
%consequence of the off--shell properties of the N-N interactions employed.
These relativistic calculations are of rather recent date, however, and the 
corresponding electromagnetic form factors have not yet been published.
 
%Some of the calculations cited above do include the Coulomb interaction for the
%case of \het. In particular, for the Faddeev calculations in coordinate
%space and the Correlated Hyperspherical   Harmonics approaches, inclusion of
%the Coulomb interaction is straightforward. 

In the following we first compare to the results of the recent calculation of 
Marcucci \et\ \cite{Marcucci98}. These authors use the trinucleon wave
functions calculated by Kievsky \et\ \cite{Kievsky94} with the Pair-correlated
Hyperspherical Harmonics (PHH) method. In this PHH calculation the AV18 
N-N interaction, which includes charge-symmetry and charge-independence breaking 
terms and gives a very good fit to N-N scattering data, is used. 

The AV18  force is complemented by the UIX three-body force which has a long-range 
part given by $2\pi$-exchange. At short range a phenomenological repulsive
term is added to simulate dispersive effects. This part is fitted to the 
binding energies of the A=3, $\infty$ systems.
\begin{figure}[htb]
%Figur mit topp/sideways hergestellt, modif boundingBox: 66 206 516 556
\centerline{\mbox{\epsfysize=9cm \epsffile{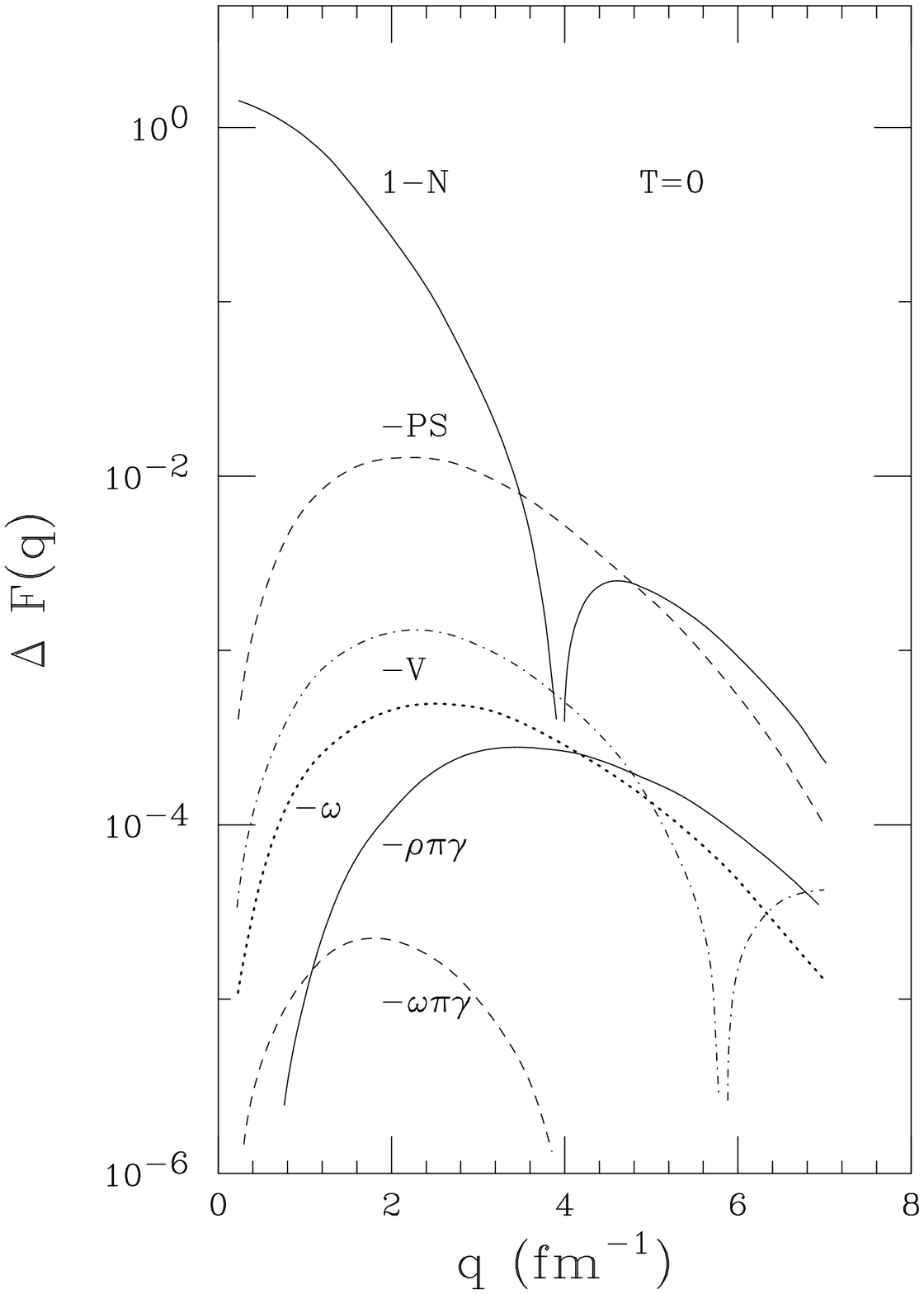}
\epsfysize=9.1cm \epsffile{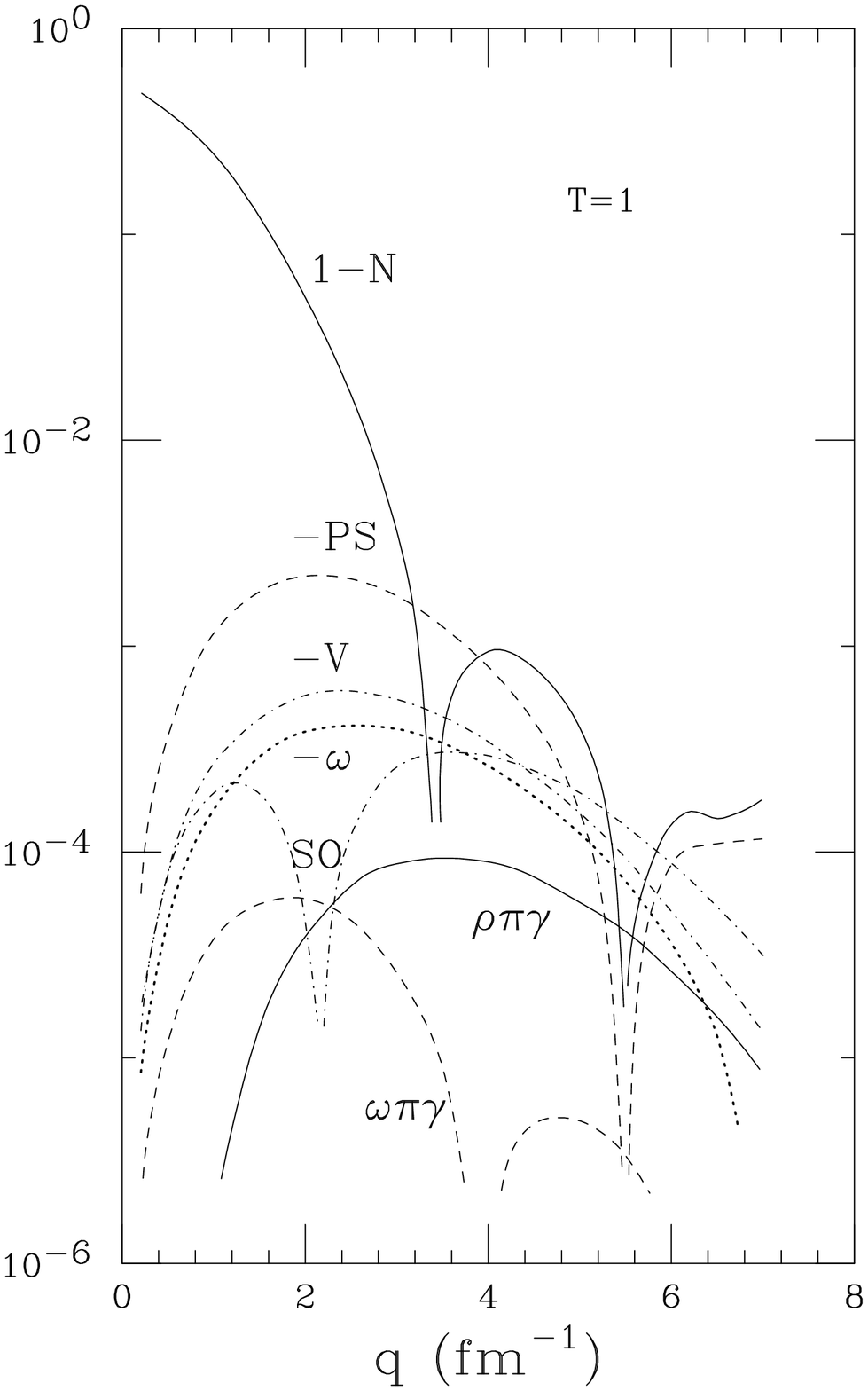}}}
\begin{center} \parbox{14cm}{\vspace*{-0.5cm} \caption{
\label{mecmar} 
Contributions of different two-body charge terms, split into T=0 and T=1 
components \protect{\cite{Marcucci98}}. The one-body part is indicated by 1-N.
} }  \end{center}
\vspace*{-0.4cm} \end{figure}

The two-body  
operators are calculated as discussed in sect. \ref{MEC}, {\em i.e.} they are 
consistently derived, as far as presently possible,  from the N-N interaction 
employed. For the two-body
currents the $\pi$- and $\rho$-like terms (PS and V in fig.~\ref{mecmar}) come 
from AV18. The model-dependent 
terms, $\pi \rho \gamma$  and $\omega \rho \gamma$, are calculated using known coupling
constants and VDM form factors. The cut-off parameters needed at short range 
have been chosen to give a good description of the deuteron magnetic
form factor $B(q)$ at high $q$. The three-body
currents resulting from the $2\pi$-exchange in the 3BF employed is accounted
for as well. For the 
two-body charge operators, which do depend on how the non-nucleonic degrees of
freedom ($N\bar{N}$, $N^*$) have been eliminated from the wave function,     
the authors follow as much as possible the derivation of the corresponding 
$\pi$- and $\rho$-like terms in the two-body currents. Again the $\pi \rho \gamma$
and $\omega \rho \gamma$ terms have to be added separately.  In fig.~\ref{mecmar} 
the different contributions of the 
two-body charge terms are shown, split into their T=0 and T=1 components.
It is obvious from the comparison to the one-body curve that two-body 
contributions will have an important effect upon the form factor in the
region of the diffraction minimum and maximum. 
\begin{figure}[htb]
%Figur mit topp/sideways hergestellt, modif boundingBox: 66 206 516 556
\centerline{\mbox{\epsfysize=7cm \epsffile{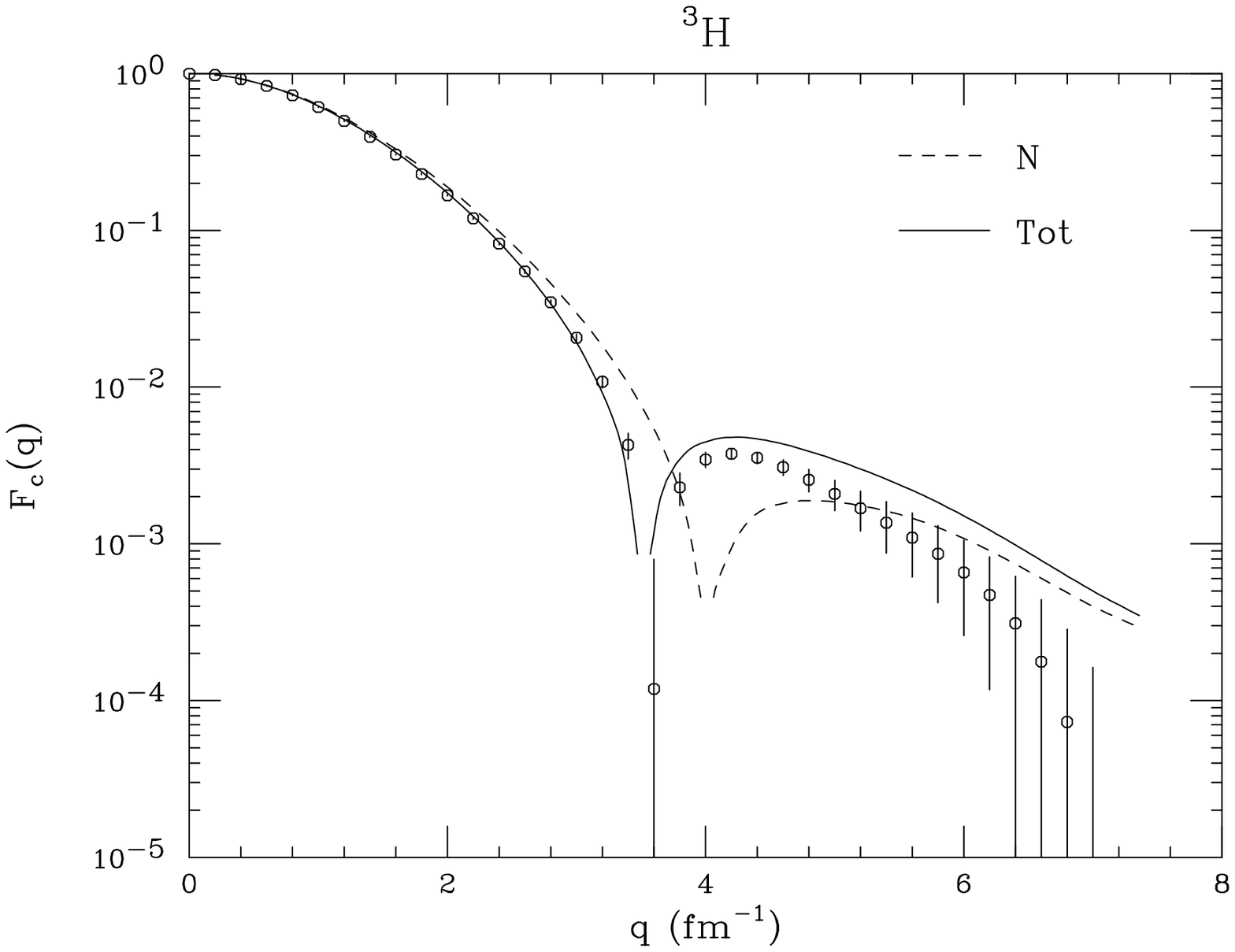}}}
\centerline{\mbox{\epsfysize=7cm \epsffile{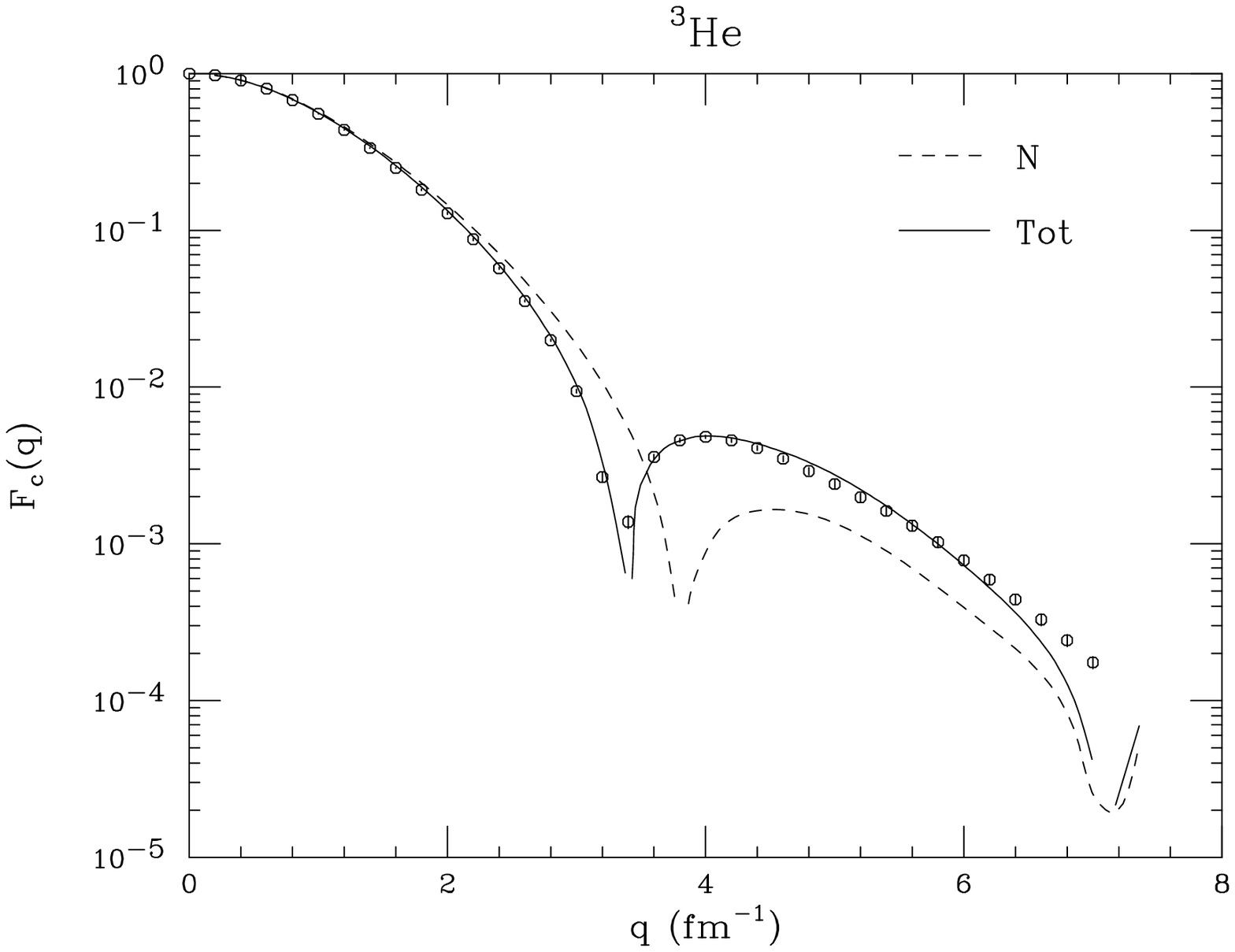}}}
\begin{center} \parbox{14cm}{\vspace*{-0.5cm} \caption{
\label{dfchhev18} 
A=3 charge form factors together with calculation of Marcucci \et\ 
\protect{\cite{Marcucci98}}.
The IA result (dashed) , and the full result including two-body terms, is given.
} }  \end{center}
\vspace*{-0.4cm} \end{figure}
\begin{figure}[htb]
%Figur mit topp/sideways hergestellt, modif boundingBox: 66 206 516 556
\centerline{\mbox{\epsfysize=6cm \epsffile{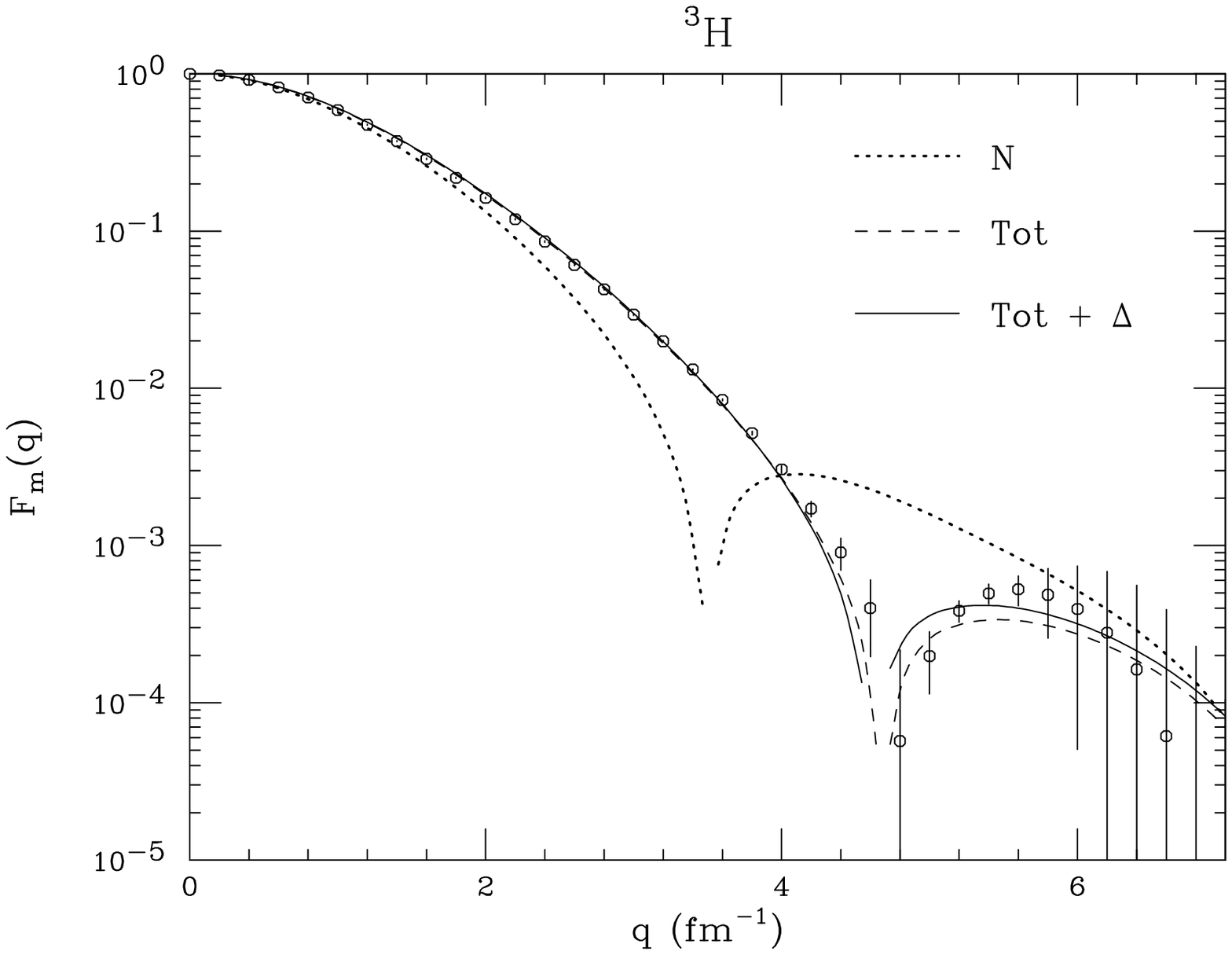}}}
\centerline{\mbox{\epsfysize=6cm \epsffile{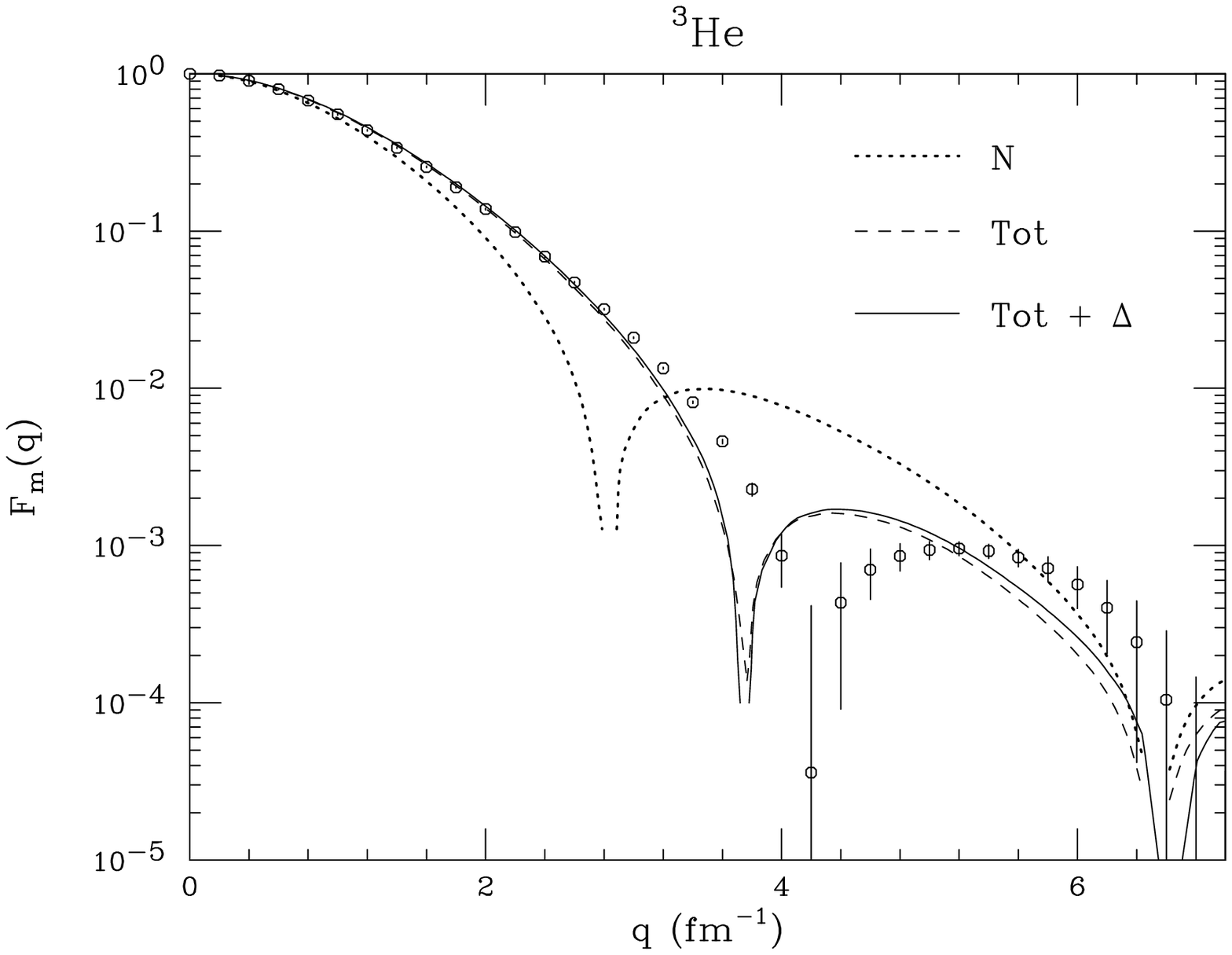}}}
\begin{center} \parbox{14cm}{\vspace*{-0.5cm} \caption{
\label{dfmhhev18} 
A=3 magnetic form factors together with calculation of Marcucci \et\
\protect{\cite{Marcucci98}}.
The IA result, the full result including two-body currents, and the result
obtained when explicitly including $\Delta$'s are given. 
} }  \end{center}
\vspace*{-0.4cm} \end{figure}

Marcucci \et\ have also explored  the role of $\Delta$-components in the A=3 ground state. 
Often these degrees of freedom are included via the two-body currents, but a 
different picture might result once they are explicitly allowed for in the
wave function, such as done 
previously by the group of  Sauer \cite{Hajduk83,Strueve87}. 
Marcucci \et\ do not carry out a full coupled
channel N-$\Delta$ calculation. Rather, they use the Transition Correlator Operator 
(TCO) method \cite{Horlacher78}, which has proven to be useful in earlier variational 
calculations. In this case, they use the AV28Q N-N interaction which allows for both
nucleons and Deltas. 

In figs. \ref{dfchhev18},\ref{dfmhhev18} we show the comparison between the 
experimental results of the A=3 form factors and the predictions of Marcucci \et. 
In each case we show both the IA-prediction, and the full calculation including 
two-body and three-body currents. For the case of the magnetic form factors, the
results obtained by explicitly including the $\Delta$ in the wave function are 
also shown.   

One observes that for all form factors the effect of two-body currents is quite 
important. In general, the full calculation does a very good job in explaining the
data. The effect of explicitly including the $\Delta$ is rather small, significantly smaller
than what was found by Strueve \et\ \cite{Strueve87}, partly because different coupling 
constants were employed.

\begin{figure}[htb]
%Figur mit topp/sideways hergestellt, modif boundingBox: 66 206 516 556
\centerline{\mbox{\epsfysize=7cm \epsffile{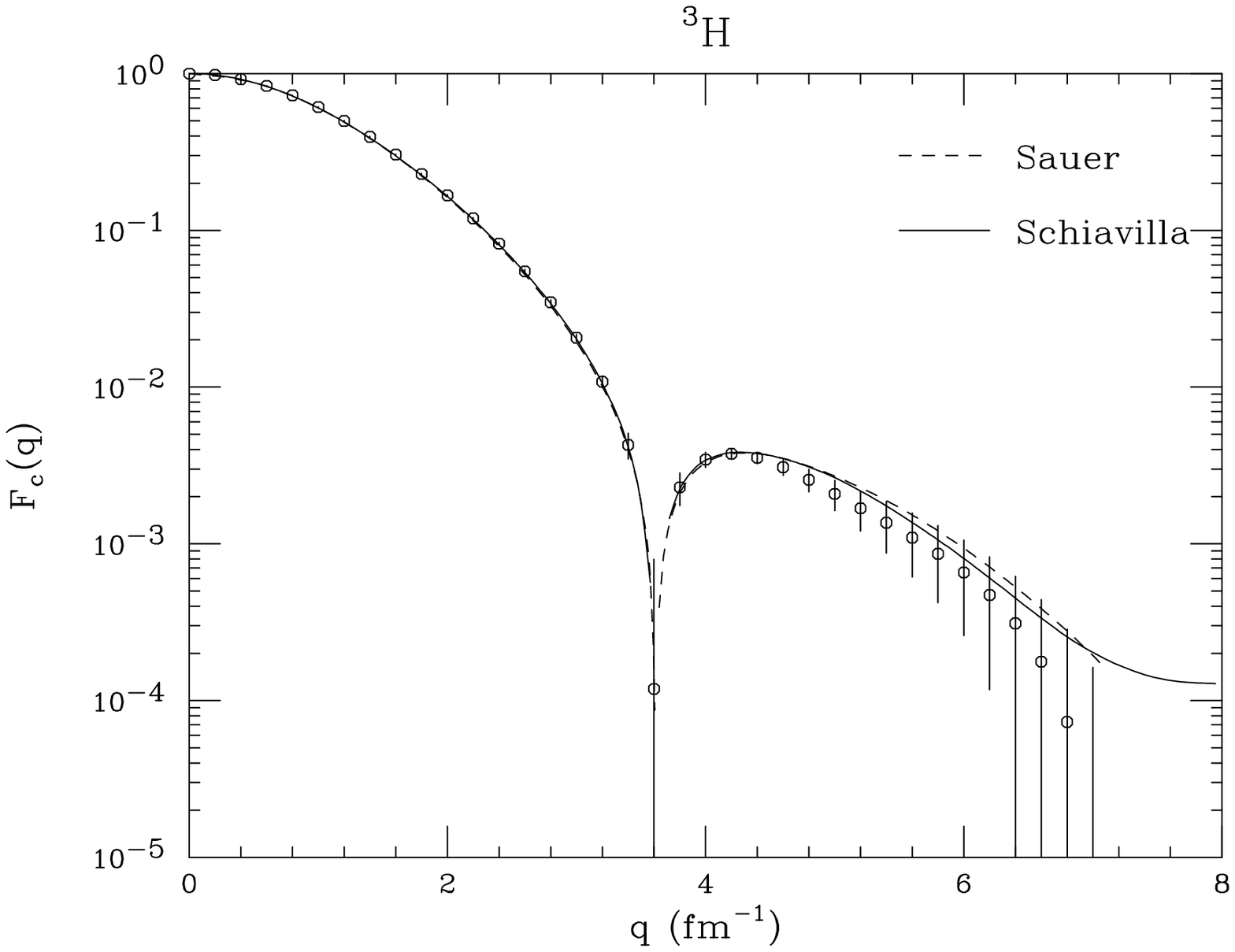}}}
\centerline{\mbox{\epsfysize=7cm \epsffile{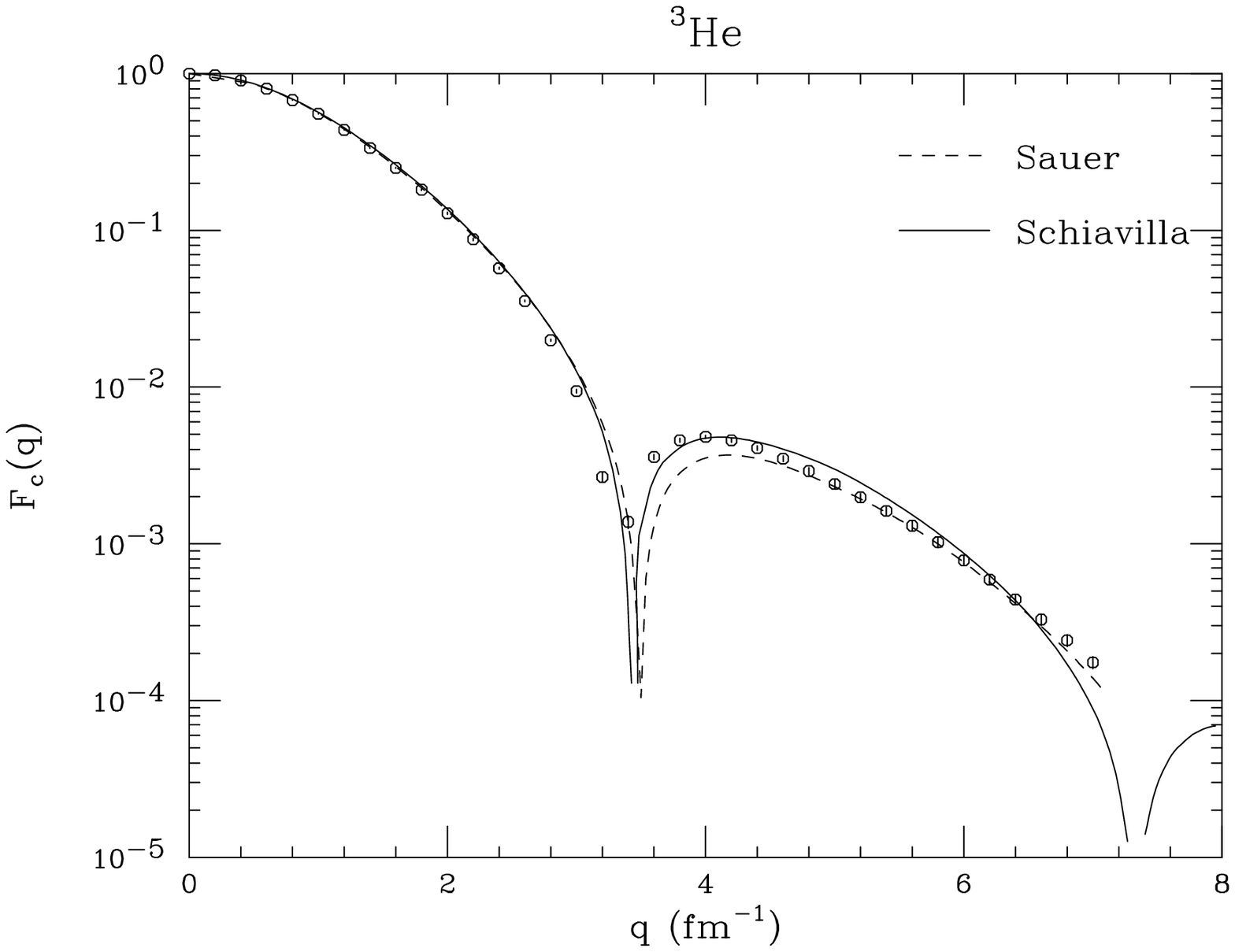}}}
\begin{center} \parbox{14cm}{\vspace*{-0.5cm} \caption{
\label{dfchhefull} 
A=3 charge form factors together with calculations of Schiavilla \et\ 
\protect{\cite{Schiavilla89,Schiavilla90}} and Sauer and collaborators 
\protect{\cite{Strueve87,Hajduk83}}.
} }  \end{center}
\vspace*{-0.4cm} \end{figure}
\begin{figure}[htb]
%Figur mit topp/sideways hergestellt, modif boundingBox: 66 206 516 556
\centerline{\mbox{\epsfysize=7cm \epsffile{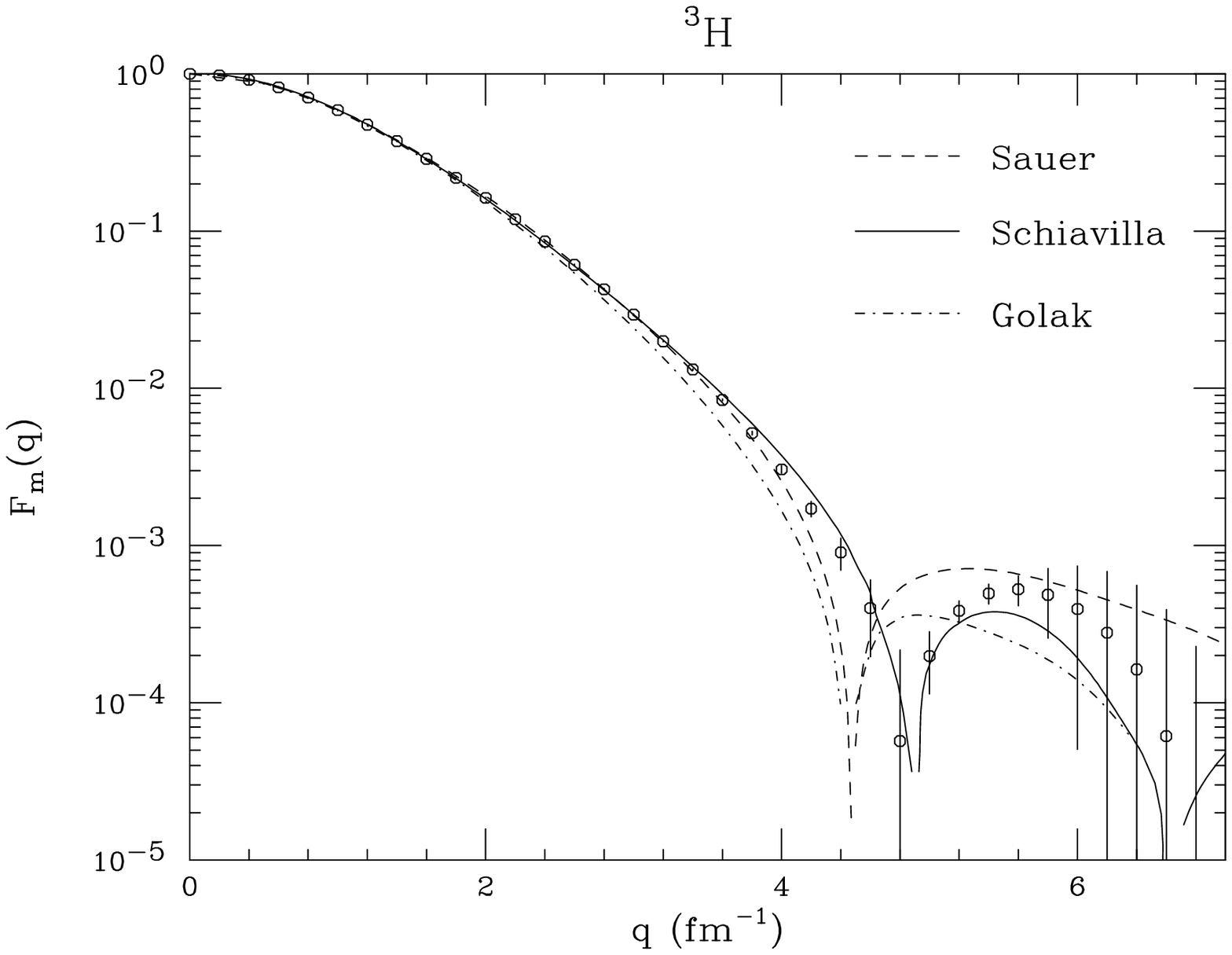}}}
\centerline{\mbox{\epsfysize=7cm \epsffile{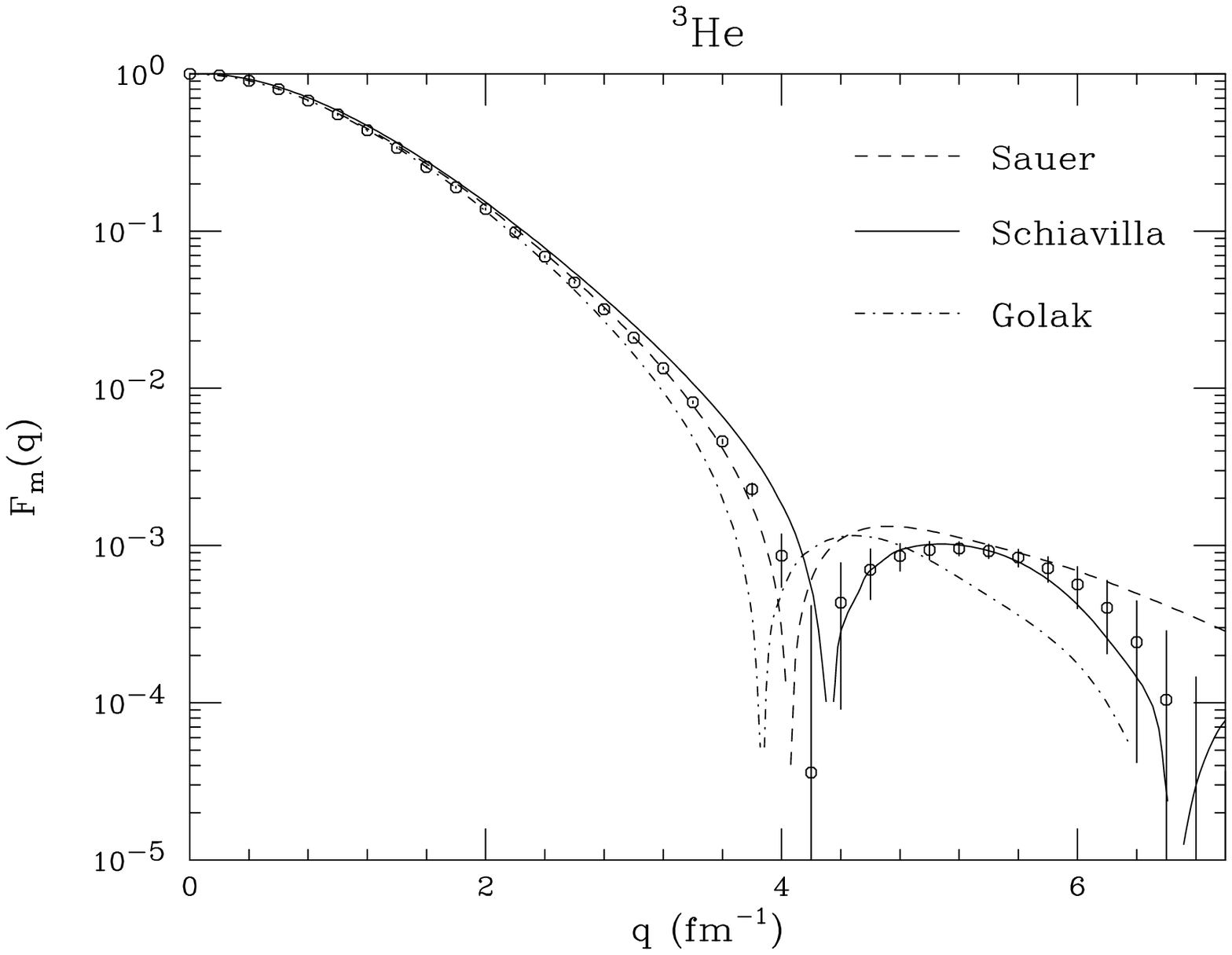}}}
\begin{center} \parbox{14cm}{\vspace*{-0.5cm} \caption{
\label{dfmhhefull} 
A=3 magnetic form factors together with calculations of Schiavilla 
\et\ \protect{\cite{Schiavilla89,Schiavilla90}}, Sauer and collaborators 
\protect{\cite{Strueve87,Hajduk83}} 
and Golak \et\ \protect{\cite{Golak00}}.
} }  \end{center}
\vspace*{-0.4cm} \end{figure}

In figs. \ref{dfchhefull},\ref{dfmhhefull} we compare the experimental data with 
further calculations of
refs.\cite{Hajduk83,Strueve87,Schiavilla89,Schiavilla90,Golak00}, which also 
are based on  nucleonic degrees of freedom.  

Sauer and collaborators \cite{Hajduk83,Strueve87} have solved the 
coupled-channel N-$\Delta$ Faddeev equations in momentum space, and 
used the Paris NN  potential  modified \cite{Hajduk83} to  
include  $\Delta$-isobar excitations via $\pi$- and 
$\rho$-meson exchanges. The presence of the $\Delta$ thus accounts for the most 
important part of the three-body force. In this calculation $\pi$-, $\rho$-, and
$\rho\gamma\pi$-meson exchange current contributions are included as  well as
the one-body Darwin-Foldy term and spin-orbit relativistic corrections.
As for the calculation of ref. \cite{Marcucci98} boost corrections
are not yet included.

Schiavilla and collaborators \cite{Schiavilla89,Schiavilla90} 
based their (older) calculation  on a variational three-body wave function computed using 
the Argonne V14 interaction and the Urbana-VII
three-nucleon force. In this calculation the $\pi$-like and $\rho$-like
meson exchange currents were derived consistently from the NN interaction used.
The authors also included the  one-body relativistic corrections mentioned 
above as well 
as $\omega$-exchange  and $\omega\gamma\rho$-contributions.

Golak \et\ \cite{Golak00} in their recent calculation 
 solve the Faddeev equations in momentum space, using
the AV18 N-N potential. The two-body currents, which up to now are only calculated 
for the  magnetic form factors, are treated according to ref. \cite{Kotlyar00}.
 
For all calculations the two-body terms 
significantly displace the diffraction minimum of the charge form factors 
to lower $q$ values  and
increase the height of the secondary  maximum by more  than a factor  of
two. The  calculations achieve a similarly good agreement with \hyt.
 This is not the case for \het\ where
ref.~\cite{Strueve87}  slightly underestimates the charge form factor, whereas 
the calculation of ref.~\cite{Schiavilla89} almost perfectly agrees with it.  
The difference could be due to the model dependence  of the one-body term, or
due to different two-body contributions.
Indeed, for the \hyt\ form factor, the calculations of refs.~\cite{Strueve87}
and \cite{Schiavilla90} predict the one-body 
minimum at 3.9 and 4.15$fm^{-1}$ respectively, while the predictions for the 
full calculation perfectly agree. Additional differences might come from  
the different parameterizations of the nucleon form factors used, particularly 
for the poorly known G$_{en}$. We do not understand why the \hyt\
form factors are better predicted than the ones for \het.

%Calculations have also been carried out by Lina and Goulard
%\cite{Lina86}.  Similar agreement  with the  data was  obtained.   In all
%these studies the $\pi$ and $\rho$ two-body contributions were found  to
%be  significantly  larger  (roughly  a  factor  of 10) than the one-body
%relativistic corrections.  More recently the relativistic effects in the
%trinucleon  were  investigated  in  a  consistent  manner  by  Rupp  and
%Tjon\cite{Rupp92}.  They  have found that,  in addition to  an increase of
%the  triton binding  energy  by ~0.3-0.4  $MeV$,  the relativistic treatment 
%tends to produce a diffraction minimum at somewhat larger momentum transfer.
 
For the magnetic form factors,  where the effect of the three-body force is also 
quite small,   the effect of the two-body currents  is
larger than for the charge form factor, and they improve considerably the 
agreement  with  experiment. However, none of the calculations perfectly 
accounts for  the shape of the \het\
form factor: the prediction of ref.~\cite{Hajduk83}  explains its 
shape below the diffraction minimum, but fails above, whereas 
ref.~\cite{Schiavilla89} shows the opposite behaviour. The latter 
calculation agrees with the \hyt\ form factor, while this is not the case
for ref.~\cite{Hajduk83}.

There  are  several  differences  between  these two  calculations.
As for the charge form factors they differently predict the one-body
contribution. The IA  predictions for the location of the diffraction
minima of \hyt\ are 3.5 and 3.7 $fm^{-1}$ for the calculations of
Sauer and Schiavilla, respectively.
The differences in the exchange current contribution are
even  larger.  

While qualitatively the role of two-body currents is understood, the situation 
for the magnetic form factors in particular is not yet entirely satisfactory 
when it  comes to a quantitative comparison. Even at low $q$, some differences 
between experiment and theory remain. While experimentally the \hyt~(\het) 
magnetic moments amount to 2.98 (--2.13) $\mu_N$, the theoretical 
predictions are {\em e.g.} 2.98 (--2.09) for V$_{18}$ and 2.87(--2.06) 
for the Paris  interaction.

\begin{figure}[htb]
%Figur mit topp/sideways hergestellt, modif boundingBox: 66 206 516 556
\centerline{\mbox{\epsfysize=9cm \epsffile{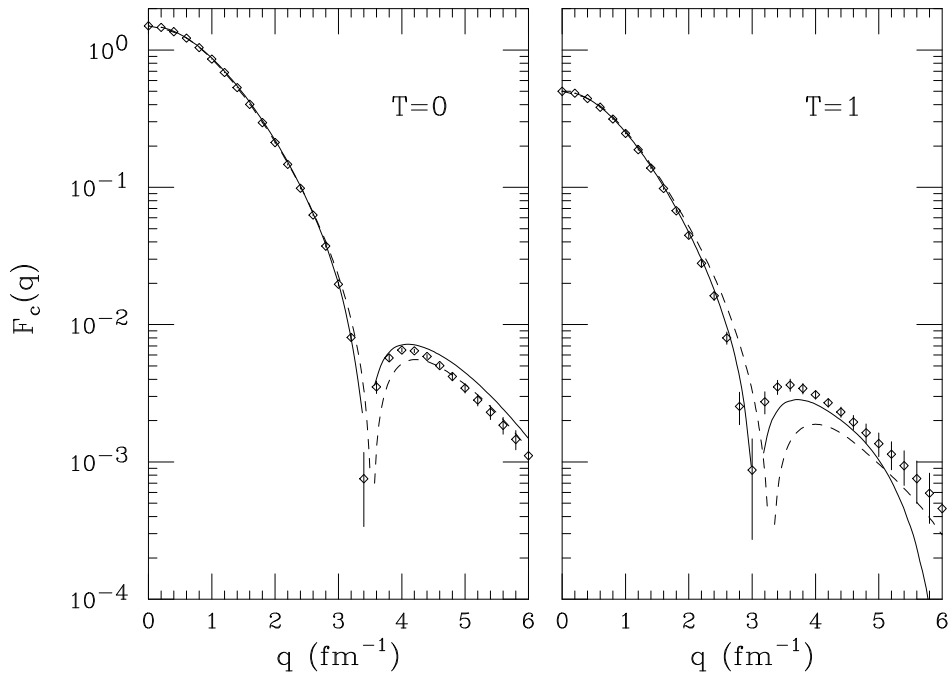}}}
\begin{center} \parbox{14cm}{\vspace*{-0.5cm} \caption{
\label{fchhe01}
T=0 and T=1 charge form factors of the A=3 systems, compared to the calculations
of Marcucci \et\ \protect{\cite{Marcucci98}} (solid) and Strueve \et\
\protect{\cite{Strueve87}} (dashed).
} }  \end{center}
\vspace*{-0.4cm} \end{figure}

Fig.~\ref{fchhe01} shows, for completeness, the  isospin-separated
$A=3$  charge form  factors  together  with  the  theoretical  predictions   of
Strueve \et\ and Marcucci \et\ \cite{Strueve87,Marcucci98}.  The  main  
observation from these
comparisons is that the  T=0 charge  form factor  is remarkably  well
predicted by both calculations.  This  agreement is quite  unexpected since
it  is reached after adding  ''model-dependent'' two-body
contributions, which lead to a sizeable
change.   For the T=1 charge form factor we observe larger differences
between the theoretical predictions.  Moreover, none of them perfectly describes
the data.  Since the agreement for the T=0 component has been found  to
be  excellent  for  all  of  them,  we  conclude that improvement in the
description of the trinucleon requires improvement of the calculation of
its isovector charge component.

For the T=0,1 magnetic form  factors the calculations are closer to the
data, so we do not separately show them.

\subsection{A=2,3 comparison}
In order to understand better whether eventual shortcomings in our understanding
of the 3N-systems are only related to the two-body input, or whether 
our understanding of genuine three-body properties is lacking, it is of 
interest to directly compare the 2- and 3-body form factors. 

To achieve this, one can extract from the \het\ and \hyt\ form factors the 
corresponding isoscalar- and isovector quantities, where the former then is 
directly comparable to the deuteron form factor. This comparison becomes
particularly interesting now that the diffraction structure of the A=2 $C0$ 
charge form factor is well determined.

The isoscalar $F_{T=0}$ and isovector $F_{T=1}$  components
of the A=3 charge form factors are defined by the following expressions:
\begin{eqnarray*}\label{isic}
 F_{0}(q) = \frac{1}{2} \left[2F_{He}(q) +
F_{H}(q)\right] \hspace*{1cm} 
F_{1}(q) = \frac{1}{2} \left[2F_{He}(q) - F_{H}(q)\right]
\end{eqnarray*}
and similarly for the magnetic form factors:
\begin{eqnarray*} \label{isim}
 F_{0}(q) =
\frac{1}{2} \left[\mu_{He}F_{He}(q) + \mu_{H}F_{H}(q)\right]
\hspace*{1cm}
 F_{1}(q) = \frac{1}{2} \left[\mu_{He}F_{He}(q) -
\mu_{H}F_{H}(q)\right]
\end{eqnarray*}
where $\mu_{H}$ and $\mu_{He}$ are the A=3 magnetic moments and all  \het\
and \hyt\  form factors are normalized to 1 at $ q=0$.  The T=0,1 form factors  can
easily be calculated from the parameters of the \het\ and \hyt\ form 
factors \cite{Amroun94}. The  effect of the  isospin impurity, due  to
the Coulomb interaction  in \het, has  been neglected.   Indeed, its
effect on the \het\  charge form  factor is very small, $\leq 1\%$  in
the region  of the  secondary diffraction  maximum and  even smaller for
lower $q$ values.

\begin{figure}[htb]
%Figur mit topp/sideways hergestellt, modif boundingBox: 66 206 516 556
\centerline{\mbox{\epsfysize=5cm \epsffile{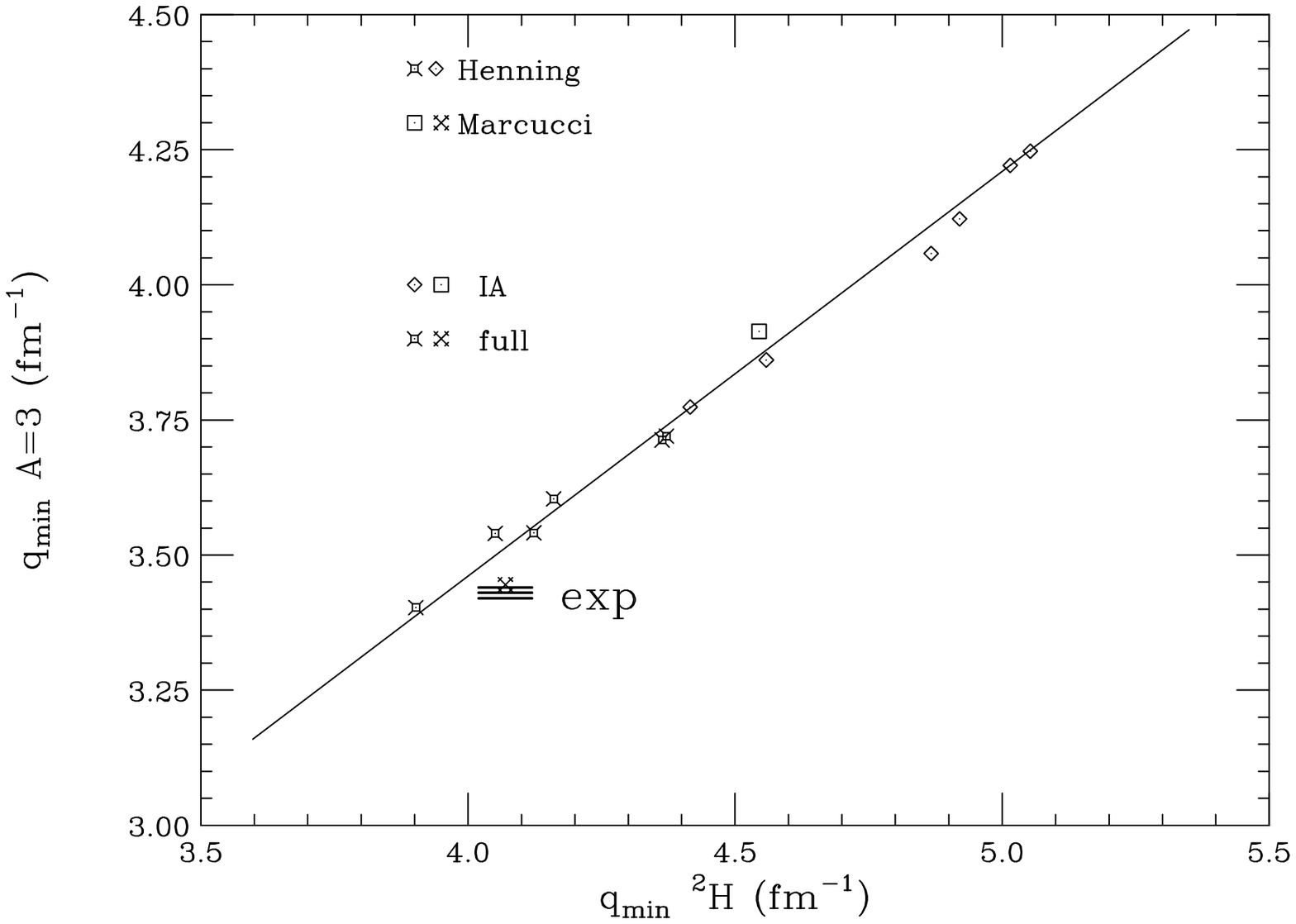}}}
\centerline{\mbox{\epsfysize=5cm \epsffile{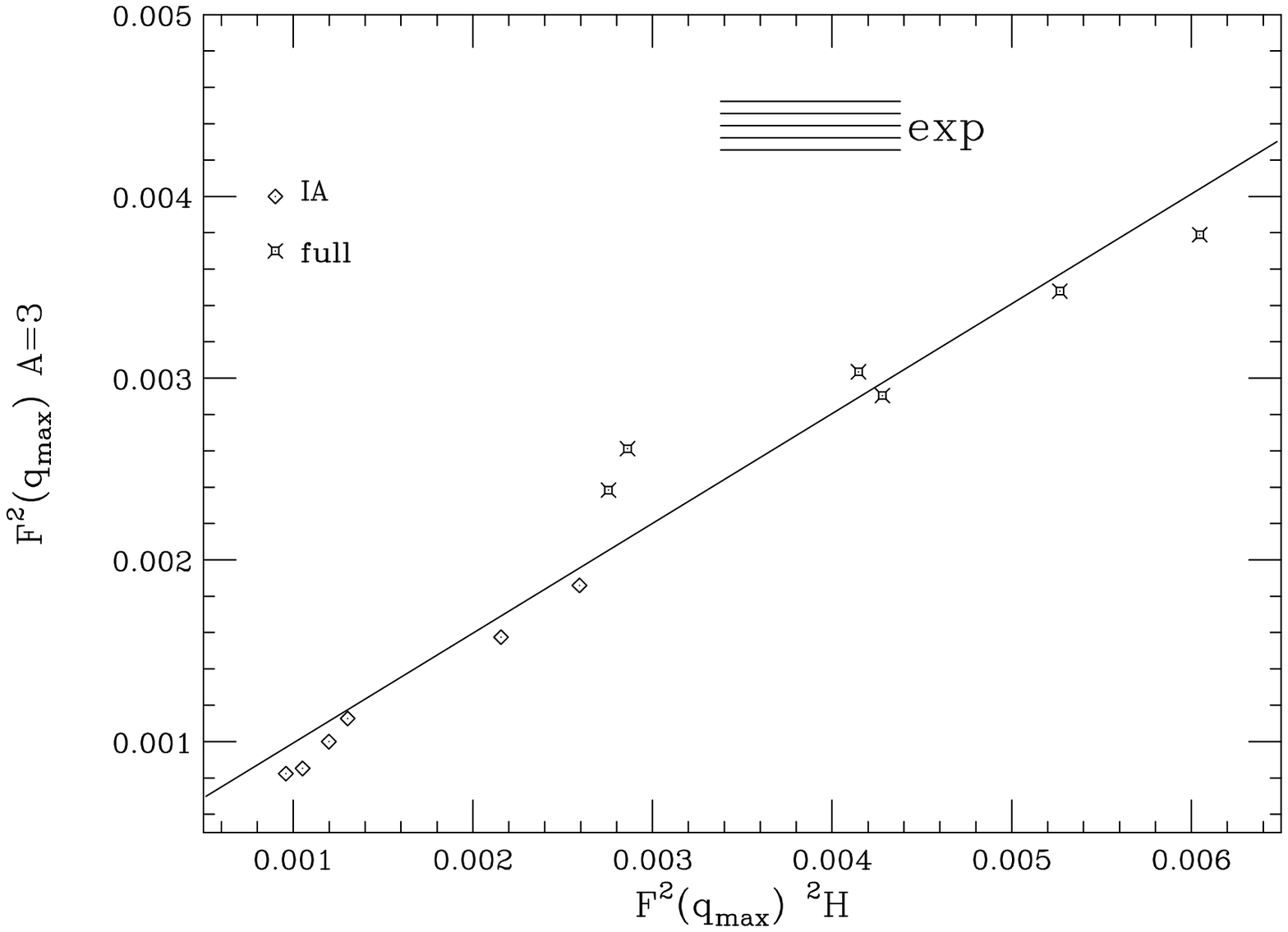}}}
\begin{center} \parbox{14cm}{\vspace*{-0.5cm} \caption{
\label{minmin} 
Correlation between the position of the diffraction minima in the A=2 and 
A=3 isoscalar charge form factors (top) and the amplitudes of the form factors 
in the diffraction maxima
(bottom). The experimental point is indicated by the shaded area (exp).
} }  \end{center}
\vspace*{-0.4cm} \end{figure}
Henning \et\ \cite{Henning95a} have made a detailed comparison between some of 
the global properties of the A=2,3 charge form factors such as the position 
of the diffraction minima, the position of the maxima and the amplitude of the
form factors in the maxima. They have done so by calculating the A=2,3 
form factors for a number of N-N interactions, with and without inclusion
of the two-body currents. The potentials employed were the venerable 
RSC potential, the Paris potential and various OBE Bonn potentials OBEPQ,
OBEPA,OBEPB and OBEPC. 

It turns out that the above-mentioned observables display a remarkable 
correlation between the A=2 and A=3 systems. There is a basically linear
relationship between the A=2 and A=3 quantities. This is demonstrated in
fig.~\ref{minmin} which gives the 
%position of the diffraction zero for the 
%A=3 isoscalar charge form factor as a function of the same quantity for
%the deuteron. 
amplitude of the A=3 isoscalar charge form factor in the diffraction 
maximum as a function of the same quantity for the deuteron. 
Not only do the values for the different potentials (to which we
have added the AV18 potential) define a basically linear correlation, the 
points with/without two-body currents also lie on the same line. 
Within the line, there is a correlation between the strength of the tensor 
interaction and the position of the minimum: N-N potentials yielding a larger
D-state produce a minimum at lower $q$. The lower part of the figure gives the
corresponding relation between the positions of the diffraction zeroes.

The inclusion of the AV18 points shows that this line is rather general; it is 
not simply a consequence of the particular way the two-body contributions are
calculated. There is some qualitative understanding for the occurrence of this
correlation \cite{Henning95a}: if the A=3 nuclei could be described in a 
quasi-deuteron model, such a correlation would be expected. A quasi-deuteron
model is obviously much too primitive for the A=3 system, but the three-nucleon 
wave functions obtained from the Faddeev equations do approximately show 
the symmetry upon interchange of the Jacobi momenta $p$ and $\sqrt{3}/2q$ which 
would result from the spectator model.

The non-understood feature of fig.~\ref{minmin} lies in the fact that the 
line describing all theoretical calculations does not pass through the 
experimental point indicated by ''exp'', particularly for the amplitudes in the 
diffraction maxima. From the comparison with experiment it would seem that 
the existing range of non-relativistic N-N potentials together
with the present understanding of two-body currents is not able to account 
simultaneously for the properties of the A = 2,3 systems. 

 Besides the difficulties with the 
\het\ magnetic form factor, the discrepancy shown in fig. \ref{minmin} is
one of the major open problems, and potentially points to the need to go
beyond the ''nuclear standard model''.  

%Fig.~\ref{fchhe01} shows, for completeness, the  isospin-separated
%$A=3$  charge form  factors  together  with  the  theoretical  predictions   of
%Strueve \et\ and Marcucci \et\ \cite{Strueve87,Marcucci98}.  The  main  
%observations from these
%comparisons is that the  T=0 charge  form factor  is remarkably  well 
%predicted by both calculations.  This  agreement is quite  unexpected since 
%it  is reached after adding  ''model-dependent'' two-body 
%contributions, which lead to a sizeable
%change.   For the T=1 charge form factor we observe larger differences 
%between the theoretical predictions.  Moreover, none of them perfectly describes
%the data.  Since the agreement for the T=0 component has been found  to
%be  excellent  for  all  of  them,  we  conclude that improvement in the
%description of the trinucleon requires improvement of the calculation of
%its isovector charge component.
%%
%\begin{figure}[htb]
%%Figur mit topp/sideways hergestellt, modif boundingBox: 66 206 516 556
%\centerline{\mbox{\epsfysize=9cm \epsffile{fchhe01.ps}}}
%\begin{center} \parbox{14cm}{\vspace*{-0.5cm} \caption{
%\label{fchhe01} 
%T=0 and T=1 charge form factors of the A=3 systems, compared to the calculations
%of Marcucci \et\ \protect{\cite{Marcucci98}} (solid) and Strueve \et\
%\protect{\cite{Strueve87}} (dashed).
%} }  \end{center}
%\vspace*{-0.4cm} \end{figure}
%%%%
%
%For the T=0,1 magnetic form  factors the calculations are quite close to the 
%data, so we do not separately show them.

\section{Helium-4}
The four-body system is interesting as the $\alpha$-particle is tightly bound, 
 and the nucleons are dominantly in a spatially symmetric S-state configuration.
As a consequence, a high central density (the highest in nuclei) is found in 
$^4 \hspace*{-0.5mm}He$. 

The data base for the \hef\ charge form factor is reasonably complete, and the 
form factor measurements extend to unusually large $q$. The pioneering 
experiment was performed by R. Frosch \et\ at Stanford \cite{Frosch67} 
using a liquid helium target. This experiment already reached the diffraction
maximum at $4 fm^{-1}$. 

More precise data at medium momentum transfers 
were provided by an experiment carried out also at the Stanford HEPL 
by McCarthy \et\ \cite{McCarthy77}; this experiment focused on the
\het\ form factors, the \hef\ was measured mainly as a check. The 
separation of longitudinal and transverse contributions, necessary for 
\het , was performed for \hef\ as well, in order to certify that indeed
a transverse form factor compatible with zero was found for \hef. As the systematic
errors are very different for the forward angle (high energy) and backward
angle (low energy) data, this provided a very stringent and useful test of the
quality of the data. 
%This experiment was carried out using liquid Helium
%cooled by a bath at 1K; the resulting superfluid Helium made, due to the 
%large heat conductivity, for a  particularly stable target.

Data at very low $q$ are also available from the experiment of Erich \et\ 
\cite{Erich68} performed at the Darmstadt electron accelerator.
This experiment used a mixed H/He gas target in order to obtain a 
continuous normalization of the helium cross sections relative to the proton.
Low-$q$ data were also measured by Ottermann \et\ \cite{Ottermann85}
at Mainz, using a 14bar
gas target, collimated to exclude the contribution of the target windows.

The highest momentum transfers were achieved by an experiment carried out at 
SLAC by Arnold \et\ \cite{Arnold78}.  In this experiment, which used electron energies 
up to 15GeV, a long gas target at 20K and 50 bar was used to obtain a high 
target thickness and large luminosity. Due to the high electron energy 
and the correspondingly poor energy resolution,  the scattered electron and 
recoil helium had to be detected in 
coincidence using the 20GeV/c and 8GeV/c-spectrometers.
The measurements reached a momentum transfer of 8$fm^{-1}$, where the count 
rates dropped to 1 event/week; the steepness of the fall-off of the form factor at 
the highest $q$ is indicative of the presence of a second diffraction zero.
 
For \hef, as for the other few-body nuclei discussed in this review, we
have used the {\em world} data to determine the experimental form factor. The 
treatment of the data and their uncertainties is similar to what was 
discussed for the case of the A=2,3 systems. The resulting form factors are 
shown below in the figures.

Several groups have calculated the ground-state wave function of the 4-body 
system. 
As for the A=3 nuclei, \hef\ is underbound by several $MeV$ in calculations that 
only include two-body forces. The contribution of a three-body force must be added.
As a matter of fact, the \hef\ binding energy is often used to fix
the scale factor of this 3-body force. 

Gl\"ockle and collaborators \cite{Gloeckle93a,Gloeckle93b} have 
employed the Faddeev-Yaku\-bovsky technique using  a selection of  N-N 
interactions, 
Viviani \et\ \cite{Viviani95} 
have employed the Correlated Hyperspherical Harmonics (CHH) approach together with
the AV14 and AV14+TNI8  interactions. As far as ground-state properties such as
binding energies go, these calculations compare well to the Variational 
Monte Carlo (VMC) results obtained by Arriaga \et ~\cite{Arriaga95}  
and the Greens-Function Monte Carlo results of Carlson and Schiavilla 
\cite{Carlson91,Carlson94a}.

Further results for the \hef\ charge form factor are available from 
Schiavilla \et\ \cite{Schiavilla90} based on a VMC calculation using the Argonne
 V14 potential supplemented by the Urbana VII three-nucleon interaction. 
The most recent result has been obtained by Schiavilla and Carlson using the V18 potential
 supplemented by the 
Urbana IX three-nucleon interaction \cite{Schiavilla00}.
%Both this calculation and the previous one use, as far as possible, two-body 
%currents that 
%are derived consistently with the underlying N-N interaction. This is possible for
%the $\pi$- and $\rho$-like exchange terms, while the $\rho \pi \gamma$ and
%$\omega \pi \gamma$ terms have to be evaluated using empirical  (and often 
%uncertain) vertex form factors. 
%
\begin{figure}[htb]
%Figur mit topp/sideways hergestellt, modif boundingBox: 66 206 516 556
\centerline{\mbox{\epsfysize=7cm \epsffile{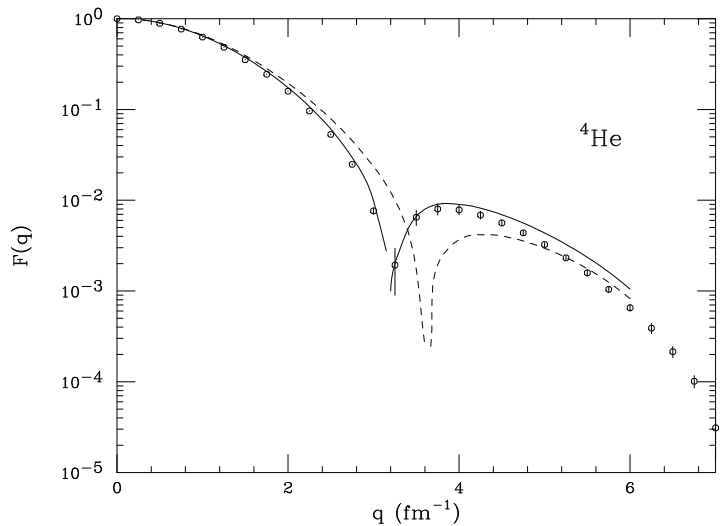}}}
\begin{center} \parbox{14cm}{\vspace*{-0.5cm} \caption{
\label{dfche4v18} 
Experimental form factor of \hef\ together with IA (dashed) and full 
calculation of 
Carlson and Schiavilla \protect{\cite{Schiavilla00}} performed using the Argonne 
V18 N-N potential together with the Urbana IX three-body force.
} }  \end{center}
\vspace*{-0.4cm} \end{figure}

In fig.~\ref{dfche4v18} we show the results of the latest calculation together 
with  the IA-prediction. The figure shows that modern calculations 
do extremely well in reproducing the experimental data. The effect of the 
two-body terms is very important in achieving this good agreement, particularly for the 
height of the diffraction maximum at $4 fm^{-1}$. This is the case despite the 
less than 
satisfactory state of the theory of isoscalar two-body contributions 
for the charge form factor. When comparing  the isoscalar charge form factors
of A=3 \cite{Amroun94} and A=4 one notes that in both cases theory somewhat 
overpredicts the form factor in the diffraction maximum.

The breakdown of the different contributions to the two-body terms is given 
in fig.~12
of ref.~\cite{Schiavilla90}. The individual contributions are quite 
similar to what we have shown in fig.~\ref{mecmar} for the isoscalar A=3 form 
factor. The $\pi$-like (PS) and $\rho$-like terms (V) give the 
largest contribution at high momentum transfer. The single-nucleon relativistic
correction, the Darwin-Foldy (DF) and Spin-Orbit (SO) terms are also significant
at large $q$, and help to get large enough a (absolute value of the) form factor 
in the diffraction maximum. The contribution of the most model-dependent 
two-body terms fortunately is relatively small. 

\begin{figure}[htb]
%Figur mit topp/sideways hergestellt, modif boundingBox: 66 206 516 556
\centerline{\mbox{\epsfysize=7cm \epsffile{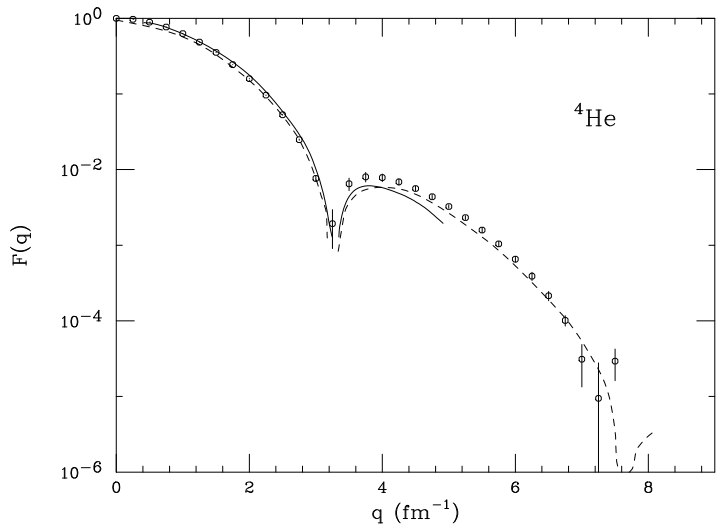}}}
\begin{center} \parbox{14cm}{\vspace*{-0.5cm} \caption{
\label{dfche4gari}
Experimental \hef\ form factor together with early results from the exp(S)-calculation
of Gari {\em et al.} \protect{\cite{Gari76a}} (solid curve) and the VMC calculation
of Schiavilla {\em et al.} \protect{\cite{Schiavilla90}} performed using the
V14 interaction (dashed). 
} }  \end{center}
\vspace*{-0.4cm} \end{figure}

In order to show the progress of the understanding of the \hef\ form factor, 
fig.~\ref{dfche4gari} displays  the result of an early calculation by Gari \et\
\cite{Gari76a}. This calculation was performed using the exp(S)-approach,
including up to 4-particle correlations. The curve shown was calculated
using the RSC N-N interaction together with a phenomenological three-body
force. The two-body contributions were calculated according to the approach by Gari and Hyuga
\cite{Gari76}. The main discrepancy to the data occurred in the region of the
diffraction maximum. The comparison to modern calculations (fig.~\ref{dfche4v18})
shows that both the IA contribution and the two-body contributions have 
increased by nearly a factor of 2, due to both a better ground-state wave function and a more
consistent calculation of the two-body processes.
In fig.~\ref{dfche4gari} we also show the results of an older calculation 
of Schiavilla \et~ \cite{Schiavilla90} obtained using the Argonne V14 plus Urbana-VII TNI 
interactions. A similar observation can be made for this calculation.

The data have also been used by Sick \cite{Sick82} to determine an accurate charge 
radius, with the goal to compare it to a precise measurement from the 2p-2s transition 
in muonic helium by Carboni \et\ \cite{Carboni77}. For \hef\ such an analysis can be done 
by including information from other observables on the large-radius behaviour of 
the proton radial wave function. This large-$r$ behaviour has a considerable 
influence on the rms-radius, but is not very well determined by
electron scattering as at small $q$, where its effect shows up, the 
experimentally determined signal, the difference of the form factor to one,
becomes small and affected by systematic errors.

For \hef\ the shape of the proton radial wave function is calculable knowing
the proton separation energy. The absolute normalization is known from a 
dispersion-relation analysis \cite{Plattner75} of the world-data on proton-helium 
elastic scattering. This constraint helps to extract from the (e,e) data
a very precise charge rms-radius, 1.676 $\pm$ 0.008 $fm$, which agrees with the
value measured using muonic atoms.  

The calculation corresponding to fig.~\ref{dfche4v18} of Schiavilla \et\ 
\cite{Schiavilla00} 
gives a charge radius of \hef\ of 1.609 $fm$. Once the contribution of two-body
currents is added, this increases to {1.634 $fm$.} In comparison with the experimental
result, this is a bit low.

Overall, the understanding of the \hef\ form factors is quite satisfactory.
This shows that the calculations of the isoscalar two-body contributions are 
quite successful even in dense systems, although the theoretical understanding 
of these terms still leaves to be desired. Other open questions on the 
structure of \hef, such as the size of the D-state contribution or the role
played by $^1 S_0$ {\em vs} $^3 S _1$ pairing, are better investigated by studying
breakup observables. 

\section{A$>$4}
The region of ''light nuclei'' up to a few years ago certainly did not go 
beyond mass number A=4: for heavier nuclei calculations of accuracy comparable to A$\leq$4
could not be performed. In this situation, one had to take recourse to 
more approximate, and much less fundamental, approaches such as the shell 
model, cluster models {\em etc}. Some of these calculations have reached
quite a high standard (see {\em e.g.}\cite{Zheng95}), but these approaches  
do not provide the direct link to the underlying N-N interaction. 

This situation has changed during the last years. Today, Greens-Function 
Monte Carlo (GFMC) calculations can be performed for nuclei up to mass A=9
\cite{Pudliner95,Pudliner97}, and Variational Monte Carlo (VMC) calculations 
have been performed up
to A=16 \cite{Forest96,Pieper90,Pieper92}. A recent exp(S)-calculation also reaches A=16 
\cite{Mihaila00}. In 
these calculations, the nuclear wave function can be obtained starting
from the N-N interaction, with no additional model assumptions. Despite the
daunting numerical effort required and some approximations that do have to
be made, the results seem to indicate that a fairly accurate understanding 
is being reached for these not-so-light nuclei.

It is of great interest to study the p-shell nuclei in terms of ''exact''
methods. For the heavier p-shell nuclei, at least, the shell model already
provides a reasonable description; in this case the comparison between the
''exact'' calculations and the model-calculations allows us to assess the 
accuracy and potential failures of the latter, and herewith lets us 
gain most valuable
insight needed when dealing with even heavier nuclei where ''exact''
calculations are out of question for quite some time to come.

GFMC calculations for nuclei up to A=9 have been performed by Pudliner \et\ 
\cite{Pudliner95,Pudliner97}. These calculations employed the AV18 N-N interaction 
together with the Urbana IX three-body force. For the binding energies, these 
calculations reach accuracies of the order of \mbox{1$MeV$} out of 40$MeV$. The calculations 
show that the N-N correlations continue to play an important role even for the
heavier nuclei; the two-body densities for these heavier nuclei actually
at N-N distances $<2~fm$ are quite similar to the ones found for the deuteron. 

\begin{figure}[htb]
%Figur mit topp/sideways hergestellt, modif boundingBox: 66 206 516 556
\centerline{\mbox{\epsfysize=9cm \epsffile{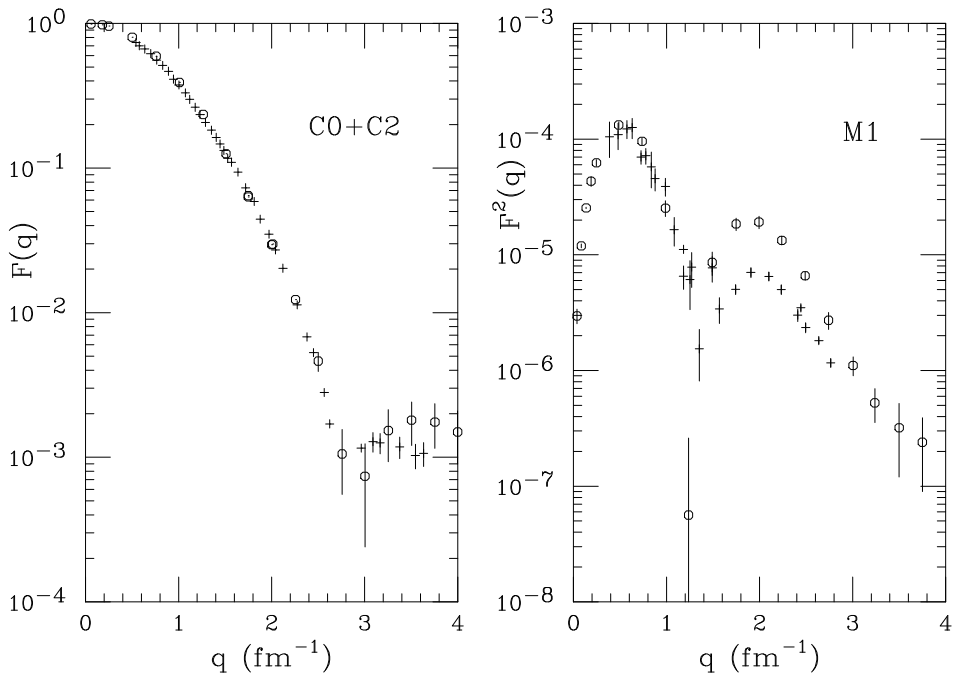}}}
\begin{center} \parbox{14cm}{\vspace*{-0.5cm} \caption{
\label{fltli6} 
$^6$Li form factors calculated by VMC by Wiringa and Schiavilla ($\circ$) 
\protect{\cite{Wiringa96}} together
with data for C0+C2 \protect{\cite{Li71,Bumiller72}} and M1 
\protect{\cite{Rand66,Lapikas78,Bergstrom82}} (+). 
} }  \end{center}
\vspace*{-0.4cm} \end{figure}
\begin{figure}[bh]
%Figur mit topp/sideways hergestellt, modif boundingBox: 66 206 516 556
\centerline{\mbox{\epsfysize=8cm \epsffile{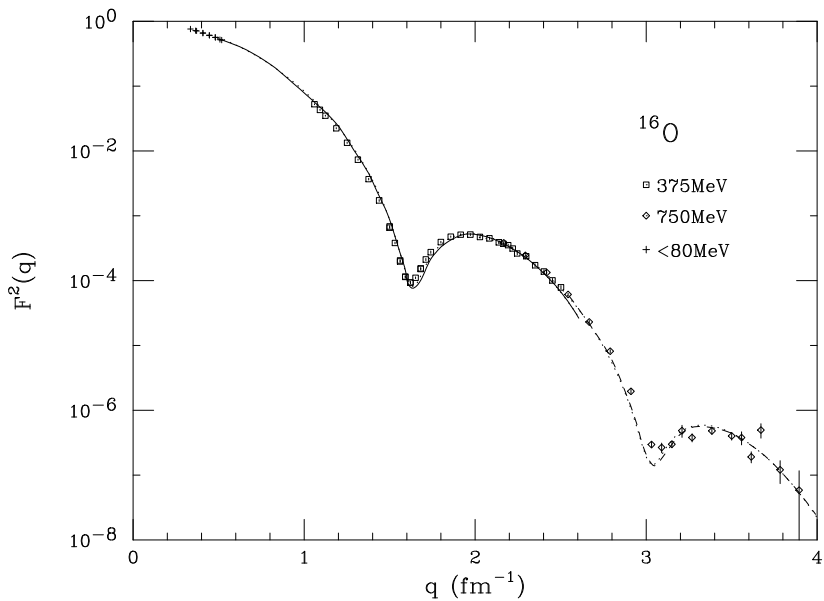}}}
\begin{center} \parbox{14cm}{\vspace*{-0.5cm} \caption{
\label{fo16} 
$Exp(S)$ calculation of the oxygen effective form factor ($\sigma/\sigma_{Mott}$)
 by Mihaila and Heisenberg 
\protect{\cite{Mihaila00}}, compared to the data of Sick and McCarthy 
\protect{\cite{Sick70}} and Bentz  \protect{\cite{Bentz71}}.  
} }  \end{center}
\vspace*{-0.4cm} \end{figure}

Only for some of the p-shell nuclei and selected states have the electromagnetic
form factors been calculated. In these calculations the two-body currents have
been included in the way discussed in sect. \ref{MEC}, {\em i.e.} as far as possible
consistently with the underlying N-N interaction. 
In figs.~\ref{fltli6},\ref{fo16} we show two examples. 
For the $^6 Li$ charge form factor the  results calculated using VMC 
agree quite well with the data of G.C. Li \et\ \cite{Li71} measured at 
Stanford, and Bumiller \et\ \cite{Bumiller72} and Suelzle \et\ \cite{Suelzle67}
at the lower momentum transfers.
The calculations include the contributions from 
two-body currents which have an effect only near the diffraction 
minimum where they shift it to somewhat smaller $q$. Both the data and the 
calculation correspond to the sum of the $C0$ and $C2$ multipolarities
($^6 Li$ has spin one)  which have not been separated.  The agreement between
calculated (2.59$ \pm $0.02$fm$) and experimental (2.56$\pm$0.05$fm$) charge 
{\em rms}-radius is quite good. 

The magnetic form factor of $^6 Li$ has first been measured by Rand \et\ 
\cite{Rand66}; the 180$^\circ$ degree data, which reach 1.4$fm^{-1}$, 
still had large statistical errors. Precise data were obtained by Lapikas \et\
\cite{Lapikas78} at low $q$, and Bergstrom \et\ \cite{Bergstrom82} by an
experiment performed using the 180$^\circ$ setup at Bates. 
The $^6 Li$ magnetic form factor calculated by Wiringa and Schiavilla, 
displayed in fig. \ref{fltli6}b, does not show as  good an
agreement with the data for momentum transfers $q>1 fm^{-1}$. This discrepancy 
with the data is not due to two-body currents, which have a relatively minor 
effect as $^6 Li$ is an isoscalar system; one therefore must assume that it is
related to deficiencies in the VMC wave function.  

For Oxygen, Mihaila and Heisenberg \cite{Mihaila00} have recently used the 
$exp(S)$ coupled-cluster expansion approach. The procedure is similar to the
$exp(S)$ method used at the time by Gari \et\ \cite{Gari76} 
(see fig.~\ref{dfche4gari}) who, in addition to \hef, also considered heavier magic 
nuclei such as oxygen and calcium.
In the coupled cluster approach, the nuclear wave function is expanded in the
many-body Hilbert space in terms of multiconfigurational creation- and 
destruction operators. The expansion coefficients carry the information on the 
nuclear correlations, and no distinction between short- and long-range 
correlation needs to be made. The wave function is expanded in a 
50$\hbar \omega$ space, and relative angular momenta up to l=11 are taken 
into account. In this calculation, the 
Argonne V18  N-N interaction together with the Urbana IX three-body force is
employed.  Two-body currents
are also included, and have a moderate effect at momentum transfers above 2 $fm^{-1}$. 
The Coulomb distortion is accounted for by Fourier-transforming the
calculated  form factor to $r$-space, and then calculating the Coulomb-distorted
cross section (effective form factor) using a phase-shift code.

The calculation is compared in fig.~\ref{fo16} to data \cite{Sick70} obtained
in an experiment that used a windowless ice-target at LN$_2$ temperature. 
%This target had been developed to allow for measurements of backward-angle data in 
%transmission geometry \cite{Sick69} and, 
%which allowed a background-free measurement
%of the small elastic form
%factors occurring at the high $q$'s. 
The low-$q$ data (0.3 -- 0.5 $fm^{-1}$) have been measured at Darmstadt by 
Bentz \cite{Bentz71}. 

The agreement between calculation and experiment is amazingly good. 
 In $r$-space, somewhat larger deviations are visible as the calculated 
$rms$-radius of 2.62$fm$ \cite{Mihaila00b} is smaller than the experimental 
one of 2.712$\pm$0.005$fm$. It will be
interesting to see whether this $exp(S)$ approach of Mihaila and Heisenberg
can with similar success predict the properties of other p-shell nuclei. 

\section{Conclusions and outlook}
In this paper we have reviewed the present status of the elastic 
form factors of nuclei with mass number A=2,3,4,6 and 16, both in terms of 
the cross section data
available, the form factors extracted and their theoretical understanding.
For the case of the deuteron, we have included the transition form 
factor to the singlet-S state. 

We have witnessed during the last years
important progress in the theory used to calculate few-body wave functions
and the corresponding electromagnetic form factors. The ''nuclear standard 
model'' --- based on the assumption that nuclei are basically non-relativistic
systems containing nucleons bound by the N-N interaction known from N-N
scattering --- is surprisingly successful. For the nuclei where the 
computational techniques are well under control, basically $A \leq 4$, 
the agreement between calculated ground state properties, and form factors
in particular, is very satisfactory. 

The most important uncertainties for $A \leq 4$ still result from the 
poorly constrained nature of the N-N interaction off-shell. There, different
N-N potentials give quite different answers. The ensuing differences mainly 
affect D-state properties.  The same quantities are somewhat uncertain 
as the data base that fixes the S-D transition in the NN-system, represented
by the 
$\epsilon_1$-parameter, is still not very satisfactory and leaves too
much freedom for the N-N potentials. Here, only double-polarization 
experiments on $\vec{n}-\vec{p}$ elastic scattering can be expected to 
lead to significant progress in the future.   

For the calculation of the electromagnetic form factors, it also has been 
very important to get the contributions of two-body terms, 
required by gauge invariance,  under control.
Much progress in this direction has been made, as many of the two-body 
terms now can be 
derived consistently with the N-N interaction employed. Diagrams not 
constrained by current conservation occasionally still involve significant
uncertainties, though. 

One obvious avenue for a deeper understanding of light nuclei consist in 
a relativistic treatment. Given the typical momenta of nucleons
in nuclei, a relativistic description is clearly desirable.
A number of different approaches have been
developed already, and we have addressed some of them when discussing the 
deuteron. For the heavier systems, the covariant calculations up to now have
provided little in terms of form factors, the quantities that yield the
perhaps most detailed test.  The covariant
approaches also give us a better control over the off-shell behaviour 
and thus have the potential to partly do away with one ingredient of the
non-relativistic calculations that still is somewhat unsatisfactory, the 
(phenomenological) three-body interaction. In this area, significant 
advances can be expected during the coming years.    

On the side of experiment, good progress has also been made. In 
particular, we now have a reasonably complete set of polarization data for
electron-deuteron scattering that allows us to separate the deuteron monopole- 
and quadrupole
form factors. For a long time, we had to contend with the unseparated
structure function $A(q)$, which is much less sensitive to the underlying 
physics.

In this review, we have  used for the nuclei A=2--4 the form factors resulting 
from a  fit to the {\em world} data, such as to base the discussion on 
the most precise experimental information presently available. The numerical
values of the various form factors, which partly have been determined in the
course of writing this review, are available upon request.   

While our knowledge of the proton charge- and the neutron magnetic form 
factors has greatly improved during the last years,  the neutron charge form 
factor $G_{en}$ remains a continued source of uncertainty for the 
understanding of few-body charge form factors.  The first double-polarization
experiments now have produced results and lead to values of $G_{en}$ that
are more reliable than what was available up to now, but the uncertainties on
$G_{en}$ are still comparatively large, and the information is limited to
moderate values of the momentum transfer. Here, important further 
progress is required in the future. With the CW-facilities today available
and the progress in polarized target and polarimeter design, 
the necessary improvements are within  reach. 

As far as the heavier nuclei go, it is very satisfactory to see the emergence 
of ''exact'' calculations (in the 
sense of the nuclear standard model) for p-shell nuclei. In the p-shell, 
these calculations can be used to ''calibrate'' the mean-field 
calculations that are standard (and often the only feasible  ones) for heavier 
nuclei.  

When looking at the overall comparison of theory and experiment, a few 
notable discrepancies remain. Even the most modern calculations do not very 
well describe the \het\ magnetic form factor; the reason for this is not known
and cannot easily be blamed on the two-body contributions which for this 
observable are largely of the model-independent kind.  The comparison between
the positions of the minima, and amplitudes of the form factor maxima, 
for the A=2,3 systems also shows a persistent discrepancy. It is not clear
if this can be taken as a signal for ''new physics'' beyond the nuclear 
standard model. It is worrisome as well that the calculations do not yet 
accurately  explain the static properties ($Q, \mu$) of the simplest nucleus, 
the deuteron.

It would evidently be of great interest to go to larger momentum transfer for some
of the form factors. An extension of the \het\ magnetic form factor to larger 
$q$ will be carried out at JLAB, a significant extension for the two Tritium form factors
is clearly feasible with available liquid-target technology. 
Pushing the $T_{20}$ measurements for 
the deuteron to higher $q$ will presumably require the development of 
a more efficient polarimeter of as yet unknown type.

Several aspects have received little attention in this review. Changes of the
nucleon upon integration in the nucleus have not been discussed, essentially 
because there is little experimental evidence for it. We have also only
occasionally discussed an understanding of the light nuclei in terms
of quark constituents; the description in terms of nucleons is very 
successful, and clear signals for the need of a description in terms of
the underlying quark degrees of freedom are hardly available. These signals
probably only can be found by going to larger momentum transfers where the 
conventional description of light nuclei in terms of nucleonic constituents
interacting via the N-N force must break down. \\[3mm]
{\bf Acknowledgements} \\[3mm]
Many people have contributed to this review. We want to thank in particular
H. Arenh\"ovel, J. Carlson, T.W. Donnelly, J. Golak, H. Henning, J. Jourdan, 
D. Rohe, P. Sauer, R. Schiavilla, W. VanOrden and R. Wiringa
who have provided help, advice or numerical results.
This work was supported by the Schweizerische Nationalfonds.

%\bibliography{rev,rev1,sick_diff,rmsdp,ymeyer3:sick}
\end{document}